\documentclass[a4paper,fleqn,11pt]{article}
\pdfoutput=1 
\usepackage{jheppub} 

\usepackage[T1]{fontenc} 

\title{Mechanical structure of the nucleon and the baryon octet: Twist-2 case}
\usepackage{graphicx}  
\usepackage{dcolumn}  
\usepackage{bm}          
\usepackage{slashed}
\usepackage{cancel}
\usepackage{float}
\usepackage{mathtools}
\usepackage{amsmath,amsfonts,amssymb,amscd,amsxtra,amsthm}
\usepackage{amsbsy}
\usepackage{amstext}

\usepackage[utf8]{inputenc} 
\usepackage{booktabs} 
\usepackage[normalem]{ulem} 
\usepackage[dvipsnames]{xcolor} 
\usepackage{tabularx}
\usepackage{enumitem}  
\usepackage{array} 
\usepackage{slashed}
\usepackage{tikz}
\usepackage{multirow}
\usepackage{cancel}
\usepackage{physics}
\usepackage{kotex}

\renewcommand\sout{\bgroup \color{red} \ULdepth=-.5ex \ULset}

\makeatletter

\preprint{INHA-NTG-05/2023}

\author[a,b]{Ho-Yeon Won}
\author[a,c]{Hyun-Chul Kim}
\author[d]{June-Young Kim}

\affiliation[a]{Department of Physics, Inha University, Incheon 22212,
 Republic of Korea}
 \affiliation[b]{CPHT, CNRS, École polytechnique, Institut
   Polytechnique de Paris, 91120 Palaiseau, France}
\affiliation[c]{School of Physics, Korea Institute for Advanced Study 
   (KIAS), Seoul 02455, Republic of Korea}
\affiliation[d]{Theory Center, Jefferson Lab, Newport News, VA 23606, USA}

\emailAdd{hoyeon.won@polytechnique.edu}
\emailAdd{hchkim@inha.ac.kr}
\emailAdd{jykim@jlab.org}

\abstract{
We investigate the gravitational form factors (GFFs) of the nucleon
and the baryon octet, decomposed into their flavor components,
utilizing a pion mean-field approach grounded in the large $N_c$
limit of Quantum Chromodynamics (QCD). Our focus is on the
contributions from the twist-2 operators to the flavor-triplet and
octet GFFs, and we decompose the mass, angular momentum, and $D$-term 
form factors of the nucleon into their respective flavors. The
strange quark contributions are found to be relatively mild for the
mass and angular momentum form factors, while providing significant
corrections to the $D$-term form factor. In the course of examining
the flavor decomposition of the GFFs, we uncover that the effects of
twist-4 operators play a crucial role. While the gluonic
contributions are suppressed by the packing fraction of the instanton 
vacuum in the twist-2 case, contributions from twist-4 operators
are of order unity, necessitating its explicit consideration. 
}
\begin{document} 

\maketitle
\flushbottom

\section{Introduction \label{sec:1}}
Strangeness in the nucleon has been one of the most crucial issues 
in comprehending the underlying structure of the nucleon. 
Since the European Muon Collaboration (EMC) announced the puzzling 
measurement that the quark intrinsic spin provides only a small
portion of the proton's spin~\cite{EuropeanMuon:1987isl,
  EuropeanMuon:1989yki}, there has been a great amount of experimental
and theoretical works (see a review~\cite{Aidala:2012mv} and
references therein). It is now known that the quark intrinsic spin carries
approximately 35~\% of the proton's spin~\cite{Aidala:2012mv}. The
rest will come from the orbital angular momenta of the quarks and the total angular momentum of the gluons inside a proton. The EMC results triggered an idea to measure
the strange contributions to the electromagnetic form factors (EMFFs)
of the proton~\cite{Kaplan:1988ku}, and the strange vector form
factors were extracted from parity-violating electron-proton
scattering ~\cite{Beise:2004py, A4:2004gdl, Maas:2004dh,
  HAPPEX:2005qax, G0:2005chy, G0:2009wvv} and theoretically (refer to
a recent review and references therein for further
details~\cite{Maas:2017snj}). Although the strange magnetic moment is
relatively small, it remains significant. For instance, the strange
magnetic form factor at $Q^2\simeq 0.1,\mathrm{GeV}^2$ was determined
to be $0.30\pm 0.17$~\cite{Maas:2017snj}. Additionally, the $\pi N$
sigma term, which contributes to the nucleon mass, incorporates  
contributions from the strange quark. Specifically, approximately
$20~\%$ of the $\pi N$ sigma term is attributed to the strange 
quark~\cite{Borasoy:1996bx}. Furthermore, investigations have been
carried out to explore the strange-quark contribution  
to the nucleon tensor charge~\cite{Kim:1996vk, Ledwig:2010tu}.  

The role of strange quarks can also extend to the gravitational
form factors~(GFFs)~\cite{Kobzarev:1962wt,Pagels:1966zza} of the
nucleon, which provide crucial insights into the properties of the
nucleon, including its mass, spin, mechanical pressure, and shear
force~\cite{Ji:1996ek, Polyakov:2002yz}. Although the concept of the
GFFs was introduced about sixty years ago~\cite{Kobzarev:1962wt,
  Pagels:1966zza}, experimental access to them had
been limited, so that they were regarded as a purely academic
interest. However, the emergence of generalized parton distributions  
(GPDs)~\cite{Muller:1994ses, Ji:1996nm, Radyushkin:1996nd,
  Radyushkin:1996ru} has paved the way for extracting experimental
information on the GFFs. It is possible to measure the obesrvables
related to the GFFs because the EMFFs and GFFs can be understood as
the first and second Mellin moments of vector GPDs,
respectively~\cite{Ji:1996ek}.   

To perform the flavor decomposition of the GFFs, we need to consider
the flavor triplet and octet energy-momentum tensor (EMT) currents,
which are not conserved~\cite{Ji:1995sv}. Consequently, they can not
be derived from the N\"other theorem. In QCD, they can still be
constructed by using the conserved singlet EMT current as a
guide~\cite{Polyakov:2018zvc}, which consist of the twist-2 and
higher-twist terms. Moreover, the gluonic degrees of freedom  
come into crucial play. While the gluonic contribution is suppressed
by the packing fraction of the instanton vacuum in the case of the
twist-2 operators, we find that the contributions from twist-4
operators are of order unity~\cite{Balla:1997hf, Diakonov:1995qy}. It
implies that, without the gluonic contributions considered, it is not
possible to perform a complete analysis of the flavor decomposition of
the GFFs. In ths current work, thus, we focus on the twist-2
contributions to the flavor decomposition of the GFFs and distributions,
and discuss why the twist-4 operators must be taken into
account. Since it is technically far involved in deriving the
effective twist-4 operators corresponding to the QCD operators, we
leave it as a future work and concentrate on the role of the twist-2
operators in the flavor decomposition of the GFFs and corresponding
distributions, and discuss why the twist-2 operators only are not
sufficient for interpreting the flavor-decomposed distributions of the
nucleon as mechanical properties. 

Bearing in mind the problems posed above, we use the chiral
quark-soliton model~($\chi$QSM), which was developed based on 
large $N_c$ QCD~\cite{Witten:1979kh, Witten:1983tx}. 
In the large $N_c$ limit of QCD, a classical baryon
can be regarded as $N_c$ valence quarks bound by a mesonic mean field
that arises as a classical solution of the saddle point equation in a
self-consistent manner, while the quantum fluctuations are suppressed
and of order $1/N_c$. Since the classical baryon has no good quantum
numbers, the zero-mode quantization is required to restore the
translational and rotational symmetries. 
These rotational and translational zero modes naturally give rise to 
the standard SU(2$N_f$) spin-flavor symmetry in the large $N_c$ limit
of QCD~\cite{Dashen:1993as, Dashen:1993jt, Dashen:1994qi}.

The $\chi$QSM as a chiral theory of the nucleon was originally derived
from the QCD instanton vacuum~\cite{Diakonov:1985eg,
  Diakonov:2002fq}. The low-energy effective partition function of QCD
was be obtained, which realizes the spontaneous breakdown of chiral
symmetry, and satisfies the relevant low-energy theorems. It is
important to note that the gluon degrees of freedom have been
integrated out through the instanton vacuum, and their effects are
incorporated into the momentum-dependent dynamical quark mass $M$. In
the $\chi$QSM, we switch off the momentum dependence of $M$ and
introduce a regularization to tame the divergent quark loops.
The $\chi$QSM has been successful in describing the breakdown of the
Gottfried sum rule~\cite{Blotz:1995tj, Pobylitsa:1998tk}, the
light-flavor asymmetry~\cite{Diakonov:1996sr, Diakonov:1997vc,
  Wakamatsu:1998rx} of polarized parton distribution functions (PDFs), and the transversity
distributions~\cite{Kim:1995bq, Schweitzer:2001sr, 
  Wakamatsu:2000fd}. It has also provided a satisfactory explanation
for the contributions of strange quarks  to axial
charges~\cite{Blotz:1993am, Blotz:1994wi} and vector
charges~\cite{Kim:1995hu, Silva:2001st}. For a comprehensive overview,
refer to the reviews~\cite{Christov:1995vm, Wakamatsu:1990ud}.  

Finally, we want to address the issue of 
three-dimensional (3D) mechanical
interpretation of the GFFs. The 3D EMT distributions have been
proposed as the Fourier transform of the GFFs~\cite{Polyakov:2002yz,
  Goeke:2007fp}. However, the 3D interpretation of the EMT
distributions has faced significant criticism~\cite{RevModPhys.29.144,
  Burkardt:2000za, Burkardt:2002hr, Miller:2007uy, Jaffe:2020ebz}.  
Due to the inability to precisely localize the nucleon wave packet
below the Compton wavelength, there are ambiguous relativistic
corrections to the 3D distributions~(see Ref.\cite{Belitsky:2005qn}).  
To deal with this ambiguity, a two-dimensional~(2D) light-front~(LF)
distribution has been used~\cite{Burkardt:2000za, Burkardt:2002hr,
  Miller:2007uy}. In the current work, however, we adhere to the 3D
interpretation of the EMT distributions. 

As discussed in a number of works considering the large $N_c$
limit~\cite{Lorce:2022cle, Kim:2023xvw}, the center of motion of the
nucleon exhibits a non-relativistic behavior (while the nucleon itself
possesses fully relativistic internal dynamics). Therefore,
information about the 3D distributions is conveyed into the 2D space
on the light cone with no change. The frame dependence of these
distributions has been explored in Ref.~\cite{Lorce:2022cle}  
in the context of the large $N_{c}$ limit. In addition, it was very
recently found that the 3D components of the EMT can be
matched with the 2D light-front components~\cite{Kim:2023xvw}. While
considering the admixture of the 3D components in this matching, the
 Melosh rotation (IMF Wigner rotation~\cite{Chen:2022smg}) effects under the Lorentz boost are
suppressed in the large $N_c$ limit. Consequently, the light-front
helicity state becomes equivalent to the canonical spin state at rest.  

The structure of the current paper is as follows: In
section~\ref{sec:2}, we recapitulate a definition of the
Belinfante-Rosenfeld EMT current in QCD and express the matrix element
of the EMT current, which is parametrized in terms of the GFFs.
We decompose the EMT current in terms of the twist, and corresponding
matrix elements. We also discuss the 3D mechanical interpretations
associated with these GFFs. In section~\ref{sec:3}, we offer a brief
explanation of the $\chi$QSM and illustrate the spin-flavor properties
of the GFFs in flavor SU(3) symmetry. In section~\ref{sec:4}, we
discuss the numerical results on the 3D EMT
distributions. Furthermore, we present the flavor-decomposed GFFs for
the baryon octet using the spin-flavor symmetry, and discuss the
incompleteness of the current scheme. Finally, in section~\ref{sec:5}
we provide a summary of our work and draw conclusions based on our
findings.   

\section{QCD energy-momentum tensor \label{sec:2}}

According to Ji's decomposition~\cite{Ji:1996nm} (see also
Refs.~\cite{Leader:2013jra, Lorce:2017wkb}), the quark ($q$) and gluon
parts ($g$) of the Belinfante-Rosenfeld-type QCD EMT currents 
are expressed as 
\begin{align}
    T^{\mu \nu}_{q}  
& = \frac{i}{4} 
    \bar{\psi}_{q} 
    \left( 
    \gamma^{ \{\mu} \overleftrightarrow{\mathcal{D}}^{\nu \}}
    \right) 
    \psi_{q}, 
    \quad    
    T^{\mu \nu}_{g}  
  = -F^{ \mu \rho,a} F^{\nu,a  }_{\ \rho} 
  + \frac{1}{4} g^{\mu \nu} F^{ \lambda \rho, a} F^{a}_{\lambda \rho},
\end{align}
where $\overleftrightarrow{\mathcal{D}}^{\mu} =
\overleftrightarrow{\partial}^{\mu} - 2 i g A^{\mu}$ is the covariant
derivative with $\overleftrightarrow{\partial}^{\mu} = 
\overrightarrow{\partial}^{\mu}- \overleftarrow{\partial}^{\mu}$, and
$a^{ \{ \mu }b^{ \nu \} } = a^{\mu} b^{\nu} + a^{\nu}
b^{\mu}$. $F^{b,\mu \nu}$ is the gluon field strength, where the superscript $b$ indicates the color index.
This EMT current
consists of the quark $(q)$ and gluon $(g)$ parts, and is a conserved
quantity:  
\begin{align}
    T^{\mu \nu} 
  = \sum_{q} T^{\mu \nu}_{q} 
  + T^{\mu
    \nu}_{g},  \quad \partial_{\mu}T^{\mu\nu} = 0.
\end{align} 
However, if we consider the separate quark and gluon EMT currents, they
are not conserved anymore. 

In addition, we will discuss the twist-projected EMT current and
emphasize the importance of the higher-twist part in the mechanical 
interpretation of the EMT distributions. The symmetric part of the EMT 
current can be divided into the twist-2 (spin-2) and twist-4 (spin-0)
parts, which are associated with the leading twist vector GPDs and
twist-4 GPDs, respectively. These twist-projected EMT currents are
given by 
\begin{align}
  T^{\mu \nu}_{a} =  \bar{T}^{\mu \nu}_{a} + \hat{T}^{\mu \nu}_{a},
  \label{eq:tw2_tw4_split}
\end{align}
where the twist-2 ($\bar{T}^{\mu \nu}_{a}$) and twist-4 ($\hat{T}^{\mu
  \nu}_{a}$) parts are defined by 
\begin{align}
    \bar{T}^{\mu \nu}_{a}  
 = T^{\mu \nu}_{a}  - \frac{1}{4} g^{\mu \nu}  T^{\alpha}_{a, \alpha},
  \quad     \hat{T}^{\mu \nu}_{a}   
 = \frac{1}{4} g^{\mu \nu}  T^{\alpha}_{a, \alpha},
\end{align}
with $a=q,g$. Note that the twist-3 part is related to the asymmetric
EMT current, which is not discussed in this work.

\subsection{Matrix element of the energy-momentum tensor current} 

The matrix element of the EMT current can be described 
by four independent Lorentz-invariant functions, 
namely $A^{a}$, $J^{a}$, $D^{a}$, and $\bar{c}^{a}$, 
which are obtained by considering all possible Lorentz structures 
and sorting them out by using the discrete symmetries 
(hermiticity, time reversal, and parity). This parameterization has
been studied extensively in previous works~\cite{Pagels:1966zza,
  Kobzarev:1962wt, Kobsarev:1970qm, Ng:1993vh}  
(For further information on the generalization of this
parametrization, interested readers may refer to
Refs.~\cite{Cotogno:2019vjb, Kim:2022bwn}). The baryon matrix element
of the EMT current is expressed as    
\begin{align}
&   \mel{B(p',J'_{3})}
    {   T_{\mu\nu}^{a}  ( 0 )  }
    {B(p,J_{3})} 
  = \bar{u}(p',J'_{3})
    \Bigg[
    A^{a}_{B}  ( t ) \frac{  P_{\mu} P_{\nu} }{  M_B } 
  + J^{a}_{B}  ( t ) \frac{  i  P_{ \{ \mu} \sigma_{\nu \} \rho}
  \Delta^{\rho} } { 2 M_B }  \cr 
&   \hspace{5.2cm}
  + D^{a}_{B}  ( t )  \frac{ \Delta_{\mu} \Delta_{\nu} -  g_{\mu\nu}
  \Delta^{2}}{4M_B}      
  + \bar{c}^{a}_{B} ( t ) M_B g_{\mu\nu}  
    \Bigg] 
    u(p,J_{3}),
\label{eq:4}
\end{align}
where $A^a$, $J^a$, $D^a$, and $\bar{c}^a$ are called the mass,
angular momentum, $D$-term, and cosmological constant term form
factors of a baryon $B$, respectively. When considering the separate
operators for quarks and gluons, they are no more conserved. This
leads to the appearance of an additional form factor, denoted by
$\bar{c}$ in the flavor decomposition. As a consequence,
the form factors of individual quarks and gluons exhibit both scale
and scheme dependence. For brevity, the dependence of the
renormalization scale of the individual quark and gluon GFFs is not
indicated in this work. The normalization of the one-particle state
for the baryon is expressed  
as $\langle B' (p', J'_{3})| B (p, J_{3}) \rangle = 2p^{0} (2\pi)^{3}
\delta_{J'_{3} J_{3}} \delta^{(3)}(\bm{p}'-\bm{p})$, 
where $J_{3}$ and $J'_{3}$ denote the spin polarizations of the initial 
and final states, respectively. 
The $M_B$ represents the mass of a baryon, while $p$ and $p'$ refer to
the initial and final momenta, respectively. We define
$P=\left(p'+p\right)/2$ and $\Delta=p'-p$, where $\Delta^{2}=t$, to
represent the average momentum and the momentum transfer between the
initial and fianl states, respectively. We express the GFFs
generically as $F^{\chi}_{B}$, where the flavor indices run over
$\chi=0,3,8$. They can be decomposed in terms of the quark components
\begin{align}
F^{\chi = 0}_{B}=F^{u}_{B}+F^{d}_{B}+F^{s}_{B}, \quad F^{\chi =
  3}_{B}=F^{u}_{B}-F^{d}_{B}, \quad F^{\chi =
  8}_{B}=\frac{1}{\sqrt{3}}\left(F^{u}_{B}+F^{d}_{B}
  -2F^{s}_{B}\right). 
\label{eq:Decomp}
\end{align}
Thus, the GFFs of a baryon are given by the sum of all quark and gluon
contributions 
\begin{align}
\sum_{a=q,g} F^{a}_{B}(t) = F_{B}(t), \quad \bar{c}_{B}(t)=0.
\end{align}
Note that the current conservation imposes the constraint that
$\bar{c}(t)$ is zero. In this respect, the $\bar{c}$ form factor
should carefully be considered when we do the flavor
decomposition, or look into the quark and gluon subsystems. 

In fact, extracting the $\bar{c}$ form factor is challenging in most
dynamical models and lattice QCD due to the higher-twist (twist-4)
corrections. To isolate this twist-4 term in a most systematic way, we can
decompose the symmetrized EMT current in terms of the twist-2 and
twist-4 components, as presented in Eq.~\eqref{eq:tw2_tw4_split}. With
these twist-projected EMT currents employed, we define their nucleon
matrix elements as follows: 
\begin{align}
&   \mel{B(p',J'_{3})}
    {   \bar{T}_{\mu\nu}^{a}  ( 0 )  }
    {B(p,J_{3})} 
  \cr
  &= \bar{u}(p',J'_{3})
    \Bigg[
    A^{a}_{B}  ( t ) \frac{  P_{\mu} P_{\nu} }{  M_B } 
  + J^{a}_{B}  ( t ) \frac{  i  P_{ \{ \mu} \sigma_{\nu \} \rho}
  \Delta^{\rho} } { 2 M_B }  
  + D^{a}_{B}  ( t )  \frac{ \Delta_{\mu} \Delta_{\nu} -  t g^{\mu \nu} }{4M_B}      
 \cr
&\hspace{1cm} - g_{\mu\nu}  
    \left\{ \frac{t}{8 M_{B}} J^{a}_{B}(t) 
  - \frac{3t}{16 M_{B}} D^{a}_{B}(t) 
  + \frac{M_{B}}{4}  
    \left(1- \frac{t}{4M_{B}^{2}}\right)
    A^{a}_{B}(t) \right\}
    \Bigg] 
    u(p,J_{3}),
\end{align}
and
\begin{align}
&   \mel{B(p',J'_{3})}
    {   \hat{T}_{\mu\nu}^{a}  ( 0 )  }
    {B(p,J_{3})} \cr
& = \bar{u}(p',J'_{3})
    \Bigg[ g_{\mu\nu}  \bigg{\{} 
    M_{B} \bar{c}_{B}^{a}(t)
  + \frac{t}{8 M_{B}} J^{a}_{B}(t) \cr
&\hspace{2.5cm}- \frac{3t}{16 M_{B}} D^{a}_{B}(t) 
  + \frac{M_{B}}{4}  
    \left(1- \frac{t}{4M_{B}^{2}}\right)
    A^{a}_{B}(t) \bigg{\}}
    \Bigg] 
    u(p,J_{3}).
\end{align}
Furthermore, since higher-twist operators are strongly constrained by
the QCD equation of motion, it is very useful to partition the QCD
operators in terms of the twist so that these constraints can
explicitly be confirmed in the nucleon matrix elements, for example, 
\begin{align}
\mel{B(p',J'_{3})}
    {   \hat{T}_{\mu\nu}^{q}  ( 0 )  }
    {B(p,J_{3})} = 0, \quad \mbox{in the chiral limit}.
\end{align}
Note that one can obviously see that the sum of the above twist-2 and
twist-4 parts is equal to the parametrization in Eq.~\eqref{eq:4}: 
\begin{align}
& \mel{B(p',J'_{3})}
    {   {T}_{\mu\nu}^{q}  ( 0 )  }
    {B(p,J_{3})} =\cr
&   \hspace{2cm}
    \mel{B(p',J'_{3})}
    {   \bar{T}_{\mu\nu}^{q}  ( 0 )  }
    {B(p,J_{3})} +\mel{B(p',J'_{3})}
    {   \hat{T}_{\mu\nu}^{q}  ( 0 )  }
    {B(p,J_{3})}.
\end{align}

\subsection{Three-dimensional distribution}

To gain insight into the mechanical interpretation of the GFFs 
in the position space, one can perform the Fourier transformation of
these form factors. This approach was first explored in
Ref.~\cite{Polyakov:2002yz} and was inspired by the concept used in
EMFFs and their charge and magnetization distributions. 
However, the interpretation of EMFFs and GFFs 
in terms of the 3D distributions has been
criticized~\cite{RevModPhys.29.144, Burkardt:2000za, Burkardt:2002hr,
  Belitsky:2005qn, Miller:2007uy, Miller:2010nz, Jaffe:2020ebz} due to
the inherent limitations imposed by the Compton wavelength, which
prevents the precise localization of the nucleon wave packet.  
Consequently, this limitation introduces ambiguous relativistic
corrections to the 3D distribution. 

One perspective suggests that if the nucleon is treated as a
non-relativistic object (where the initial and final wave packets
become equivalent and well-localized), the form
factor~\cite{Sachs:1962zzc} can be understood as a 3D distribution.  
However, if one insists on using the strict definition of the distribution, 
we can consider the following approach: if we consider the
infinite momentum frame or the light-front formalism, then ambiguous 
relativistic corrections are kinematically suppressed, effectively
rendering the system non-relativistic. However, we have to pay the
cost of losing longitudinal information, reducing the distribution to
a 2D one. Another way is to take a conceptual detour in the treatment
of 3D distributions. From a Wigner phase space
perspective~\cite{Lorce:2020onh,Chen:2023dxp, Lorce:2022jyi,
  Chen:2022smg}, the 3D distribution can be regarded as
quasi-probabilistic, reflecting the internal dynamics of the hadron,
with all ambiguous relativistic corrections encapsulated in the Wigner
distributions. Furthermore, recent developments have introduced the
definition of 3D spatial distributions  in the zero average momentum
frame~\cite{Epelbaum:2022fjc, Panteleeva:2022khw, Alharazin:2022xvp}.  

Since the 2D IMF provides clear and unambiguous definitions of EMT
distributions, we can choose to work within that frame. However, in
the context of the large $N_{c}$ limit, it is both natural and
sufficient to focus on the 3D distribution. While the internal
dynamics of the nucleon is fully described within a relativistic
framework (including all the relativistic motions of the quarks), the
center of motion of the nucleon is treated in a non-relativistic
manner due to the $1/N_{c}$ suppression. 
This means that translational corrections, such as $\bm{P}^{2}/2M_{N}
\sim \mathcal{O}(N^{-1}_{c})$, to the nucleon energy are
parametrically suppressed, and the same suppression applies to the
nucleon GFFs. Consequently, the soliton nature of the nucleon
is inherently static and collectively non-relativistic. 
A related discussion of this topic can be found in
Ref.~\cite{Lorce:2022cle}. Moreover, the large $N_{c}$ approximation
causes the equivalence between the light-front helicity state and the
canonical spin state at rest. This allows one to perform
{\it{matching}}~\cite{Kim:2023xvw} between the 3D components of the
EMT and the 2D LF ones. 

In the Breit frame, the quark and gluon components of the GFFs
are determined by taking the Fourier transforms of the matrix element
of the EMT current between the initial and final states of the
baryon. This definition is found in Ref.~\cite{Polyakov:2002yz}: 
 \begin{align}
 \mathcal{O}^{a,B}_{\mu\nu}(\bm{r},J'_{3},J_{3})=\int
   \frac{d^{3}\Delta}{(2\pi)^{3}2P^{0}} e^{-i\bm{\Delta}\cdot \bm{r}}
   \langle B(p',J'_{3}) | \mathcal{O}^{a}_{\mu\nu}(0) | B(p, J_{3}) \rangle,
   \label{eq:Mother_def_dis}
\end{align}
with $\mathcal{O}= \{T,\bar{T}, \hat{T}\}$. 

In the following subsections, we will mainly discuss the role of the
twist-2 and twist-4 GFFs in the 3D mechanical interpretation, focusing
on the importance of the twist-4 pieces in the EMT distributions.  For
example, when performing flavor decomposition and partitioning the
quark gluon subsystems, it is inappropriate to discuss the mechanical
interpretations in terms of the twist-2 part only, such as local
stability conditions, mechanical raidus, and mass radius.

\subsection{Mass distribution}
The temporal components $\mathcal{O}=\{T,\bar{T},\hat{T}\}$ of the EMT
currents are related to the quark and gluon contributions to the mass
distributions $f=\{ \varepsilon,\bar{\varepsilon},\hat{\varepsilon}
\}$ inside a baryon   
\begin{align}
&   f^{a}_{B}(r)  \delta_{J'_{3} J_{3}} 
 := \mathcal{O}_{00}^{a,B}(\bm{r},J'_{3},J_{3}) =M_{B}
        \int
        \frac{d^{3}\Delta}{(2\pi)^{3}} 
        e^{-i\bm{\Delta} \cdot \bm{r}} 
        F^{a}_{B}(t) \delta_{J'_{3} J_{3}},
        \label{eq:Edensity_full}
\end{align}
where the 3D mass (monopole) form factors $F=\{ \mathcal{E},
\bar{\mathcal{E}}, \hat{\mathcal{E}} \}$ are given by 
\begin{subequations}
\begin{align}
& \mathcal{E}^{a}_{B} (t) = \left[
    A^{a}_{B}(t) + \overline{c}^{a}_{B}(t) - \frac{t}{4M_{B}^{2}}
    \left(
    A^{a}_{B}(t) - 2J^{a}_{B}(t) + D^{a}_{B}(t)
    \right) 
    \right], \\
& \bar{\mathcal{E}}^{a}_{B} (t) =  \frac{3}{4}
    \left[ A^{a}_{B}(t)
  - \frac{t}{4M_{B}^{2} } 
    \left( A^{a}_{B}(t) - 2 J^{a}_{B}(t) 
  + \frac{1}{3} D^{a}_{B}(t)\right) 
    \right], \\
& \hat{\mathcal{E}}^{a}_{B} (t) = \frac{1}{4} 
    \left[ A^{a}_{B}(t) 
  + 4\bar{c}^{a}_{B}(t)
  - \frac{t}{4M_{B}^{2} } 
    \left( A^{a}_{B}(t) 
  - 2 J^{a}_{B}(t) 
  + 3 D^{a}_{B}(t)
    \right) \right].
\end{align}
\end{subequations}
 Note that we have obvious relations
\begin{align}
\mathcal{E}^{a}_{B}(t)=
  \bar{\mathcal{E}}^{a}_{B}(t)+\hat{\mathcal{E}}^{a}_{B}(t), \quad
  \varepsilon^{a}_{B}(r)=\bar{\varepsilon}^{a}_{B}(r) +
  \hat{\varepsilon}^{a}_{B}(r). 
\end{align}
In the forward limit, they are connected to the $A$ and $\bar{c}$ form
factors: 
\begin{align}
&\mathcal{E}^{a}_{B}(0) = A^{a}_{B}(0),  \quad
                \bar{\mathcal{E}}^{a}_{B}(0) =
                \frac{3}{4}A^{a}_{B}(0), \quad
                \hat{\mathcal{E}}^{a}_{B}(0) = \frac{1}{4}A^{a}_{B}(0)
                + \bar{c}^{a}_{B}(0). 
\end{align}
By integrating the spatial components of the EMT currents
($\mathcal{O}^{a,B}_{00}$) over space, the mass of a spin-1/2 baryon
at rest can be calculated as  
\begin{subequations}
\begin{align}
   & \int d^{3}r  \varepsilon^{a}_{B}(r)  
  =M_{B} \mathcal{E}_{B}(0)= M_{B} [A^{a}_{B}(0) + \bar{c}^{a}_{B}(0) ],
  \label{eq:tw2_mass_0}\\
   & \int d^{3}r  \bar{\varepsilon}^{a}_{B}(r)  
  =M_{B} \bar{\mathcal{E}}_{B}(0)= M_{B} \frac{3}{4} [A^{a}_{B}(0)  ],
\label{eq:tw2_mass}
\\
   & \int d^{3}r  \hat{\varepsilon}^{a}_{B}(r)  
\int d^{3}r  \hat{\varepsilon}^{a}_{B}(r)  
  =M_{B} \hat{\mathcal{E}}_{B}(0)= M_{B} \frac{1}{4} [A^{a}_{B}(0) + 4
     \bar{c}^{a}_{B}(0)  ], 
  \label{eq:tw2_mass_2}
\end{align}
\end{subequations}
with 
\begin{align}
\int d^{3}r f_{B}(r)= \int d^{3}r \sum_{a=q,g}f^{a}_{B}(r) =
  M_{B}\left\{1,\frac{3}{4},\frac{1}{4} \right\}, 
\end{align} 
and the normalized mass form factor $A_{B}(0)=1$, where the
contribution of $\overline{c}_{B}$ to $\varepsilon_{B}$ is zero due to
the conservation of the EMT current. However, the
equations~\eqref{eq:tw2_mass_0} and \eqref{eq:tw2_mass_2} imply that
when we split the energy distributions into the quark and gluon parts,
they start to be subject to the $\bar{c}^{a}_{B}$ form factor. In
other words, the twist-2 distribution~\eqref{eq:tw2_mass} is only
independent of the $\bar{c}$ form factor. 

The size of the mass distribution can be expressed in terms of the
mass radius. Intuitively, we impose the following condition 
\begin{align}
\varepsilon^{a}_{B} > 0,
\end{align}
which can be confirmed numerically within the current work. 
This positivity of the energy distribution allows us to define the
mass radius for separate quark and gluon contributions. It is given by
either the integral of the mass distribution or the derivative of the 
form factors $6 \left[
    A_{B}(t)
  - \frac{t}{4m_{B}^{2}}   D_{B}  ( t ) 
    \right]$ with respect to the momentum squared,  
\begin{align}
   \sum_{a=q,g} \expval{r_{\mathrm{mass}}^{2}}^{a}_{B}
& =  \frac{\sum_{a=q,g}\int d^{3}r \, r^{2}  
    \varepsilon^{a}_{B}(r)}{ \sum_{a=q,g}\int d^{3}r \,
    \varepsilon^{a}_{B}(r)}   
  = 6 \frac{d}{dt}
    \left[
    A_{B}(t)
  - \frac{t}{4 M_{B}^{2}}   D_{B}  ( t ) 
    \right]_{t=0}.
\end{align}
Once we do the flavor decomposition of the mass radius, we then have
the contribution from $\bar{c}^{a}_{B}$, which must not be
neglected. 

\subsection{Angular momentum distribution}

The mixed components of the EMT current ($T^{a,B}_{0i}$)  
are associated with the total angular momentum~(AM) distributions 
[sum of spin and orbital angular momentum~(OAM)] by the Belinfante and
Rosenfeld construction. The definition of the total AM
distributions inside a baryon is given by the AM 
operator in QCD as follows: 
\begin{align}
    J^{a,B}_{i}  (\bm{r},J_{3}^{\prime},J_{3})
& :=\epsilon_{ijk} r_{j}
    T^{a,B}_{0k}  (\bm{r},J_{3}^{\prime},J_{3}) \cr
&  
  = 2\left(\hat{S}_{j}\right)_{J_{3}^{\prime}J_{3}}
    \int \frac{d^{3}\Delta}{(2\pi)^{3}}
    e^{-i\bm{\Delta} \cdot \bm{r}} 
    \left[ 
    \left(
    J^{a}_{B}(t)  
  + \frac{2}{3}t  \frac{dJ^{a}_{B}(t)}{dt}
    \right) \delta_{ij} \right. \cr
&   \hspace{4.6cm}
    \left. 
  + \left(\Delta_{i}\Delta_{j}
  - \frac{1}{3}\bm{\Delta}^{2} \delta_{ij} \right)
  \frac{dJ^{a}_{B}(t)}{dt} 
    \right]. 
\end{align}
In the following discussion, we will separate it into its monopole and
quadrupole parts. Note, however, that the quadrupole distribution is
related to the monopole distribution~\cite{Lorce:2017wkb,
  Schweitzer:2019kkd, Polyakov:2002yz}.  
For the purpose of this discussion, we refer to the monopole
distribution~\cite{Polyakov:2002yz} as the AM distribution, which  
can be expressed as follows
\begin{align}
    \rho^{a}_{J,B}(r) 
  :=\int \frac{d^{3}\Delta}{(2\pi)^{3}}
    e^{-i\bm{\Delta} \cdot \bm{r}} 
    \left[ 
    \left(J^{a}_{B}(t)+ \frac{2}{3}t
    \frac{dJ^{a}_{B}(t)}{dt}\right)
    \right].
\end{align}
Integrating both $J^{a,B}_{i}  (\bm{r},J_{3}^{\prime},J_{3})$ 
and $\rho^{a}_{J,B}(r)$ over 3D space yields the spin of the baryon as
follows 
\begin{align}
    \int d^{3}r \sum_{a=q,g} 
    J^{a,B}_{i}  (\bm{r},J_{3}^{\prime},J_{3})
  = 2 \left(\hat{S}_{i}\right)_{J_{3}^{\prime}J_{3}} 
    J_{B}(0) 
  = \left(\hat{S}_{i}\right)_{J_{3}^{\prime}J_{3}},
\end{align}
with $\rho_{J,B}(r)=\sum_{a=q,g}\rho^{a}_{J,B}(r)$. 
The AM form factor $J_{B}(0)$ is normalized to $1/2$ 
to ensure that the integral of the AM distribution 
$J^{a,B}_{i}  (\bm{r},J_{3}^{\prime},J_{3})$ 
over space is equivalent to the spin operator of a baryon. 
Note that the quadrupole component has no effect on the spin
normalization. 

The decomposition of the angular momentum into the OAM and the quark
spin requires the twist-3 component of the EMT (the antisymmetric part
of the EMT current), which is not discussed in this paper. For more
information on the separation of the OAM and the quark spin using the
QCD equation of motion, see Refs.~\cite{Lorce:2017wkb, 
  Leader:2013jra, Kim:2023pll, Kim:2024cbq}.  
In addition, the $J(t)$ form factor (angular momentum distributions)
comes from the off-diagonal components of the EMT, so it is not
affected by the twist-4 contribution. Therefore, the definition of the 3D
distributions remains intact, unaffected by the twist
classification. In other words, the flavor decomposition can be
performed unambiguously without higher-twist contribution. 

\subsection{Mechanical properties}
The spatial components of the EMT, denoted by $T^{a,B}_{ij}$, 
give information about the mechanical properties of a baryon. 
They include the pressure and shear-force distributions, i.e., $f=\{p,
\bar{p}, \hat{p} \}$ and $s$, respectively, inside a baryon. By
decomposing $\mathcal{O}=\{ T, \bar{T}, \hat{T}\}$ into irreducible
tensors, the pressure and shear-force distributions are related to
rank 0 and rank 2 tensors, respectively: 
\begin{align}
    \mathcal{O}^{a,B}_{ij}  (\bm{r},J'_{3},J_{3}) 
  = f^{a}_{B}(r) \delta^{ij} \delta_{J'_{3}J_{3}}
  + s^{a}_{B}(r)  \left(  \frac{r^{i}r^{j}}{r^{2}} - \frac{1}{3}
  \delta^{ij}  \right)  \delta_{J'_{3}J_{3}}.
\label{eq:tij}
\end{align}
The pressure and shear-force distributions are defined as
\begin{align}
&   f^{a}_{B}(r)
  =  M_{B}     \int     \frac{d^{3}\Delta}{(2\pi)^{3}}
                e^{-i\bm{\Delta} \cdot  \bm{r}}
                F^{a}_{B}(t), \cr 
&   s^{a}_{B}(r)
  = - \frac{1}{4M_{B}}  r \frac{d}{dr}  \frac{1}{r} \frac{d}{dr}
                                               \int \frac{d^{3}\Delta}{(2\pi)^{3}}
    e^{-i\bm{\Delta}\cdot \bm{r}} D^{a}_{B}(t), 
\label{eq:psdef}
\end{align}
where the pressure form factors $F=\{ \mathcal{P}, \bar{\mathcal{P}},
\hat{\mathcal{P}} \}$ are listed as
\begin{align}
    \mathcal{P}^{a}_{B}(t)
& = \left[ - \bar{c}^{a}_{B}(t)
  + \frac{t}{6M_{B}^{2} } 
     D^{a}_{B}(t)
    \right], \cr
    \bar{\mathcal{P}}^{a}_{B}(t)
& = \frac{1}{4}
    \left[ A^{a}_{B}(t)
  - \frac{t}{4M_{B}^{2} } 
    \left( A^{a}_{B}(t) - 2 J^{a}_{B}(t) 
  + \frac{1}{3} D^{a}_{B}(t)\right) 
    \right], \cr
    \hat{\mathcal{P}}^{a}_{B}(t)
& = -\frac{1}{4} 
    \left[ A^{a}_{B}(t) 
  + 4\bar{c}^{a}_{B}(t)
  - \frac{t}{4M_{B}^{2} } 
    \left( A^{a}_{B}(t) 
  - 2 J^{a}_{B}(t) 
  + 3 D^{a}_{B}(t)
    \right) \right].
\end{align}
Note that we have obvious relations
\begin{align}
\mathcal{P}^{a}_{B}(t)=
  \bar{\mathcal{P}}^{a}_{B}(t)+\hat{\mathcal{P}}^{a}_{B}(t), \quad
  p^{a}_{B}(r)=\bar{p}^{a}_{B}(r) + \hat{p}^{a}_{B}(r). 
\label{eq:relation_btw_ep}
\end{align}
At the zero momentum transfer $t\to 0$, they are reduced to
\begin{align}
    \mathcal{P}^{a}_{B}(0)
 =  - \bar{c}^{a}_{B}(0)
, \quad
    \bar{\mathcal{P}}^{a}_{B}
 = \frac{1}{4}
     A^{a}_{B}(0)
    , \quad
    \hat{\mathcal{P}}^{a}_{B}
 = -\frac{1}{4}  A^{a}_{B}(0) 
  -\bar{c}^{a}_{B}(0) .
\end{align}
The pressure and shear-force distributions can be
expressed as the Fourier transforms of the $F=\{ \mathcal{P},
\bar{\mathcal{P}}, \hat{\mathcal{P}} \}$ and $D$-term form factors 
\begin{align}
&  F^{a}_{B}(t)
  =  \frac{1}{M_{B}} \int d^{3}r \, 
    j_{0}(r\sqrt{-t}) f^{a}_{B}(r), \cr
&   D^{a}_{B}(t) 
  = 4M_{B} \int d^{3}r  \, 
    \frac{j_{2}(r\sqrt{-t})}{t} s^{a}_{B}(r). 
\label{eq:d-term2}
\end{align}
Interestingly, the twist-projected 3D multipole energy and pressure
form factors are related to each other: 
\begin{align}
    \bar{\mathcal{P}}^{a}_{B}(t)  = \frac{1}{3}
  \bar{\mathcal{E}}^{a}_{B}(t), \quad 
    \hat{\mathcal{P}}^{a}_{B}(t) = - \hat{\mathcal{E}}^{a}_{B}(t). 
\end{align}
This means that the twist-projected energy and pressure distributions
are also related to each other as follows: 
\begin{align}
\bar{\varepsilon}^{a}_{B}(r) = 3 \bar{p}^{a}_{B}(r), \quad
  \hat{\varepsilon}^{a}_{B}(r) = - \hat{p}^{a}_{B}(r). 
\end{align}
While the distributions of the gluon and quark shear forces, which
arise from the off-diagonal component of the EMT, do not depend on 
$\overline{c}^{a}(t)$ form factor, the knowledge of $\overline{c}^{a}(t)$ form factor is
required to determine the pressure distributions. Therefore, the
unambiguous definition of the 3D pressure distribution is not possible
without the twist-4 form factors considered.  

Now we are in a position to discuss the stability conditions
The distributions of the stress tensors $p_{B}$ and $s_{B}$, 
which represent the sum of each parton contribution 
($p_{B}:=\sum_{a=q,g}p^{a}_{B}$, $s_{B}:=\sum_{a=q,g}s^{a}_{B}$),  
are strongly constrainted by the conservation of the EMT current. This
constraint is expressed by the equilibrium equation:   
\begin{align} 
    \frac{\partial }{ \partial r} 
    \left(\frac{2}{3}s_{B}(r) + p_{B}(r)\right)
  + \frac{2s_{B}(r)}{r} 
  = 0,
\label{eq:diffEq1}
\end{align}
which connects the pressure distribution to the shear-force one. 
Analyzing the individual contributions of the partons to these
distributions, we discover an intriguing equilibrium equation that
relates the quark and gluon subsystems, which is expressed by the
continuity equation: 
\begin{align} 
    \sum_{a=q,g}     \partial^{i} T^{a,B}_{ij} 
  = \sum_{a=q,g} \frac{r_{j}}{r}
    \left[   \frac{2}{3}  \frac{\partial s^{a}_{B}(r)}{ \partial r}
  + \frac{2s^{a}_{B}(r)}{r}   + \frac{ \partial p^{a}_{B}(r)}{ \partial  r}
    \right] 
  = \sum_{q=u,d,s}\tilde{f}^{q}_{B,j}+\tilde{f}^{g}_{B,j} 
  = 0,
\label{eq:diffEq2}
\end{align}
where the internal force between the quarks and gluon inside the
baryon is represented as 
\begin{align}
    \tilde{f}^{a}_{B,j} 
  = - M_{B} \frac{\partial}{ \partial r^{j}} \int
    \frac{d^{3}\Delta}{(2\pi)^{3}}   e^{-i\bm{\Delta} \cdot  \bm{r}}
    \overline{c}^{a}_{B}(t).
\label{eq:diffEq3}
\end{align}
As a consequence of Eq.~\eqref{eq:diffEq2} for a mechanically stable
baryon, the sum of the internal forces $\tilde{f}^{a}_{B,j}$ between the
partons must cancel out each other. Additionally, integrating
Eq.~\eqref{eq:psdef} over space leads to a critical stability
criterion known as the von Laue stability condition: 
\begin{align}
  \int^{\infty}_{0} dr \   r^{2} p_{B}(r)=0.
  \label{eq:stability}
\end{align}
This condition implies that the pressure distribution must have at
least one nodal point where it becomes null. Furthermore, another
stability criterion, proposed in several
works~\cite{Perevalova:2016dln, Polyakov:2018rew, Lorce:2018egm}, is  
worth mentioning. Perevalova et al.~\cite{Perevalova:2016dln}
introduced a local stability criterion that states a specific
combination of the pressure and shear-force distributions must be
positive (outward) at any given distance $r$: 
\begin{align}
\frac{2}{3} s_{B}(r) + p_{B}(r) > 0.
\label{eq:mecstab1}
\end{align}
This function can be interpreted as the normal force field, while the
tangential force can be expressed as $-\frac{1}{3} s_{B}(r) +
p_{B}(r)$. Furthermore, the positivity of the shear-force distribution
over $r$ in Eq.~\eqref{eq:diffEq1}, i.e., $s_{B}(r)>0$, implies that
$\frac{2}{3} s_{B}(r) + p_{B}(r) > 0$ is a monotonically decreasing
function. To quantify the mechanical size of a baryon system, the
mechanical radius is defined as: 
\begin{align}
    \sum_{a=q,g}\langle r^{2}_{\mathrm{mech}} \rangle^{a}_{B} 
  = \frac{\sum_{a=q,g}\int d^{3} r~r^{2}
    \left(\frac{2}{3} s^{a}_{B}(r) + p^{a}_{B}(r) \right) }
    {\sum_{a=q,g}\int d^{3} r~
    \left(\frac{2}{3} s^{a}_{B}(r) + p^{a}_{B}(r) \right)} 
  = \frac{6D_{B}(0)}{\int^{0}_{-\infty}   D_{B}(t) dt}.
\label{eq:mecradius}
\end{align}
Regardless of knowing the $\bar{c}$ form factor, the
stability condition is satisfied, since it is a quantity that adds up
all partonic contributions, i.e. $\sum_{a=q,g}\bar{c}^{a}_{B}=0$. Thus,
the higher-twist form factor plays no role in the stability
condition. However, when the quark and gluon subsystems are
decomposed, the higher-twist form factor comes into essential play.

\section{Chiral quark-soliton model \label{sec:3}}
In this section, we briefly review the $\chi$QSM. 
\subsection{Classical nucleon}
The $\chi$QSM is based primarily on two fundamental principles: 
chiral symmetry breaking and the large $N_c$ limit of QCD. 
This model is constructed, based on the effective partition function
of QCD, which is applicable in the low-energy regime. 
In Euclidean space, the partition function is expressed as
\begin{align}
    Z_{\mathrm{eff}} 
  = \int 
    \mathcal{D} \psi^{\dagger} 
    \mathcal{D} \psi 
    \mathcal{D} U 
    \exp(-S_{\mathrm{eff}}), \quad 
    S_{\mathrm{eff}} =  \int d^{4} x \, 
    \psi^{\dagger} 
    \left(
    i\slashed{\partial} + i M U^{\gamma_{5}} + i \hat{m}
    \right) 
    \psi,
\label{eq:parti}
\end{align}
where $M$ denotes the dynamical quark mass. It is orginally given as
a momentum-dependent one $M(k)$, where $k$ stands for the quark
momentum or quark virtuality. For simplicity, we switch off the
momentum dependence of $M(k)$, and consider it as a free parameter.
We fix its value by reproducing various nucleon form factors 
and the mass differences between the nucleon and the $\Delta$ baryon. 
The most favorable value of $M$ is found to be $420$~MeV, since
various observables computed with that value were well
reproduced~\cite{Christov:1995vm}. $\hat{m}$ represents the diagonal
matrix of the current quark masses in the SU(3) flavor space.  We
assume isospin symmetry, setting $ \bar{m}= m_{u}= m_{d}$. While the
strange current quark mass is typically treated perturbatively, its
contributions to the GFFs are found to be small. Thus, we impose the
flavor SU(3) symmetry, fixing $m_s=\bar{m}$.

Since we use the constant $M$, we have to tame the divergences arising
from the quark-loop integrals. To deal with them, we introduce 
the proper-time regularization. We fix the cutoff mass
$\Lambda$ by fitting the pion decay constant $f_{\pi}=93$~MeV,   
and determine the current quark mass $\bar{m}$ by reproducing the pion
mass $m_{\pi}=139$ MeV~ (see Ref.~\cite{Goeke:2005fs} for more
details).  

The chiral field $U^{\gamma_{5}}$ is represented by the $U$ field:
\begin{align}
    U^{\gamma_{5}} 
  = \frac{1+\gamma_{5}}{2} U 
  + \frac{1-\gamma_{5}}{2} U^{\dagger},
\end{align}
with $U = \exp(i \pi^{a} \lambda^{a})$. The $\pi^{a}$ denote the
pseudo-Nambu-Goldstone (pNG) fields, and $\lambda^{a}$ designate the 
Gell-Mann matrices. In the pion mean-field approach, we consider the
hedgehog symmetry, which is a minimal symmetry that align the spatial
vector with the isospin vector in the mean field: 
\begin{align}
    U^{\gamma_{5}}_{\mathrm{SU(2)}} 
  = \exp[i\gamma_{5} \hat{\bm{n}}\cdot  \bm{\tau} P(r)],
\end{align}
where $\pi^{a}(\bm{r})= \hat{n}^{a} P(r)$ with
$\hat{n}^{a}=r^{a}/|\bm{r}|$ for $a=1,2,3$,  and $\pi^{a}(\bm{r})= 0$
for $a=4,...8$.  This symmetry ensures the invariance of the pion mean 
field under  
$\mathrm{SU}(2)_{\mathrm{flavor}} \otimes
\mathrm{SU}(2)_{\mathrm{spin}}$ rotations.  
The SU(3) chiral field in Eq.~\eqref{eq:parti} is constructed by
using the trivial embedding~\cite{Witten:1983tx}:
\begin{align}
    U^{\gamma_{5}} 
  = \left( 
    \begin{array}{c c}
      U^{\gamma_{5}}_{\mathrm{SU(2)}} & 0 \\ 
      0                               & 1  
    \end{array}\right),
\end{align}
where it contains the chiral field SU(2) as a subgroup: 
$\mathrm{SU}(2)_{\mathrm{flavor}}\otimes
\mathrm{SU}(2)_{\mathrm{spin}}\otimes \mathrm{U}(1)_{Y}\otimes
\mathrm{U}(1)_{Y_{R}}$. Here, $Y$ and $Y_{R}$ denote the hypercharge
and right hypercharge, respectively. 

The SU(2) one-particle Dirac Hamiltonian is defined as: 
\begin{align}
    h(U) 
  = \gamma_{4} \gamma_{k} \partial_{k} 
  + \gamma_{4} M U^{\gamma_{5}}_{\mathrm{SU(2)}}
  + \gamma_{4} \bar{m},
\end{align}
where the strange part is obtained by replacing the chiral field by unity, 
i.e., $U^{\gamma_{5}} \to 1$. The eigenfunctions and eigenenergies are
obtained by diagonalizing  $h(U)$:
\begin{align}
    h(U) \psi_{n} (\bm{r}) 
  = E_{n} \psi_{n} (\bm{r}), \quad 
    h(1) \psi_{n^{0}} (\bm{r}) 
  = E_{n^{0}} \psi_{n^{0}} (\bm{r}).
\end{align}
The Dirac spectrum $E_{n}$ consists of the upper and lower Dirac
continuum, which are polarized by the pion mean field from the free
Dirac spectrum $E_{n^{0}}$, and the bound state level energy (or
valence quark energy $E_{v}$), which emerges when the chiral field is
sufficiently strong. 

To compute properties such as the mass, spin and electromagnetic
properties in the baryonic sector, it is necessary to evaluate the
corresponding correlation function with a pion background field. This
is done by performing the functional integral described in
Eq.\eqref{eq:parti}. Having integrated the fermionic fields, we obtain
the fermionic determinant. The bosonic field can only be solved
approximately by using the saddle-point approximation, which holds in
the large $N_c$ approximation. In this approach, the result is
determined by the integrand evaluated in the classical mesonic  
configuration. It is important to note that quantum fluctuations are
suppressed in the $1/N_c$ expansion~\cite{Witten:1979kh}. 

The classical configuration of the pion field $P_{\mathrm{cl}}(r)$ is
obtained by solving the following saddle point equation: 
\begin{align}
\frac{\delta S_{\mathrm{eff}}}{\delta P(r)}
  \bigg{|}_{P(r)=P_{\mathrm{cl}}(r)} = 0, 
  \label{eq:cla_meson}
\end{align}
which yields 
\begin{align}
M_{\mathrm{sol}} = N_{c} E_{\mathrm{val}} + E_{\mathrm{sea}},
\end{align}
where $N_cE_{\mathrm{val}}$ denote the $N_c$ valence-quark
(level-quark) contribution, and $E_{\mathrm{sea}}$ represents the sum
of the negative Dirac continuum energy with the vacuum energy
subtracted. This quantity is logarithmically divergent and requires
a regularization. The specific regularization functions employed are
provided in Appendix~\ref{app:A}. 

\subsection{Collective quantization}
The classical soliton does not have the well-defined momentum and
spin-flavor quantum numbers. To restore the corresponding symmetries,
we introduce translational and rotational zero modes. These modes
allow us to replace the functional integral over the mean field $U$ in
the presence of a background pion field with the integrals over the
center of mass (CM) coordinates $\bm{X}$ and the rotational matrix $R$ in
flavor space: 
\begin{align}
\int \mathcal{D}U \mathcal{F}[U(\bm{x})] \to  \int d^{3}\bm{X} \int
  \mathcal{D} R \, \mathcal{F}\left[ T R U_{\mathrm{cl}}(\bm{x})
  R^{\dagger} T^{\dagger}\right], 
\end{align}
where the unitary transformation $T$ represents the translational
symmetry. It is important to note that both the CM coordinates 
$\bm{X}(t)$ and the rotation matrix $R(t)$ depend weakly on time. The 
translational zero modes endow the classical soliton with the
momentum, while the rotational zero modes furnish it with the  
spin-flavor quantum numbers. The slow rotation and displacement of 
the soliton give rise to kinetic corrections that are suppressed
in the $1/N_{c}$ expansion. When considering a baryon in the Breit
frame (i.e., $\bm{P}=(\bm{p}'+\bm{p})/2=0$), the translational
kinetic correction does not contribute to the GFFs. Therefore, in this
study we focus on the rotational zero modes to order of $\Omega \sim
1/N_{c}$ and the translational ones to the zeroth order. Having
performed the collective quantization, we obtain the collective
Hamiltonian:  
\begin{align}
  H_{\mathrm{coll}}  &= M_{\mathrm{sol}}  + \frac{1}{2I_{1}}
                       \sum_{i=1}^{3}  \hat{J}_{i}^{2} +
                       \frac{1}{2I_{2}}   \sum_{p=4}^{7}
                       \hat{J}_{p}^{2}, 
\end{align}
where $I_{1}$ and $I_{2}$ represent the moments of inertia, and their
explicit expressions can be found in Appendix~\ref{app:A}. 
The hedgehog symmetry of the mean field implies that baryon states
emerge with the selection rules: $\bm{J}+\bm{T}=0$ and
$Y_{R}=N_{c}/3$. Consequently, diagonalizing $H_{\mathrm{coll}}$, we
derive the rotational wave function for a baryon with spin and flavor
indices: 
\begin{align}
  \Psi_{(YTT_{3})(Y_{R}JJ_{3})}^{(\mu)}(R) & =
                                             \sqrt{\mathrm{dim}(\mu)}
                                             ( - 1 )^{J_{3}-Y_{R}/2}
                                             D_{(YTT_{3})(Y_{R}J-J_{3})}^{(\mu)*}
                                             ( R  ), 
\end{align}
where $D_{ab}^{(\mu)}$ denotes the SU(3) Wigner $D$
function with the corresponding
SU(3) representation $\mu$.

\subsection{Effective EMT operator \label{sec:Eff_op}} 
\begin{figure}[htp]
\centering
\includegraphics[scale=0.55]{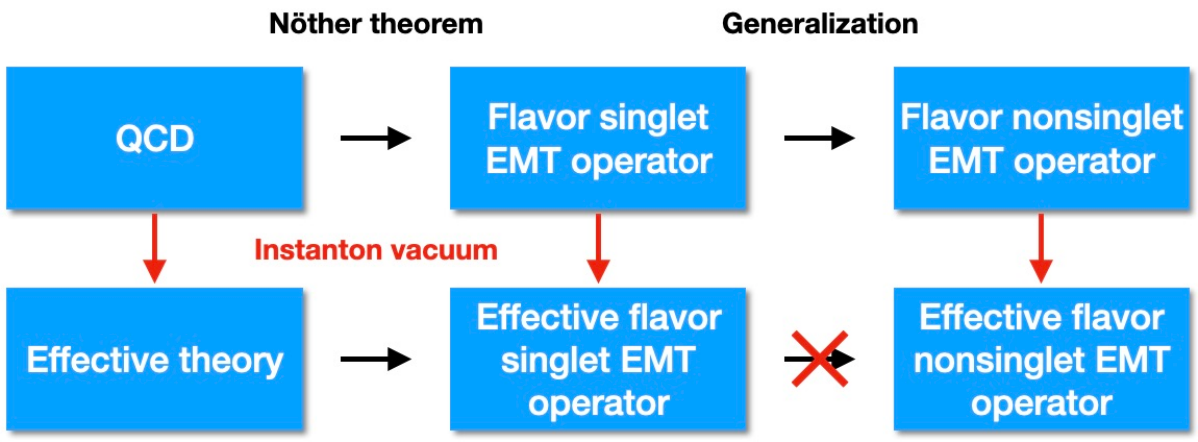}
\caption{The logical chain that the effective operator derived from the QCD instanton vacuum must be used in the study of flavor-decomposed GFFs is drawn.}
\label{fig:1}
\end{figure}

Now we are in a position to derive the effective EMT operator. Given a
global symmetry, a conserved current can be derived by 
the N\"other theorem. However, a non-conserved current such as the
flavor nonsinglet EMT current in QCD can be constructed, guided by the
singlet EMT current. The effective operator corresponding to the
flavor nonsinglet EMT current in QCD can be obtained from the
instanton vacuum, which can be coherently adapted to effective
quark-gluon dynamics (see Fig.~\ref{fig:1}).  

\subsubsection{EMT current from the global symmetry}

This effective dynamics~\eqref{eq:parti} includes two degrees
of freedom: the dynamical quark and the pion field. In the systematic
treatment, i.e. the dilutness of the instanton liquid and the $1/N_{c}$
expansion, all gauge-dependent parts are integrated out through the
instanton vacuum. Using the effective action derived from QCD, one can
easily obtain a conserved quantity such as the EMT current from the
N\"other theorem. The expression for the EMT current in Minkowski
space is written as follows~\cite{Goeke:2007fp}: 
\begin{align}
    T^{\mu\nu}_{\chi=0}(x) 
  = \frac{i}{4}   \bar{\psi} (x)
    \gamma^{ \{\mu} \overleftrightarrow{\partial}^{\nu\} }
    \lambda_{0} 
    \psi (x), 
\label{eq:EMT_current}
\end{align}
where $\lambda_{\chi}$ are the SU(3) Gell-Mann matrices with the
flavor singlet $\lambda_{0}= \mathrm{diag}(1,1,1)$ Gell-Mann
matrix. Since there is a kinetic term only for the quarks in the
effective action, it is obvious to have the EMT operator with the
ordinary derivative. In any models, such as the bag  
model~\cite{Neubelt:2019sou}, it is also possible to derive a similar
EMT operator without explicit connection to the QCD operator. 

As shown in Eq.~\eqref{eq:EMT_current}, it is no longer possible to  
distinguish between the quark and gluon parts of the EMT current, 
since the gluon degrees of freedom are absent in the effective action.
Thus, this effective operator should be understood as a
total~(quark+gluon) contribution: 
\begin{align}
T^{\mu \nu}_{q} (x) + T^{\mu \nu}_{g} (x)~[\mathrm{QCD}]
  \xrightarrow{\mathrm{eff}} T^{\mu \nu}_{\chi=0}
  (x)~[\mathrm{Eq}.~\eqref{eq:EMT_current}]. 
  \label{eq:total_eff}
\end{align}

We briefly discuss the various sum rules in QCD. The nucleon momentum
and spin sum rules are satisfied by the sum of the quark and gluon
contributions  
\begin{align}
    \sum_{a=q,g}A_{N}^{a}(0)=A_{N}(0)=1, \quad
    \sum_{a=q,g}J_{N}^{a}(0)=J_{N}(0)=1/2, 
    \quad [\mathrm{QCD}] .
\label{eq:QCDsr}
\end{align}
In addition, the nucleon matrix element of the trace part of the EMT
current is normalized to the nucleon mass in QCD in the forward limit: 
\begin{align}
    \sum_{a=q,g}\frac{\langle N | T^{\mu}_{a,\mu} |  N \rangle}{2M_{N}} 
  = M_{N} A_{N}(0), \quad \mathrm{with} \quad T^{\mu}_{q,\mu} 
  = O(m_{q}),
    \quad [\mathrm{QCD}]. 
  \label{eq:QCDtrace}
\end{align}
This indicates that the nucleon mass mainly comes from the gluon
component of the EMT current, due to the trace anomaly. 
Similar to the momentum and spin sum rules, the von Laue condition is
known as the normalization of the stress tensor, which is closely
related to the conservation of the EMT current: 
\begin{align}
    \sum_{a=q,g} \int d^{3} r\;
    p_{N}^{a}(r)=0, 
    \quad \sum_{a=q,g}
    \bar{c}_{N}^{a}(t)
  = 0, \quad
    [\mathrm{QCD}]. 
\end{align}
As a result, the sum of both the quark and the gluon yields the
correct sum rules in QCD.  

In the effective theory, as explained in Eq.~\eqref{eq:total_eff}, the
dynamical quark (+ background pion mean field) alone accounts for the
internal dynamics of the nucleon. For example, the momentum
and spin sum rules are given only in terms of the quark degrees of
freedom~\eqref{eq:QCDsr}  
\begin{align}
    \sum_{q=u,d,\ldots}
    A_{N}^{q}(0)=A_{N}(0)=1, \quad
    \sum_{q=u,d,\ldots}
    J_{N}^{q}(0)=J_{N}(0)=1/2,
    \quad
    [\mathrm{Eq}.~\eqref{eq:EMT_current}].  
\end{align}
the trace anomaly~\eqref{eq:QCDtrace} is also expressed in terms of
the dynamical quarks, which are effectively dressed by gluons: 
\begin{align}
    \frac{\langle N | T^{\mu}_{\chi=0,\mu} |  N \rangle}{2M_{N}} = M_{N}
    \sum_{q=u,d,\ldots} A_{N}^{q}(0), \quad \mathrm{with} \quad
    T^{\mu}_{\chi=0,\mu} =  T^{00}_{\chi=0}, 
    \quad [\mathrm{Eq}.~\eqref{eq:EMT_current}] .
\end{align}
Here, we have used the fact that the spatial part of the EMT current
vanishes in the forward limit, i.e. $ \hat{T}^{ii}_{\chi=0}=0$. This
relation holds true only if the equation of motion of the pion
fields~\eqref{eq:cla_meson} is satisfied~\cite{Diakonov:1996sr,
  Goeke:2007fp}. Owing to the hedgehog symmetry of the self-consistent 
pion field, one can generalize it to the spatial $ij$-component of the
EMT in the forward limit 
\begin{align}
    \frac{i}{2}
    \bar{\psi}(x) \gamma^{i} \partial^{j} \psi(x) 
  = \delta^{ij} f, \quad \mathrm{with} \quad f=0. 
\label{eq:stab_op}
\end{align}
Interestingly, Eq.~\eqref{eq:stab_op} leads to the von Laue condition
(see also Ref.~\cite{Goeke:2007fp}) 
\begin{align}
    \sum_{a=q,g} \int d^{3} r \;
    p_{N}^{a}(r) = 0
    \overset{\mathrm{eff}}{\longrightarrow} 
    \sum_{q} \int d^{3} r \;
    p_{N}^{q}(r)=0, \quad [\mathrm{Eq}.~\eqref{eq:EMT_current}],
\end{align}
and the null result of the $\bar{c}$ form factor, induced by the
conservation of the EMT current, 
\begin{align}
    \sum_{a=q,g}\bar{c}^{a}_{N}(t)
  = 0 \overset{\mathrm{eff}}{\longrightarrow}
    \sum_{q}\bar{c}^{q}_{N}(t)
  = 0, \quad [\mathrm{Eq}.~\eqref{eq:EMT_current}].
\end{align}
As a result, using the flavor-singlet EMT current derived from the
global symmetry~\eqref{eq:EMT_current}, we have succeeded in
satisfying all QCD sum rules by means of the dynamical quark and pion
fields in the effective theory.  

So far, we have discussed the flavor-singlet EMT current. Since we aim
at examining the quark flavor decomposition of the GFFs,
we need to construct the effective flavor-triplet and -octet EMT
currents by introducing the flavor SU(3) Gell-Mann matrices:
\begin{align}
    T^{\mu\nu}_{\chi}(x) 
  = \frac{i}{4}   \bar{\psi} (x)
    \gamma^{ \{\mu} \overleftrightarrow{\partial}^{\nu\} }
    \lambda_{\chi} 
    \psi (x).
\label{eq:EMT_current_flav}
\end{align}
Having employed Eq.~\eqref{eq:EMT_current_flav}, we already performed
the flavor decomposition of the GFFs~\cite{Won:2023cyd, Won:2023ial}. 
It was also implemented in Refs.~\cite{Wakamatsu:2005vk,
  Wakamatsu:2006dy, Wakamatsu:2007uc}.  

Strictly speaking, it is only possible to derive the
flavor-singlet EMT current~\eqref{eq:EMT_current} as a N\"other 
current. This means that there is no proper way for deriving the
flavor-triplet and -octet EMT currents in the effective theory. Thus, 
we are not able to guarantee that the current given in
Eq.~\eqref{eq:EMT_current_flav} is consistent 
with the effective quark-gluon dynamics. Consequently, the only way to
construct the currents is that starting from the QCD nonsinglet EMT
currents we derive the corresponding effective operator from the QCD
instanton vacuum. By doing that, we preserve the effective quark-gluon
dynamics.

\subsubsection{EMT current from QCD instanton vacuum}

In the systematic expansion (expansions in $1/N_{c}$ and dilutness of
the instanton liquid), the QCD gluon operator can be converted into
the effective quark operators in the effective
theory~\cite{Diakonov:1995qy, Balla:1997hf}. Using this method, we are
able to construct the effective operators corresponding to the QCD
operators consistently and systematically. 
Moreover, the low-energy theorems of QCD such as the chiral anomaly
and the trace anomaly are correctly satisfied~\cite{Diakonov:1995qy}. 

The chiral-even twist-2 local operator is generated by expanding
the non-local vector current, which measures the vector GPDs, with
respect to the space-time distance 
\begin{align}
    O^{\mu \nu_{1} \ldots \nu_{n}}_{q}
&:= \bar{\psi} (x) \gamma^{ \{\mu}
    \overleftrightarrow{D}^{\nu_{1}}
    \overleftrightarrow{D}^{\nu_{2}} \ldots
    \overleftrightarrow{D}^{\nu_{n} \} } \lambda_{\chi}
    \psi(x) 
  - \mathrm{traces}, \cr 
    O^{\mu \nu_{1} \ldots \nu_{n}}_{g}
&:= - F^{ \{\mu \rho,a}
    \overleftrightarrow{D}^{\nu_{1}} \ldots
    \overleftrightarrow{D}^{\nu_{n-1}} F^{ \nu_{n} \},a}_{\ \rho} 
  - \mathrm{traces}. 
\end{align}
For example, the first and second Mellin moments of the GPDs are known
as the leading-twist electromagnetic form factors and GFFs,
respectively. The corresponding local currents are given by 
\begin{align}
&J^{\mu}_{\chi}=\bar{\psi} (x) \gamma^{ \mu }  \lambda_{\chi} \psi(x), \cr
&\bar{T}^{\mu \nu}_{\chi}= \frac{i}{4}\bar{\psi} (x) \gamma^{ \{\mu }
       \overleftrightarrow{D}^{\nu \} }   \lambda_{\chi} \psi(x)
      - \mathrm{traces}, \quad \bar{T}^{\mu \nu}_{g} =
    -\frac{1}{2}F^{ \{\mu \rho,a} F^{\nu \},a}_{\ \rho} - \mathrm{traces}. 
\end{align}

In Ref.~\cite{Balla:1997hf}, it was shown that in the twist-2 local
operators the effect of the instanton field~(gauge field in the
covariant derivative) is always of order $(M\bar{\rho})^{2}$. Thus,
when working on leading order in $(M\bar{\rho})$, it is consistent to
keep only the ordinary derivative part [order unity $\sim O (1)$] in
the twist-2 quark operators 
\begin{align}
&   \bar{\psi} (x) \gamma^{ \{\mu} \overleftrightarrow{D}^{\nu_1} \overleftrightarrow{D}^{\nu_2} \ldots \overleftrightarrow{D}^{\nu_n \} } \lambda_{\chi} \psi(x) - \mathrm{traces} \cr
&   \hspace{1cm}  
    \overset{\mathrm{eff}}{\longrightarrow} 
    \frac{i}{4}\bar{\psi} (x)
    \gamma^{ \{\mu} \overleftrightarrow{\partial}^{\nu_1} \overleftrightarrow{\partial}^{\nu_2} \ldots \overleftrightarrow{\partial}^{\nu_n \} } \lambda_{\chi} \psi(x) - \mathrm{traces}, 
\end{align}
and to set the twist-2 gluon operators by zero: 
\begin{align}
& - F^{ \{\mu \rho,a} \overleftrightarrow{D}^{\nu_{1}} \ldots
    \overleftrightarrow{D}^{\nu_{n-1}} 
    F^{\nu_{n} \},a }_{\ \rho} -
    \mathrm{traces} \quad
    \overset{\mathrm{eff}}{\longrightarrow} \quad 0 \quad
  + \quad O(M^{2} \bar{\rho}^{2}). 
\end{align}
With this parametric order $O(1)$ in the expansion with respect to the
packing fraction, we then have the following EMT
operators for the effective theory: 
\begin{align}
    \bar{T}^{\mu\nu}_{\chi}(x) 
  = \frac{i}{4}   \bar{\psi} (x)
    \gamma^{ \{\mu} \overleftrightarrow{\partial}^{\nu\} }
    \lambda_{\chi} 
    \psi (x) - \mathrm{traces}, \quad     \bar{T}^{\mu\nu}_{g}(x) 
  = 0.
\label{eq:EMT_current_tw2}
\end{align}
Consequently, this method for deriving the effective operators
validates the use of the ordinary derivative operator with trace 
subtracted. For the purpose of the flavor decomposition, we will use
the twist-2 operator~\eqref{eq:EMT_current_tw2} derived from the QCD,
instead of the Eq.~\eqref{eq:EMT_current_flav}. The key
difference lies in the subtraction of traces (or twist
classification). 

For the twist-3 part, the gauge-dependent parts in the covariant
derivative lead to spin-flavor dependent quark operators to parametric
order $O(1)$. The contributions from these operators have been found
to be crucial for satisfying the QCD equation of motion, as discussed
in a recent study~\cite{Kim:2023pll}. Furthermore, they have 
a significant influence on the numerical values of the orbital
angular momentum and the spin-orbit correlation~\cite{Kim:2024cbq}.  

Similar to the twist-3 part~\cite{Kim:2023pll}, the gauge-field
dependent part of the twist-4 operators should be 
replaced by the spin-flavor dependent quark operators. We expect that
this twist-4 EMT current will gain the large gauge field
contributions of parametric order $O(1)$. It is also essential to
consider the gauge-field dependent part of the twist-4 EMT current
such that the QCD equation of motion~\eqref{eq:QCDtrace} is satisfied,
i.e., $T^{\mu}_{q,\mu}=0$ in the chiral limit. The explicit expression of
the twist-4 effective operator is currently under derivation, and
will appear elsewhere~\cite{Twist4}. In this study we will focus on
the leading-twist GFFs by using Eq.~\eqref{eq:EMT_current_tw2}.

\subsection{Matrix element of the EMT current in the large $N_{c}$
  limit of QCD} 

The matrix element of the symmetrized EMT current in Euclidean
space can be calculated as follows: 
\begin{align}
  &\mel{B (p',J_{3}')}
    {
    \bar{T}_{\mu\nu,\chi}(0)}
    {B(  p,J_{3})}
  = \lim_{T\to\infty} 
    \frac{1}{Z_{\mathrm{eff}}} 
    \mathcal{N}^{*}(p')\mathcal{N}(p)  
    e^{ip_{4}\frac{T}{2}-ip'_{4}\frac{T}{2}} 
   \int d^{3}\bm{x} \, d^{3}\bm{y}  \cr
&\times e^{(-i\bm{p}' \cdot \bm{y} + i\bm{p}  \cdot \bm{x})} \int  \mathcal{D} \psi 
    \mathcal{D} \psi^{\dagger}
    \mathcal{D} U 
    J_{B}(\bm{y},T/2) 
    \bar{T}_{\mu\nu,\chi}(0)
    J_{B}^{\dagger}(\bm{x},-T/2)
\exp\left[ - S_{\mathrm{eff}}\right],
\end{align}
where $J_{B}$ represents the Ioffe-type current consisting of 
the $N_{c}$ valence quarks~\cite{Ioffe:1981kw} and 
$\bar{T}_{\mu\nu,\chi}(0)$ denotes 
the twist-2 EMT current~\eqref{eq:EMT_current_tw2} derived from the
effective chiral theory in Euclidean space. Note that
$\mathcal{N}^{*}(p')\mathcal{N}(p')$ yields the non-relativistic
normalization $2M_{\mathrm{sol}}$, and the baryon state carries the
spin, isospin, and hypercharge quantum numbers $B=\{
J,J_{3},T,T_{3},Y\}$. 

To understand the behavior of the GFFs, we need to discuss the large
$N_{c}$ limit of the kinematic variables. The large $N_c$ behavior for
the baryon mass is given as $M_{B}\sim \mathcal{O}(N_{c})\sim
M_{\mathrm{sol}}$. The three-momentum shows $p^{k}\sim
\mathcal{O}(N^{0}_{c})$, and the energy scales as $p^{0}\sim
\mathcal{O}(N^{1}_{c})$.  
Therefore, the average momentum and the momentum transfer behaves as  
\begin{align}
    \Delta^{0}  \sim O(N^{-1}_{c}), \quad 
    \Delta^{i}  \sim O(N^{0}_{c}), \quad 
    P^{0}       \sim O(N^{1}_{c}), \quad 
    P^{i}       \sim O(N^{0}_{c}).
\end{align}
In addition, the moments of inertia are given in the following order 
\begin{align}
    I_{1} \sim \mathcal{O}(N^{1}_{c}), \quad 
    I_{2} \sim \mathcal{O}(N^{1}_{c}).
\end{align}

Defining the static EMT distribution in the large $N_{c}$ limit
\begin{align}
    \bar{T}_{\chi}^{\mu\nu} (\bm{r}, J'_{3}, J_{3})
& = \int \frac{d^{3}\Delta}{(2\pi)^{3} 
    2M_{\mathrm{sol}}}
    e^{-i\bm{\Delta}\cdot\bm{r}} 
    \mel{B(p',J_{3}')}{\bar{T}_{\chi}^{\mu\nu}(0)}{B(p,J_{3})},
\end{align}
we obtain the final expressions for the GFFs\footnote{Instead of using
  the $A(t)$ form factor, we adopt the 3D multipole form factor
  $\bar{\mathcal{E}}$, which reveals the clear and natural $N_{c}$
  scaling. This is also true for the electromagnetic form factors. The
  Sach form factors $G_{E}$ and $G_{M}$ exhibit a clearer and more
  natural $N_{c}$ scaling than the Pauli and Dirac form factors
  $F_{1,2}(t)$. These facts are deeply rooted in the non-relativistic
  nature of the collective nucleon motion in the large $N_{c}$ limit
  of QCD.}  
as the 3D Fourier transforms of the EMT distributions: 
\begin{align}
&   \bar{\mathcal{E}}^{\chi}_{B}(t) \delta_{J_{3}^{\prime}J_{3}} = \frac{1}{M_{\mathrm{sol}}} \int d^{3}r
    j_{0}(r\sqrt{-t})\bar{\varepsilon}^{\chi}_{B}(r),\cr 
&   \bar{\mathcal{P}}^{\chi}_{B}(t) \delta_{J_{3}^{\prime}J_{3}}
  =  \frac{1}{M_{\mathrm{sol}}} \int  d^{3}r  j_{0}(r\sqrt{-t}) \bar{p}^{\chi}_{B}(r),
    \cr  
&   D_{B}^{\chi}(t) \delta_{J_{3}^{\prime}J_{3}}
  = 4 M_{\mathrm{sol}} \int d^{3}r \frac{j_{2}(r\sqrt{-t})}{t} s^{\chi}_{B}(r),   \cr
&   2 S_{J_{3}^{\prime}J_{3}}^{3} J_{B}^{\chi}(t) 
  = 3\int d^{3}r \frac{j_{1}(r\sqrt{-t})}{r\sqrt{-t}} 
    \rho_{J,B}^{\chi}(r),
    \label{FFs}
\end{align}
where the respective distributions $\bar{\varepsilon}^{\chi}_{B}$,
$\rho_{J,B}^{\chi}$, $s^{\chi}_{B}$,  and $\bar{p}^{\chi}_{B}$ are given by
\begin{align}
      \bar{\varepsilon}^{\chi}_{B}  ( r  )  
& = \frac{1}{\sqrt{3}}  \expval{D_{\chi 8}}_{B} \mathcal{M} (  r )
  - \frac{2}{I_{1}}  \expval{D_{\chi i}J_{i}}_{B}  \mathcal{J}_{1} ( r  )
  - \frac{2}{I_{2}}  \expval{D_{\chi a}J_{a}}_{B}  \mathcal{J}_{2} ( r  ),  \cr
    \rho_{J,B}^{\chi} ( r  )  
& = \expval{D_{\chi 3}}_{B} \left(  \mathcal{Q}_{0} ( r  ) + \frac{1}{I_{1}} \mathcal{Q}_{1} (  r ) \right) 
  - \frac{1}{\sqrt{3}}  \expval{D_{\chi 8}J_{3}}_{B}  \frac{1}{I_{1}}  \mathcal{I}_{1}(r)
  - \expval{d_{ab3}D_{\chi a}J_{b}}_{B} \frac{1}{I_{2}} \mathcal{I}_{2} ( r  ),   \cr
    s^{\chi}_{B}  ( r  )
& = \frac{1}{\sqrt{3}}  \expval{D_{\chi 8}}_{B} \mathcal{N}_{1} ( r  )
  - \frac{2}{I_{1}} \expval{D_{\chi i}J_{i}}_{B}  \mathcal{J}_{3} ( r  )
  - \frac{2}{I_{2}} \expval{D_{\chi a}J_{a}}_{B}  \mathcal{J}_{4} ( r  ), \cr
    \bar{p}^{\chi}_{B}  ( r  )
& = \frac{1}{3} \bar{\varepsilon}^{\chi}_{B}  ( r  ).
    \label{eq:model_EMT}
\end{align}
The $\langle ... \rangle_{B}$ denotes the matrix element of the
SU$(2N_{f})$ spin-flavor operators between the initial and final
rotational wave functions 
\begin{align} 
    \langle ... \rangle_{B} 
  = \int dR \, 
    \Psi^{(\mu)*}_{(Y'T'T'_{3})(Y_{R}J'J'_{3})} (R) \, 
    ... \, 
    \Psi^{(\mu)}_{(YTT_{3})(Y_{R}JJ_{3})} (R). 
\end{align}
The detailed expressions for the distributions $\mathcal{M}, \
\mathcal{J}_{1}, \ \mathcal{J}_{2}$, etc. are given in Appendix~\ref{app:A}.  
In the limit $\chi\to 0$, the results for the flavor singlet EMT 
distributions~\cite{Goeke:2007fp,Won:2022cyy} are recovered as
follows: 
\begin{align}
    \bar{\varepsilon}^{0}_{B}  ( \bm{r}  )  
& =  \mathcal{M} (  \bm{r} ), \quad
    \rho^{0}_{J,B} ( \bm{r}  )   
  =  -    \frac{1}{2I_{1}}  \mathcal{I}_{1}(\bm{r}), \cr
    s^{0}_{B}  ( \bm{r}  )
& = \mathcal{N}_{1} ( \bm{r}  ), \quad
    \bar{p}^{0}_{B}  ( \bm{r}  )
  = \frac{1}{4} \mathcal{M} (  \bm{r} ).
\end{align}
The integrals of the individual EMT distributions over 3D space
satisfy the following relations: 
\begin{align}
&   \int d^{3} r \, \mathcal{M}(\bm{r}) 
  = \frac{3}{4}
    M_{\mathrm{sol}}, \quad
    \int d^{3} r \, \mathcal{I}_{1}(\bm{r}) 
  = - I_{1}, 
\label{eq:Nor}
\end{align}
Using Eq.~\eqref{eq:Nor}, we find that the mass and AM form factors
are properly normalized to its mass [$A_{B}(0)=1$] and AM [$J_{B}(0)=1/2$]
\begin{align}
    \int d^{3} r \,  \bar{\varepsilon}^{0}_{B}(\bm{r})  = \frac{3}{4} M_{\mathrm{sol}}, \quad 
    \int d^{3} r \,  \rho^{0}_{J,B}(\bm{r})  
  = 1/2.
\end{align}
In addition, the first relation in Eq.~\eqref{eq:Nor} results in break down of the von Laue condition: 
\begin{align}
    \int d^{3} r \, \bar{p}^{0}_{B}(\bm{r}) = \frac{1}{4} M_{\mathrm{sol}}.
\end{align}
The inclusion of the twist-4 contribution restores the von Laue
condition. It can be inferred indirectly from the estimation of the
EMT distribution using the EMT current~\eqref{eq:EMT_current} derived
from the N\"other theorem. When we use the EMT
current~\eqref{eq:EMT_current}, we find that the 
presence of the $s$ quark has no effect on the normalizations of the
GFFs and the von Laue condition (see Refs.~\cite{Won:2023ial,
  Won:2023rec}). It is noteworthy that the mass and AM normalizations
hold true regardless of the configuration of the pion mean
field. However, the von Laue condition is only satisfied when the pion
mean field assumes a classical configuration. This emphasizes the
importance of considering the dynamical nature of the system when
describing properties related to the stress tensor.    

Furthermore, while the conserved EMT current ensures the normalization 
of the total AM, the distinction between intrinsic spin and OAM 
remains an important issue for careful investigation (see
Refs.~\cite{Kim:2023pll, Kim:2024cbq}).

\section{Numerical results \label{sec:4}}

Before delving into the numerical results for the flavor
decompositions of the EMT distributions and form factors,  
it is crucial to acknowledge the limitations of the current approach. 
We have made certain assumptions regarding the rotational and
translational zero modes, treating them up to corrections of $1/N_{c}$
and zero, respectively. Additionally, we have considered the flavor
SU(3) symmetry where the strange current quark mass, $m_{s}$, is set
to $m_{u}=m_{d}=m_{s}$. We have previously investigated the impact of
$m_{s}$ on the GFFs and EMT distributions, and found that while these
contributions introduce some differences in the octet baryon GFFs,  
they are ultimately negligible, with $m_{s}$ corrections approximately
10\%~\cite{Won:2022cyy}. Moreover, if we were to include the $m_{s}$
corrections in the stress tensor $T^{ij}$, the von Laue condition
would be violated. Consequently, we would have to artificially
reconstruct the pressure distribution by solving the differential
equation~\eqref{eq:diffEq1} in terms of the shear-force distributions.   
Therefore, in the context of examining flavor structures, 
it is reasonable to ignore these contributions in order to clearer 
understand the GFFs and distributions.

\subsection{Mass distribution: Twist-2 part \label{sec:4_1}} 
As discussed in the previous section, we will discuss the twist-2 mass
distribution decoupled from the $\bar{c}$ form factor. By taking the
linear combinations of the $\chi=0,3, 8$ components of 
Eq.~\eqref{eq:model_EMT}, we can derive the flavor-singlet, -triplet,
and -octet components of the mass distributions of the 
nucleon. It is important to note that the 3D mass distribution, 
defined in the instant form quantization at the rest frame, is
normalized as follows: 
\begin{align}
   \frac{3}{4} A^{\chi}_{p}(0) 
  = \frac{1}{M_{\mathrm{sol}}} 
    \int d^{3}r \,
    \bar{\varepsilon}^{\chi}_{p}(r).
\label{eq:IF_mass}
\end{align}
The values of each component $\chi=0,3,8$ for the mass form factors 
(or normalization of the mass distribution) are listed as follows:
\begin{align}
    A^{0}_{p}(0)  
&=1, \quad 
    A^{3}_{p}(0)  
  = 0.25, \quad   
    A^{8}_{p}(0)  
    = 0.47, \quad [\, \mathrm{SU}(3) \,] \cr 
    A^{0}_{p}(0)  
&=1, \quad 
    A^{3}_{p}(0)  
  = 0.24 \hspace{3.27cm} [\, \mathrm{SU}(2) \,].  
\end{align}
The proton twist-2 mass distribution, or the flavor-singlet mass distribution
$\bar{\varepsilon}^{0}_{p}$, is normalized to its mass $3M_{\mathrm{sol}}/4$, 
which ensures that the mass form factor is normalized to
$A^{0}_{p}(0)=1$. Since the spin-flavor operator
of the flavor-singlet component is proportional to unity,  
the masses of the octet baryons are all degenerate. It is worth noting
that as discussed in Eq.\eqref{eq:EMT_current_tw2} the gluon
contributions to the leading-twist operators are parametrically
suppressed with respect to the instanton packing
fraction~\cite{Balla:1997hf, Polyakov:2018exb}, allowing us to
consider them negligible at the low normalization point 
of $\mu\sim 600$ MeV. Therefore, the gluon contributions to the
leading-twist GFFs can be ignored throughout this study:
\begin{align}
A^{g}_{B} =0, \quad J^{g}_{B}=0.
\end{align}
 Thus, the normalization of
the nucleon mass is solely determined by the quark
contributions. Furthermore, we observe that the flavor-triplet mass
distribution is smaller than the flavor-singlet one. It suggests that
the parametric suppression of the flavor triplet in SU(2) symmetry
remains valid in flavor SU(3) symmetry. 
In addition, the flavor octet component of the mass form factor is
approximately half of the singlet component. We can determine the
individual quark contributions to the mass distribution of the proton. 
\begin{figure}[htp]
\centering
  \includegraphics[scale=0.147]{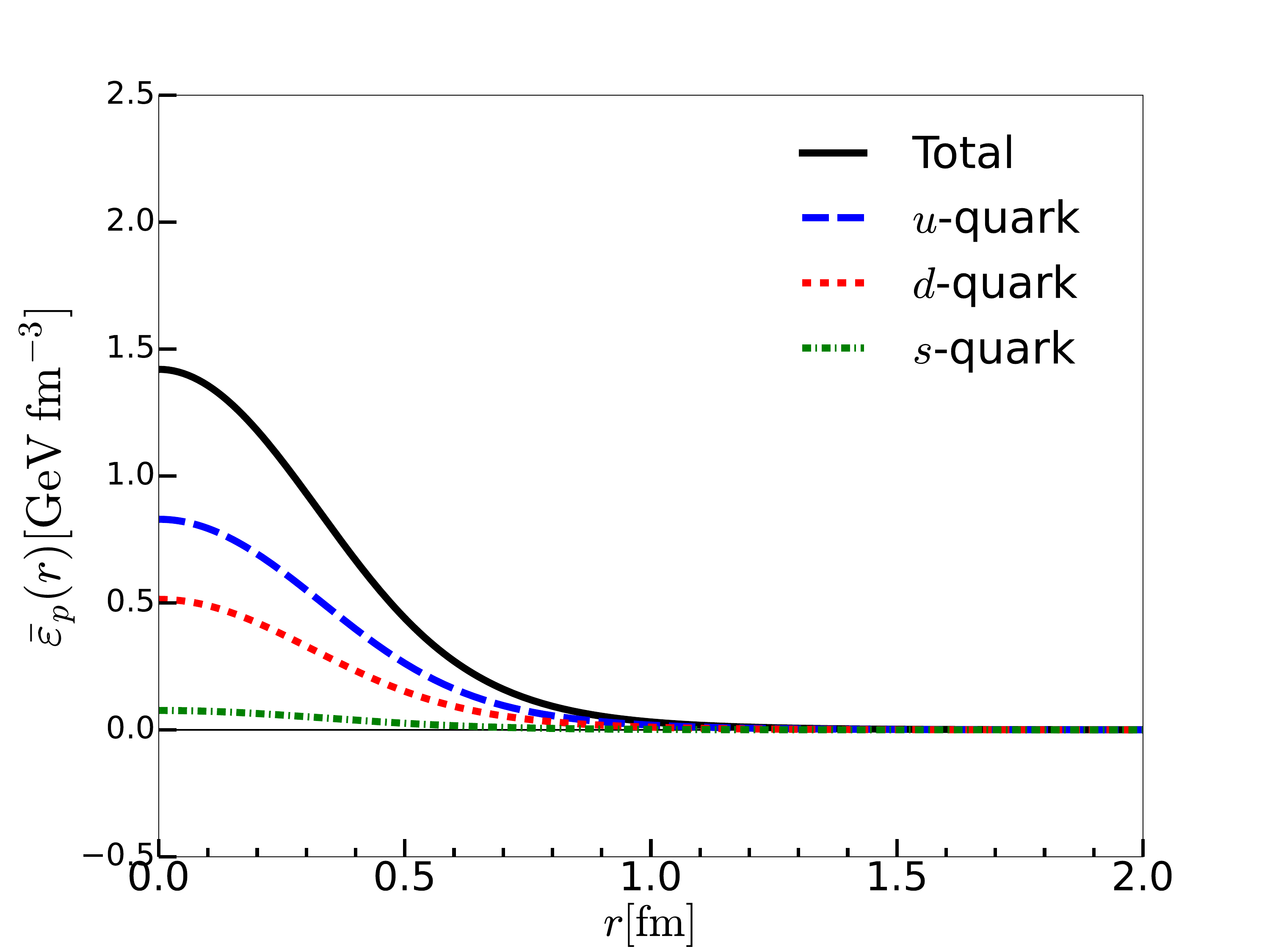}
  \includegraphics[scale=0.147]{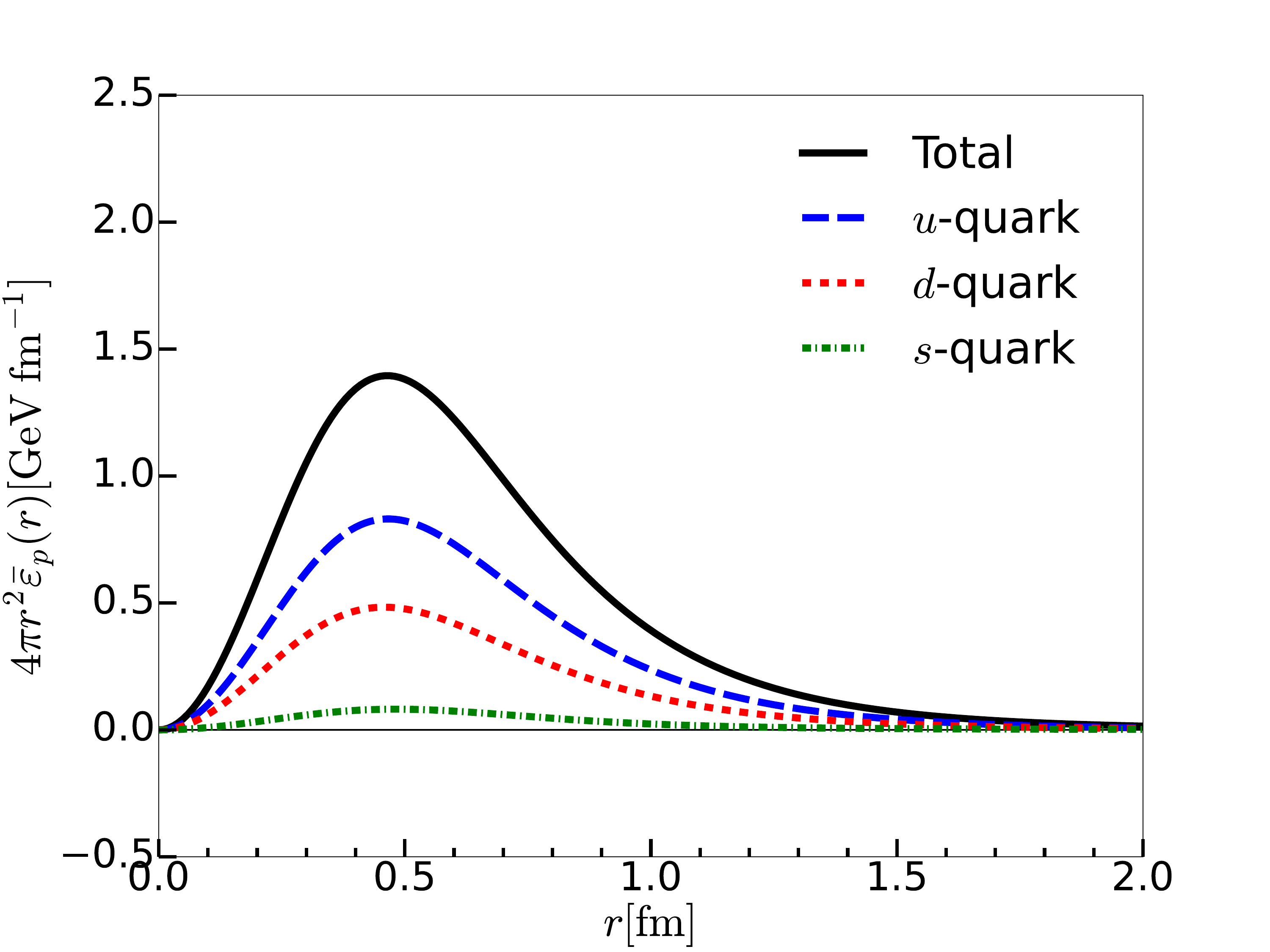}
  \caption{The 3D mass distribution of the proton and its flavor
    decomposition with the flavor SU(3) symmetry are plotted.  
  The solid (black), long-dashed (blue), short-dashed (red), and
  dashed-dotted (green) curves denote the total, $u$-, $d$-, and
  $s$-quark contributions to the mass distributions, respectively.}   
\label{fig:2}
\end{figure}
The left panel of Fig.~\ref{fig:2} shows the 3D mass distribution of
the nucleon and its flavor decomposition with the flavor SU(3)
symmetry.  First, they are kept positive definite at any given $r$
\begin{align}
    \bar{\varepsilon}^{u,d,s}_{p}(r) >0. 
    \label{eq:positive_energy}
\end{align} 
Numerically, we find that the sum of the $u$- and $d$-quark
contributions,  as well as the $s$-quark contribution, is normalized
to $3M_{\mathrm{sol}}/4$ when integrated over $r$. At the origin of
the proton, the magnitudes of the mass distributions for the $u$-,
$d$-, and $s$-quarks are found to be: 

\begin{align}
    &\bar{\varepsilon}^{u}_{p}(0) 
  = 0.83~\mathrm{GeV/fm^{3}}, \quad 
    \bar{\varepsilon}^{d}_{p}(0) 
  = 0.51~\mathrm{GeV/fm^{3}}, \cr
    &\bar{\varepsilon}^{s}_{p}(0) 
  = 0.08~\mathrm{GeV/fm^{3}}. 
\label{eq:center_energy}
\end{align}
We observe that the $u$-quark contributions to the mass distribution  
are approximately twice as large as those of the $d$-quark for
the proton. This can be intuitively understood by considering the
number of valence quarks inside the proton. Additionally, the
$s$-quark contribution is approximately $10\%$ of the $u$-quark
contribution. Notably, while the role of the $u$-quark inside a proton
is taken by the $d$-quark inside a neutron, i.e.,
$\bar{\varepsilon}^{u}_{p}(r)=\bar{\varepsilon}^{d}_{n}(r)$, the $s$-quark
contribution remains unchanged, i.e.,
$\bar{\varepsilon}^{s}_{p}(r)=\bar{\varepsilon}^{s}_{n}(r)$. 

In the right panel of Figure~\ref{fig:2}, we depict the
$r^{2}$-weighted mass distributions. To quantify how far 
the mass distributions spread over coordinates space,  
we introduce the 3D mass radius. Since the total mass radius is
independent of the $\bar{c}$ form factor, we unambiguously obtain it
in the flavor SU(3) symmetry 
\begin{align}
    \langle r^{2}_{\mathrm{mass}} \rangle_{p} = 0.54~\mathrm{fm}^{2},  
\end{align}
which is equal to the radius in the flavor SU(2) symmetry. 
It is important to note that, as discussed in
Refs.~\cite{Polyakov:2018rew, Won:2022cyy},  the mass radius is
smaller than the charge radius. 

Next, we consider the individual quark contributions to the proton
momentum fraction carried by the quark $A^{q}_{p}(0)$. This quantity
can be accessed directly from the twist-2 GFF
$\mathcal{E}^{q}_{p}(0)=\frac{3}{4}A^{q}_{p}(0)$ as follows: 
\begin{align}
    A^{u}_{p}(0)
& = 0.59, \quad 
    A^{d}_{p}(0)
  = 0.35, \quad 
    A^{s}_{p}(0)
  = 0.06, \quad [\, \mathrm{SU}(3) \,] \cr    
    A^{u}_{p}(0)
& = 0.62, \quad 
    A^{d}_{p}(0)
  = 0.38, \hspace{3.1cm} 
    [\, \mathrm{SU}(2) \,].
\end{align}
In other words, these numbers can be understood as the second Mellin
moments of the PDFs. We list the predictions of the proton momentum
fraction carried by the $u$-, $d$-, and $s$-quarks: 
\begin{align}
[ \langle x \rangle_{u} : \langle x \rangle_{d} : \langle x
  \rangle_{s} ] = [59\% : 35\% : 6\%]. 
\end{align}

\subsection{Angular momentum distribution \label{sec:4_2}}

By taking the components $\chi=0,3, 8$ from
equation~\eqref{eq:model_EMT}, we obtain the flavor-singlet, -triplet,
and -octet AMs.  While the flavor-singlet AM is appropriately
normalized to $J^{0}_{p}(0)=1/2$, given by 
\begin{align}
    J^{0}_{p}(0) 
  = \int d^{3}r \, 
    \rho^{0}_{J,p}(r) 
  = \frac{1}{2},
\end{align}
the flavor-triplet and -octet components are not constrained by
conserved quantities and are estimated as follows:
\begin{align}
&   J^{0}_{p} = 0.50, \quad   
    J^{3}_{p} = 0.58, \quad 
    J^{8}_{p} = 0.22, \quad [ \, \mathrm{SU}(3) \, ] \cr
&   J^{0}_{p} = 0.50, \quad 
    J^{3}_{p} = 0.55, \hspace{2.56cm} [ \, \mathrm{SU}(2) \, ].
\end{align}
The parametrically large value of the flavor-triplet AM in the flavor
SU(2) symmetry is retained in the flavor SU(3) symmetry. Furthermore,
the flavor-octet component exhibits the same order of $N_{c}$ as the
flavor-triplet component, but numerically it is approximately a half
of its magnitude. 

\begin{figure}[htp] 
\centering
\includegraphics[scale=0.147]{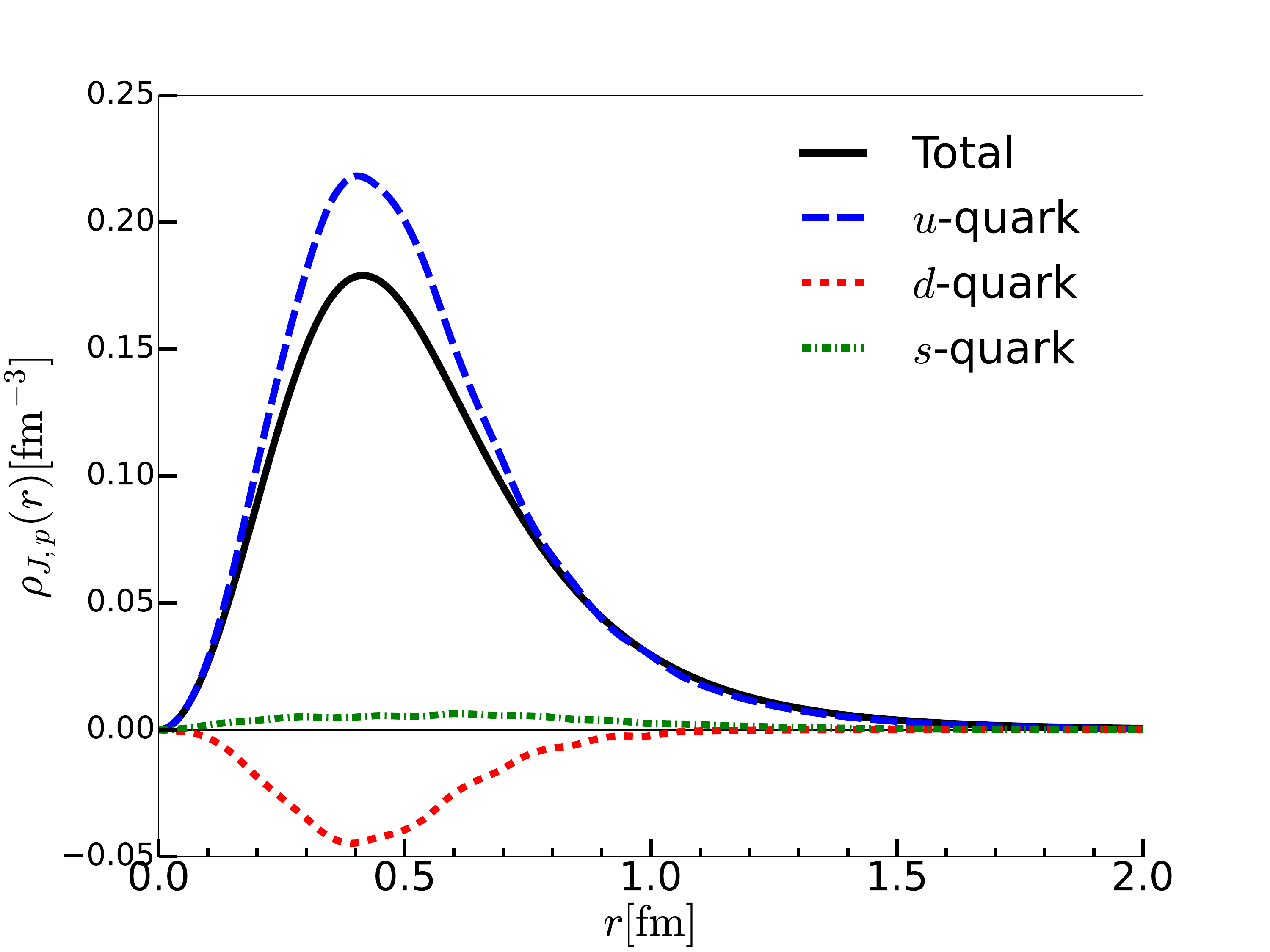}
  \caption{The 3D AM distribution of the nucleon and its flavor
    decomposition with the flavor SU(3) symmetry are drawn. The
    solid~(black), long-dashed~(blue), short-dashed~(red), and
    dashed-dotted~(green) curves denote the total, $u$-, $d$-, and
    $s$-quark contributions to the AM distributions, respectively.}  
\label{fig:3}
\end{figure}

Figure~\ref{fig:3} illustrates the individual flavor-decomposed AM
distributions inside the proton, utilizing the relations provided in
Eq.~\eqref{eq:Decomp}. Notably, the $u$- and $s$-quark
distributions exhibit positive values throughout the range of $r$,
while the $d$-quark distribution is negative. It implies that the
polarization of the $s$-quark aligns parallel to that of the
$u$-quark, whereas the $d$-quark polarization is arranged 
in the opposite direction to that of the $u$-quark. 
When these 3D AM distributions are integrated over $r$, the resulting
values are shown below: 
\begin{align}
&   J^{u}_{p} =  0.52, \quad 
    J^{d}_{p} = -0.06, \quad 
    J^{s}_{p} =  0.04, \quad [ \, \mathrm{SU}(3) \, ] \cr 
&   J^{u}_{p} =  0.53, \quad 
    J^{d}_{p} = -0.03, \hspace{2.55cm} [ \, \mathrm{SU}(2) \, ].
\end{align}
As expected, the total AM is mostly carried by the $u$-quark,  
whereas the $d$-quark and $s$-quark give only minor contributions. 
These results are in line with the predictions from the SU(2) version
of the $\chi$QSM~\cite{Won:2023rec} and are compatible with findings
from lattice QCD simulations~\cite{LHPC:2010jcs}. A comprehensive
analysis of the scale evolution of the AM form factors can be found  
in Refs.~\cite{Goeke:2007fq, Wakamatsu:2007ar}.

In comparison with the results obtained from the SU(2)
$\chi$QSM~\cite{Goeke:2007fp,Won:2023rec,Kim:2020nug},  
we observe that while the contribution of the $u$-quark to the total
AM remains nearly unchanged, the polarization of the $d$-quark
contribution is slightly enhanced. This suggests that the $s$ quark is
polarized in the opposite direction to the $d$ quark, effectively
canceling each other out and keeping the total AM at $1/2$.
Interestingly, the magnitude of the $s$ quark contribution to the
total AM is nearly equal to that of the $d$ quark
contribution. However, a non-trivial question arises regarding the 
decomposition of AM into spin and OAM. According to the Ji's
relation~\cite{Ji:1996ek}, the total AM can be expressed as the sum of
the intrinsic spin and the OAM: 
\begin{align}
    J  = \frac{1}{2}   \sum_{q} \Delta q 
  + \sum_{q} L^{q}+J_{g}, \quad J_{g}=0,
\label{eq:Ji}
\end{align} 
where we focus on the quark contributions, 
since the gluon contributions are parametrically suppressed 
in the QCD instanton vacuum~\cite{Balla:1997hf, Diakonov:1995qy}. In
the $\chi$QSM, the antisymmetric part of the $0k$ component of the
Ji's EMT current captures the spin of the $s$-wave quarks, while the
non-symmetric part accounts for the quark AM with OAM $L=1$.  
This implies that the static quark spin and the relativistic motion of
the quark explain the intrinsic spin and OAM,
respectively. Remarkably, we find that 50\% of the flavor-singlet AM  
is due to the relativistic motion of the quarks inside the nucleon:
\begin{align}
    \frac{1}{2}
  = \frac{1}{2} \sum_{q} \Delta q + \sum_{q} L^{q} = 0.23 + 0.27.
\end{align}
It is worth noting that the effect of corrections due to the strange
quark mass ($m_s$) on the AM decomposition has been estimated in
Ref.~\cite{Won:2022cyy} and found to be negligible, with only a few
percent effect on the proton. Furthermore, in the $\chi$QSM, the
validity of Ji's relation for the flavor singlet component has been
analytically proven in Refs.~\cite{Ossmann:2004bp, Goeke:2007fp}, even
in the presence of flavor SU(3) symmetry 
breaking~\cite{Won:2022cyy}. However, a careful treatment is required
to study the separate contributions of the quark flavors to the OAM
and the intrinsic spin for the following two reasons: different UV
divergence patterns between the total angular momentum and the
separate spin and the OAM~\cite{Kim:2023yhp}; lack of knowledge of the 
proper matching between the QCD operator and the twist-3 effective
operator~\cite{Kim:2023pll}. 

\subsection{Mechanical properties: Twist-2 part \label{sec:4_3}}
The $ij$ component of the EMT is related to the pressure and
shear-force distributions by the 3D Fourier transform and provides
crucial information for understanding the stability conditions of the
nucleon. To fully interpret these mechanical properties, knowledge of
both the $\bar{c}$ and $D$ term form factors is required. A well-known  
stability condition, known as the von Laue condition, arises from the
conservation of the EMT current. Similar to the normalization of the
mass and spin of the baryon, the von Laue condition serves as a
normalization condition for the stress tensor. 

First, if we use the EMT current~\eqref{eq:EMT_current} derived from
the N\"other theorem, all the quark and gluon parts are effectively
included in Eq.~\eqref{eq:EMT_current}. Using this effective current,
the analytical proof of the global stability condition was carried out
in Ref.~\cite{Goeke:2007fp}, which considers only $u$ and $d$
quarks. Importantly, this result holds even in the case of flavor
SU(3) symmetry, since the expression for the SU(2) isoscalar pressure
distribution~\cite{Goeke:2007fp,Won:2022cyy, Won:2023rec} coincides
with that of the SU(3) flavor-singlet pressure
distribution~\cite{Won:2022cyy}: 
\begin{align}
    \int d^{3}r \,   
    p^{u+d+s}_{p}(r) 
  = 0.
  \label{eq:st_p}
\end{align}
However, as emphasized in section~\ref{sec:Eff_op}, there is no
proper way of constructing the effective flavor-triplet and -octet EMT
currents by a global symmetry. Thus, instead of using
\eqref{eq:EMT_current_flav}, we will adopt the effective operator
\eqref{eq:EMT_current_tw2} derived from the QCD instanton vacuum and
concentrate on the twist-2 part. The integral of the twist-2 part of
the pressure distribution over $r$ is not zero:
\begin{align}
    \int d^{3}r \, \bar{p}^{u+d+s}_{p}(r)
  = \frac{1}{4} M_{N}, \quad  \int d^{3}r \, \hat{p}^{u+d+s}_{p}(r)
  = -\frac{1}{4} M_{N} .
\label{eq:hoho2}
\end{align}
Compared to Eq.~\eqref{eq:st_p}, the amount of energy $\frac{1}{4}
M_{N}$ in Eq.\eqref{eq:hoho2} leaks away to the twist-4 quark and gluon
part, of which the amount is $1/3$ times the normalization for the
twist-2 energy distribution. Obviously, there is no role for the
$\bar{c}$ form factor in the flavor-singlet twist-2 and twist-4
pressure distributions. On the other hand, as discussed in 
Eq.~\eqref{eq:d-term2}, each quark contribution to the von Laue
condition is due to the $\bar{c}$ form factor, which is beyond the
scope of the present work. What we can at least discuss is the flavor
decomposition of the twist-2 part 
\begin{align}
    &\int d^{3}r \, \bar{p}^{u}_{p}(r)
  = \frac{1}{4} M_{N} A^{u}_{p}(0), \quad      \int d^{3}r \, \bar{p}^{d}_{p}(r)
  = \frac{1}{4} M_{N} A^{d}_{p}(0), \cr      
  &\int d^{3}r \, \bar{p}^{s}_{p}(r)
  = \frac{1}{4} M_{N} A^{s}_{p}(0).
\end{align}
They are only proportional to the $A^{q}_{p}(0)$ form factors, which
in turn are canceled out by a part of the twist-4 part. 

Unlike the pressure distribution, the shear-force distribution is the
off-diagonal part of the EMT. Thus, regardless of the twist
classification, we can extract the $D$-term form factor from the
shear-force distribution. 
For the quark part, we obtain the flavor singlet, triplet, 
and octet $D$-term form factors by Fourier transform 
of the shear-force distributions:
\begin{align}
    D^{0}_{p}(0)
& = - 2.531, \quad 
    D^{3}_{p}(0)
  = 0.063, \quad 
    D^{8}_{p}(0)
  = - 0.697, \quad [ \, \mathrm{SU}(3) \,] \cr 
    D^{0}_{p}(0)
& = - 2.531, \quad
    D^{3}_{p}(0)
  = 0.295, \hspace{3.62cm} [ \, \mathrm{SU}(2) \,].
\end{align}
Note that the gluon contributions to the $D$-term form factor are
suppressed at low normalization points 
\begin{align}
D^{g}=0.
\end{align}
Burkert {\it{et al.}}~\cite{Burkert:2018bqq} analyzed the experimental
data on DVCS, and extracted the $D$-term form factor assuming the
large $N_{c}$ while neglecting $s$-quark contributions. In the flavor
SU(2) sector, the isovector component of the $D$-term is
parametrically small in the large $N_{c}$, but we find that it is
numerically non-negligible. On the other hand, in the flavor SU(3) case,
the flavor-triplet $D$-term is almost zero, i.e., $D^{3}_{p}\sim
0$. This suggests that the large $N_{c}$ assumption is more
appropriate for the flavor SU(3) sector.  In lattice QCD
simulations~\cite{Gockeler:2003jfa, LHPC:2007blg} in the flavor SU(2)
sector, the value of the isovector $D$-term is found to be very
small. However, lattice results suffer from substantial uncertainties,
leaving the sign of the isovector $D$-term undetermined. Very
recently, the lattice simulation~\cite{Hackett:2023rif} performed in
the flavor SU(3) sector predicts an almost zero value of the isovector
$D$-term form factor, which is consistent with the prediction of the
current work.

\begin{figure}[htp]
\centering
\includegraphics[scale=0.147]{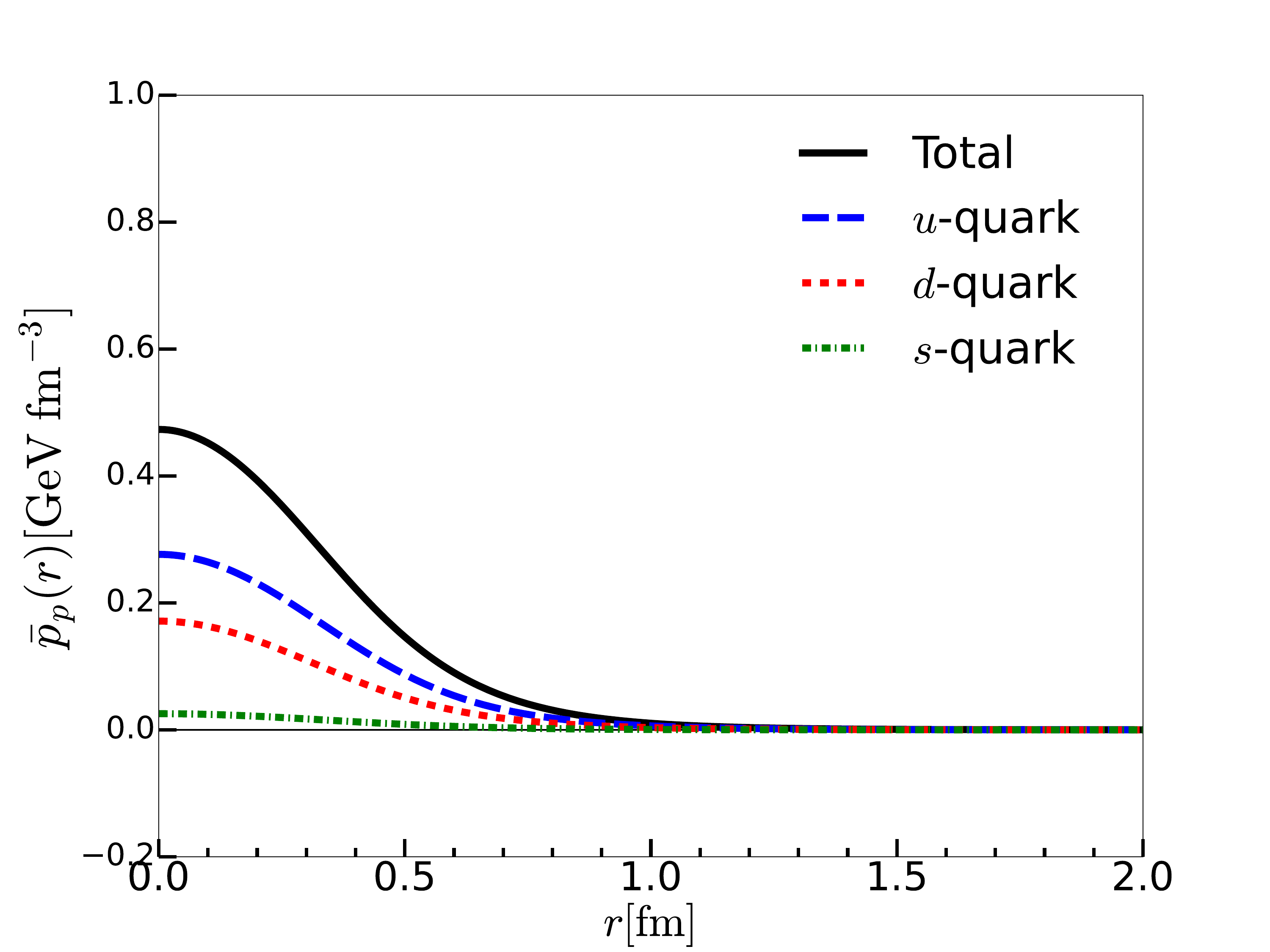}
\includegraphics[scale=0.147]{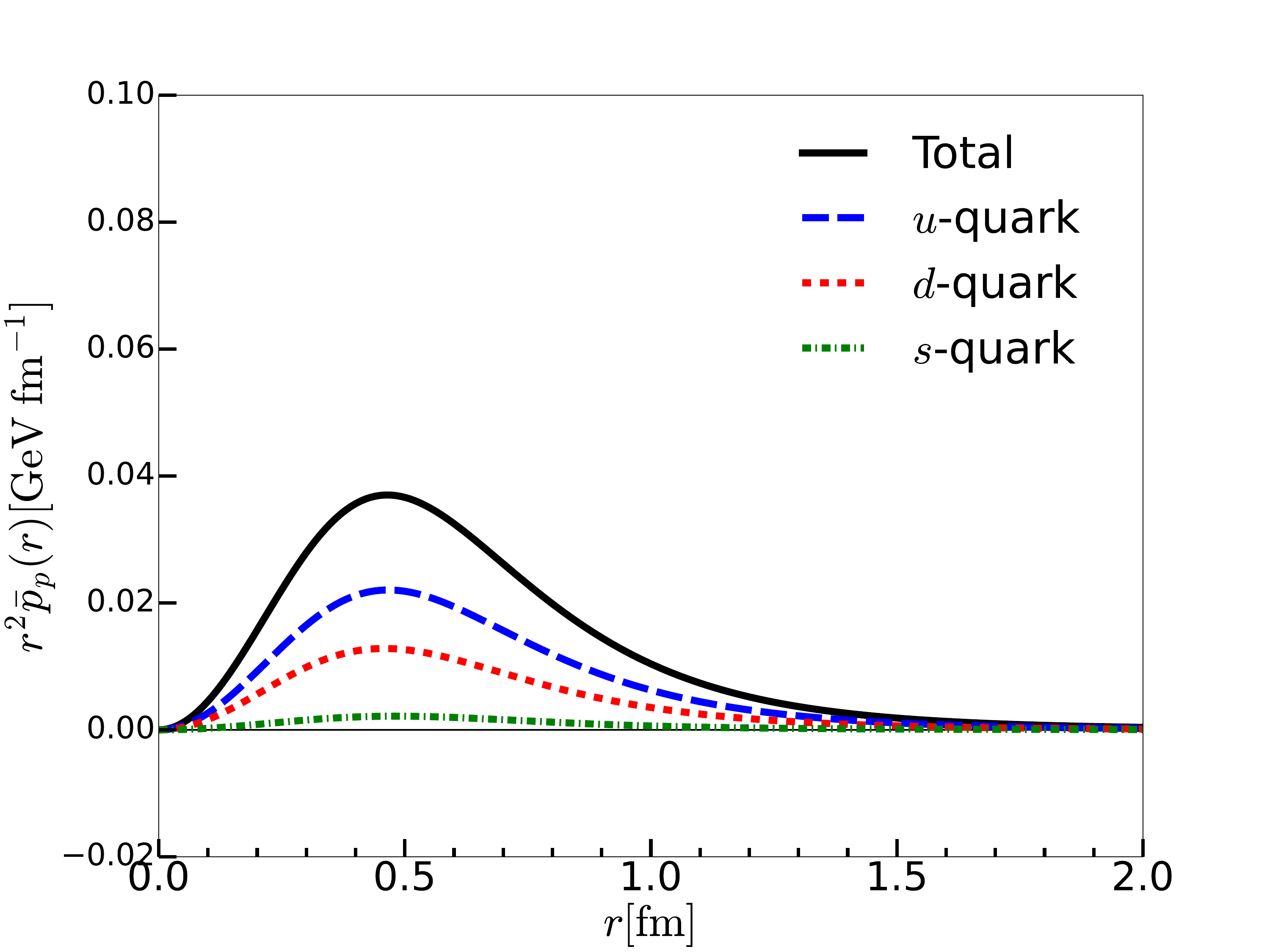}
\caption{The 3D pressure distribution of the nucleon and its flavor
  decomposition with SU(3) symmetry are plotted. 
The solid (black), long-dashed (blue), short-dashed (red), 
and dashed-dotted (green) curves denote the total, 
$u$-, $d$-, and $s$-quark contributions to the pressure distributions,
respectively.}   
\label{fig:4}
\end{figure}

We obtain the flavor-decomposed pressure and shear-force distributions   
by linearly combining the $\chi=0,3,8$ components. 
The resulting distributions are depicted in Figure~\ref{fig:4}.
The values of the flavor-decomposed pressures at the center of the
proton are given by 
\begin{align}
    \bar{p}^{u}_{p}(0)
& = 0.28~\mathrm{GeV/fm}^{3}, \quad     
    \bar{p}^{d}_{p}(0)
  = 0.17~\mathrm{GeV/fm}^{3}, \cr      
    \bar{p}^{s}_{p}(0)
& = 0.03~\mathrm{GeV/fm}^{3}.        
\label{eq:center_pressure}
\end{align}
As discussed in Eq.~\eqref{eq:relation_btw_ep}, the pressure
distribution is three times smaller than the energy distribution,
which is reflected in the numbers listed in ~\eqref{eq:center_energy}
and~\eqref{eq:center_pressure}. The dominance of the $u$-quark in the
central region is evident, as valence quarks tend to cluster around the
core. To satisfy the von Laue condition, the total pressure
distribution $p^{u+d+s}$ must have at least one nodal
point. However, the twist-2 pressure distribution for each quark is
always positive over $r$ [see \eqref{eq:positive_energy}]. This
implies that it has no nodal point at all. As a result, since the
twist-2 part shows only the repulsive force, we can expect that the
twist-4 part would provide attractive forces to stabilize the
nucleon system.   

\begin{figure}[htp]
\centering
\includegraphics[scale=0.147]{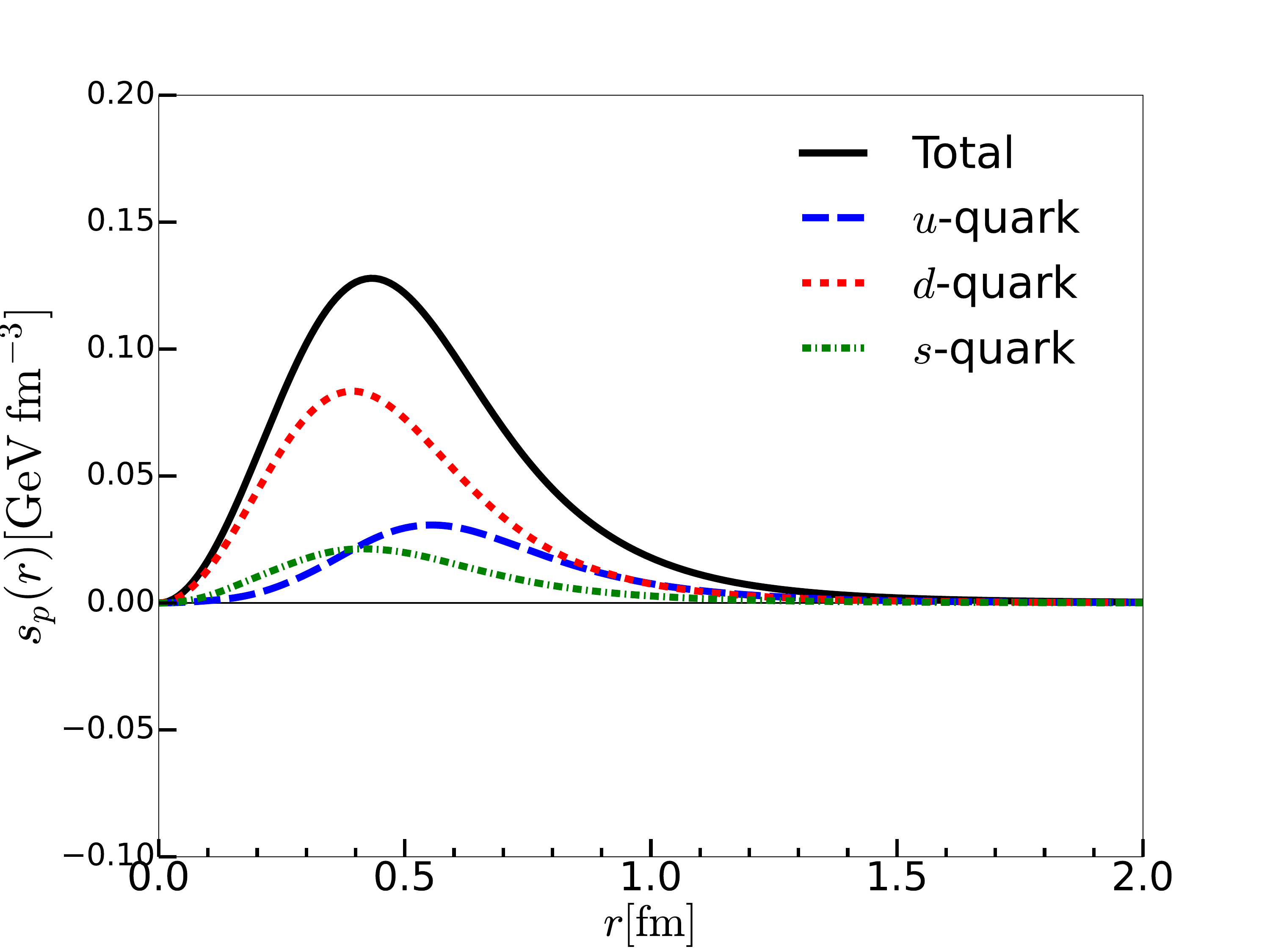}
\includegraphics[scale=0.147]{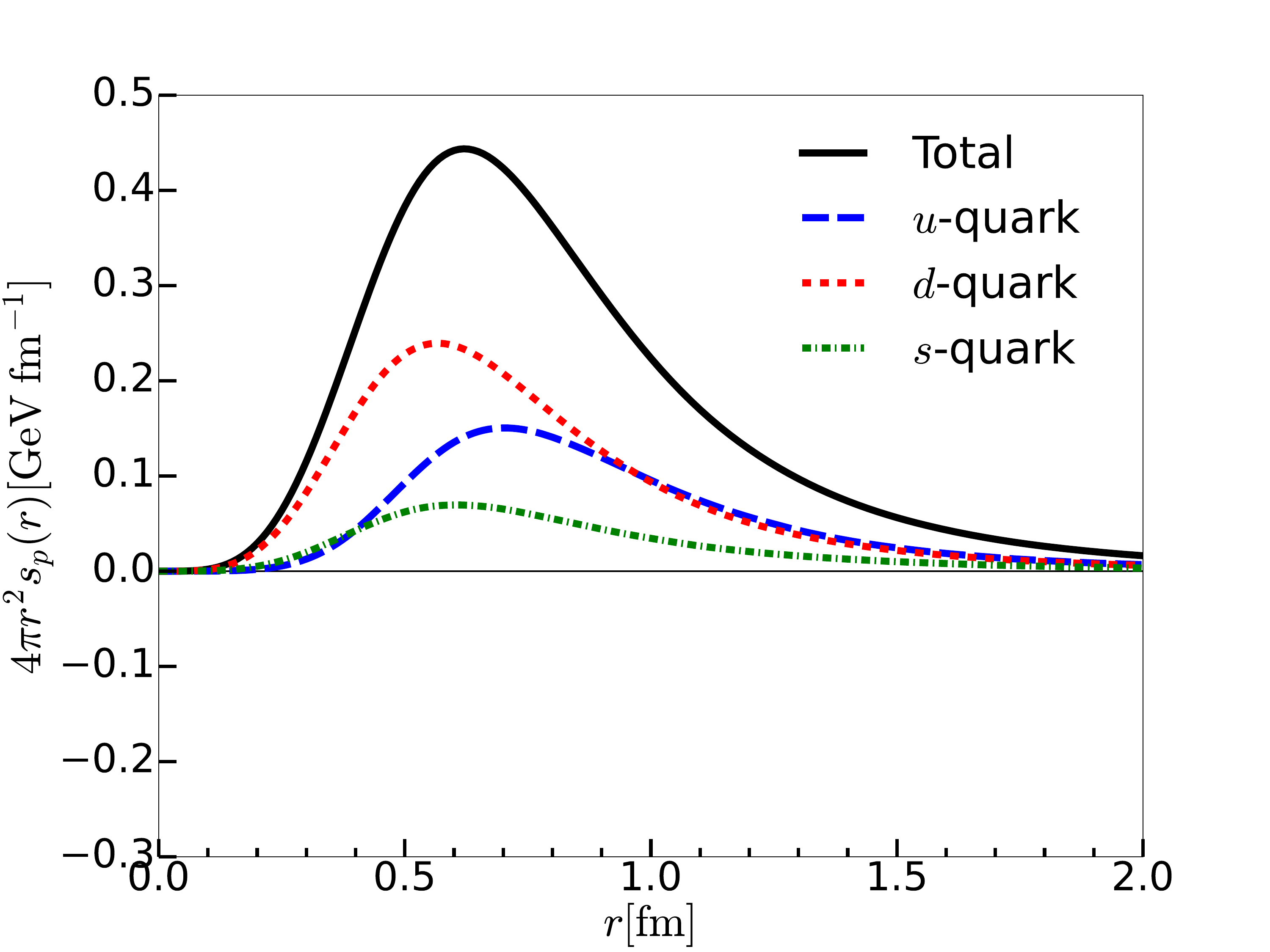}
  \caption{The shear-force distribution of the nucleon and its flavor
    decomposition with the flavor SU(3) symmetry are drawn. The
    solid~(black), long-dashed~(blue), short-dashed~(red), and
    dashed-dotted~(green) curves denote the total, $u$-, $d$-, and
    $s$-quark contributions to the pressure distributions,
    respectively.}   
\label{fig:5}
\end{figure}

In Fig.~\ref{fig:5}, the flavor-decomposed shear-force distributions
are drawn. 
All the flavor-decomposed shear-force distributions have been 
determined to be positive throughout the range of $r$. 
This positive definiteness of the shear force distribution, 
$\sum_{q} s^{q}_{p}(r)>0$, leads to the 
inequality~\cite{Goeke:2007fp, Perevalova:2016dln, Polyakov:2018zvc,
  Lorce:2018egm}: 
\begin{align}
\frac{2}{3} s_{p}(r) + p_{p}(r) > 0,
\end{align}
which comes from the equilibrium equation~\eqref{eq:diffEq2} 
relating the pressure distribution to shear-force one.

\subsection{Flavor-decomposed GFFs of the proton  \label{sec:4_4}}
\begin{figure}[htp]
\centering
\includegraphics[scale=0.147]{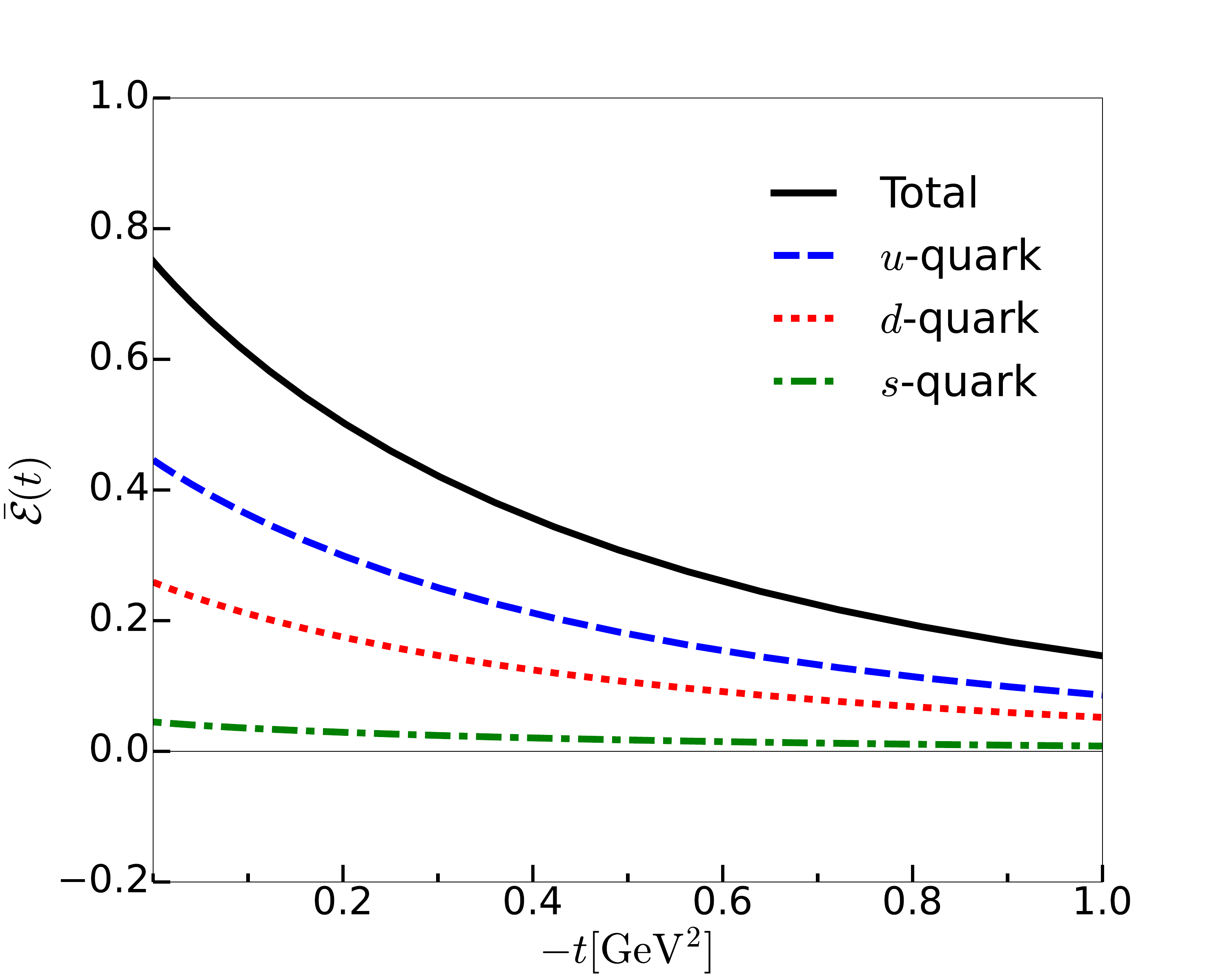}
\includegraphics[scale=0.147]{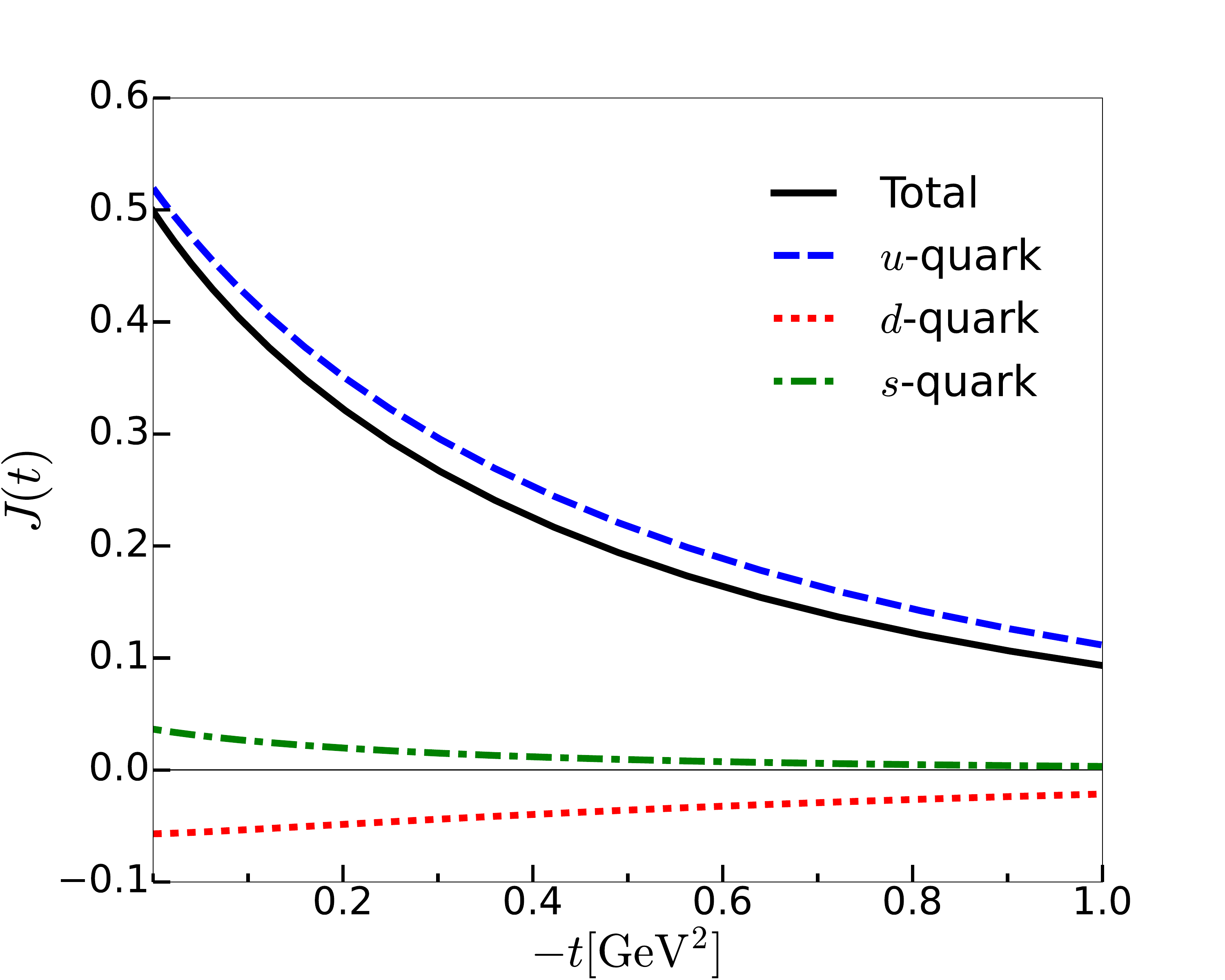}
\includegraphics[scale=0.147]{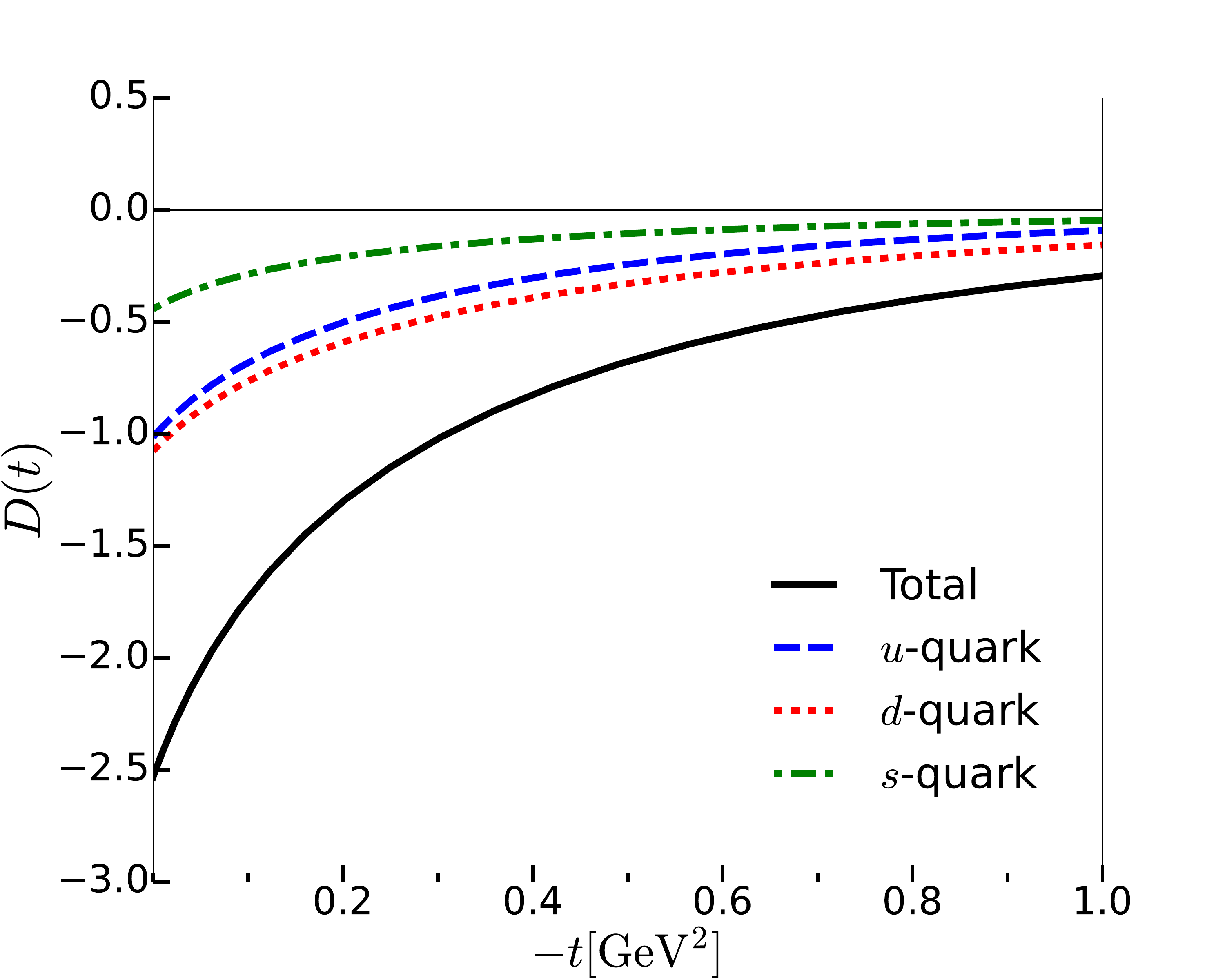}
  \caption{The flavor-decomposed gravitational form factors of the
    proton are drawn. The solid~(black), long-dashed~(blue),
    short-dashed~(red), and dashed-dotted~(green) curves denote the
    total, $u$-, $d$-, and $s$-quark contributions to the GFFs,
    respectively.}  
  \label{fig:6}
\end{figure}

We are now in a position to examine the $t$-dependence of the nucleon
GFFs in the flavor SU(3) symmetry. By performing a 3D Fourier
transform of the EMT distributions, we obtain the GFFs, which are
depicted in Fig.~\ref{fig:6}. As discussed in the previous subsection, 
we have observed that the $s$-quark contributions to the $A$ and $J$
form factors are marginal. However, the $s$-quark's influences on the
$D$ form factor is found to be non-negligible.  
Consequently, the $s$-quark plays an important role in the mechanical 
interpretation of the proton. For additional insights into the contributions 
of valence and sea quarks to the GFFs, refer to Ref.~\cite{Won:2023ial}.

\subsection{SU(3) spin-flavor sturcture and the hyperon GFFs  \label{sec:4_4}}
In the large $N_{c}$ limit of QCD, the relation between the
lowest-lying baryons can be understood in a model-independent manner
using spin-flavor symmetry. While the GFFs in the flavor SU(2)
symmetry were investigated in Ref.~\cite{Kim:2023xvw}, we aim to
extend this analysis to the flavor SU(3) sector in our current
work. The chiral soliton approach describes the spin-flavor symmetry
using collective operators. The matrix elements of these 
operators, which are listed in Tables~\ref{tab:0} and ~\ref{tab:01},  
provide insights into the spin-flavor structure.

Utilizing the matrix elements of the spin-flavor operators, 
we establish the following spin-flavor relations in the flavor SU(3)
symmetry: 
\begin{itemize}
\item The flavor-singlet GFFs for the octet baryons are degenerate. 
\item The flavor-triplet GFFs are propotional to the isospin
  projection $T_{3}$, i.e., $F^{3}_{B} \propto T_{3}$. Consequently,
  we find the relations: 
      \begin{align}
      &    \sum_{B \in \mathrm{octet}}   F^{3}_{B}  = 0, \ \  \sum_{ B=p, n} F^{3}_{B} = 0,  \ \
      &    \sum_{ B=\Sigma^{+}, \Sigma^{0}, \Sigma^{-}} F^{3}_{B} = 0,  \ \
      &    \sum_{ B=\Xi^{0}, \Xi^{-}} F^{3}_{B} = 0.
      \end{align}
\item The flavor-octet GFFs for the iso-multipolets are
  degenerate. Additionally, we obtain: 
      \begin{align}
      &    \sum_{B \in \mathrm{octet}}   F^{8}_{B}  = 0, \quad   \sum_{B = \Lambda, \Sigma} F^{8}_{B} = 0.
      \end{align}  
\end{itemize}
Remarkably, our numerical calculations confirm that these spin-flavor
relations are indeed satisfied; see Tab.~\ref{tab:1}.
\begin{table}[htb] 
\centering
\caption{Flavor-decomposed gravitational form factors for the octet
  baryons}   
\begin{center}
  \renewcommand{\arraystretch}{1.7}
\scalebox{0.80}{%
\begin{tabular}{crrrrrrrrrrrr} 
  \hline
  \hline
  $B$    
  & $A_{B}^{u}  ( 0 ) $  & $A_{B}^{d} ( 0 ) $ & $A_{B}^{s} ( 0 ) $  
  & $J_{B}^{u}  ( 0 ) $  & $J_{B}^{d} ( 0 ) $ & $J_{B}^{s} ( 0 ) $  
  & $D_{B}^{u}  ( 0 ) $  & $D_{B}^{d} ( 0 ) $ & $D_{B}^{s} ( 0 ) $  \\
  \hline
  $p$          
  &  $0.595$ &  $0.345$ &  $0.060$ 
  &  $0.520$ & $-0.057$ &  $0.036$ 
  & $-1.014$ & $-1.076$ & $-0.441$  \\
  $n$          
  &  $0.345$ &  $0.595$ &  $0.060$ 
  & $-0.057$ &  $0.520$ &  $0.036$ 
  & $-1.076$ & $-1.014$ & $-0.441$  \\
  $\Lambda$    
  &  $0.339$ &  $0.339$ &  $0.321$ 
  &  $0.055$ &  $0.055$ &  $0.390$ 
  & $-0.960$ & $-0.960$ & $-0.611$  \\
  $\Sigma^{+}$ 
  &  $0.595$ &  $0.060$ &  $0.345$ 
  &  $0.520$ &  $0.036$ & $-0.057$ 
  & $-1.014$ & $-0.441$ & $-1.076$  \\
  $\Sigma^{0}$ 
  &  $0.327$ &  $0.327$ &  $0.345$ 
  &  $0.278$ &  $0.278$ & $-0.057$ 
  & $-0.727$ & $-0.727$ & $-1.076$  \\
  $\Sigma^{-}$ 
  &  $0.060$ &  $0.595$ &  $0.345$ 
  &  $0.036$ &  $0.520$ & $-0.057$ 
  & $-0.441$ & $-1.014$ & $-1.076$  \\
  $\Xi^{0}$    
  &  $0.345$ &  $0.060$ &  $0.595$ 
  & $-0.057$ & $-0.036$ &  $0.552$ 
  & $-1.076$ & $-0.441$ & $-1.014$  \\
  $\Xi^{-}$    
  &  $0.060$ &  $0.345$ &  $0.595$ 
  & $-0.036$ & $-0.057$ &  $0.520$ 
  & $-0.441$ & $-1.076$ & $-1.014$  \\
  \hline
  \hline
\end{tabular}}
\end{center}
\label{tab:1}
\end{table}

\begin{figure}[htp]
\centering
\includegraphics[scale=0.147]{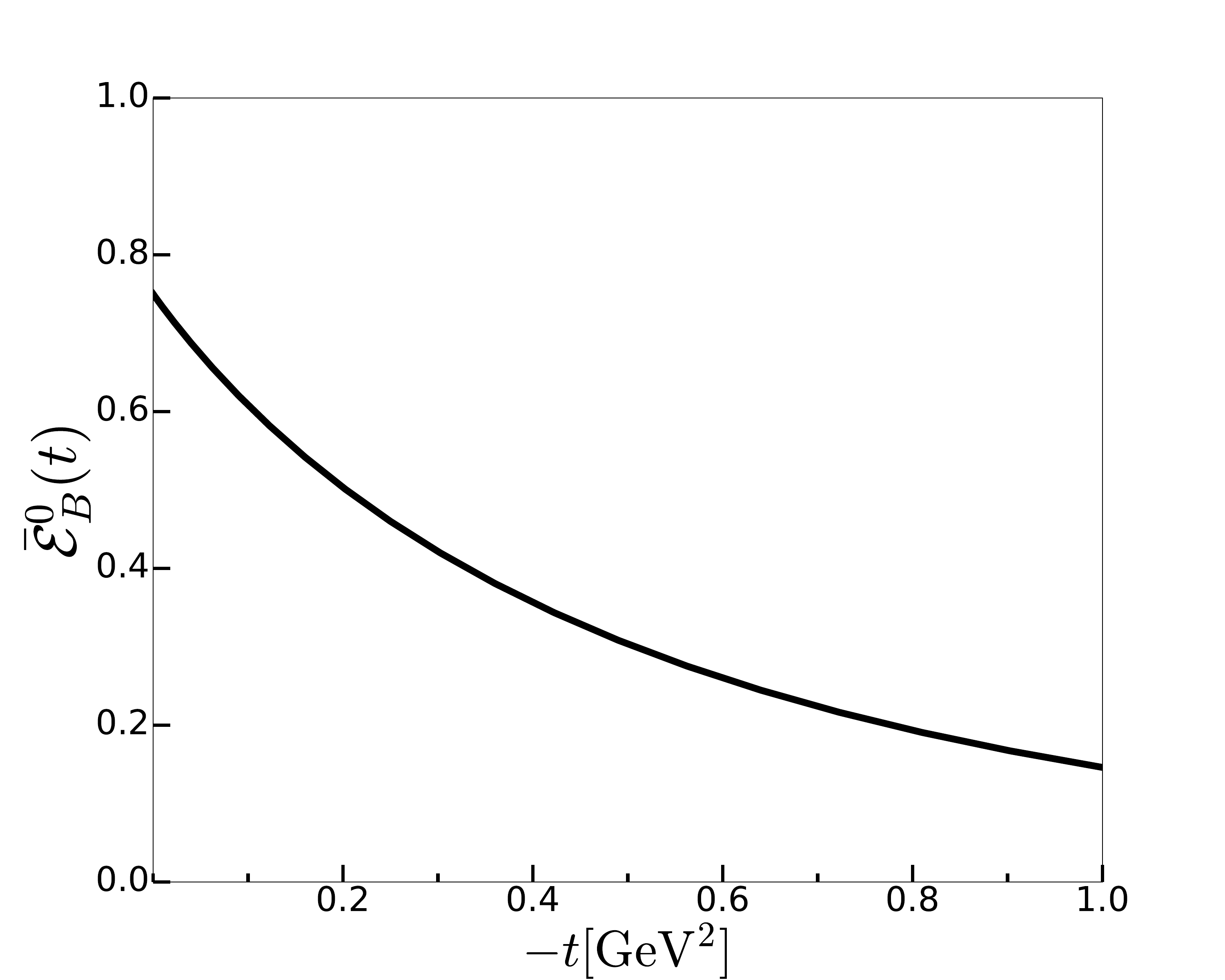}
\includegraphics[scale=0.147]{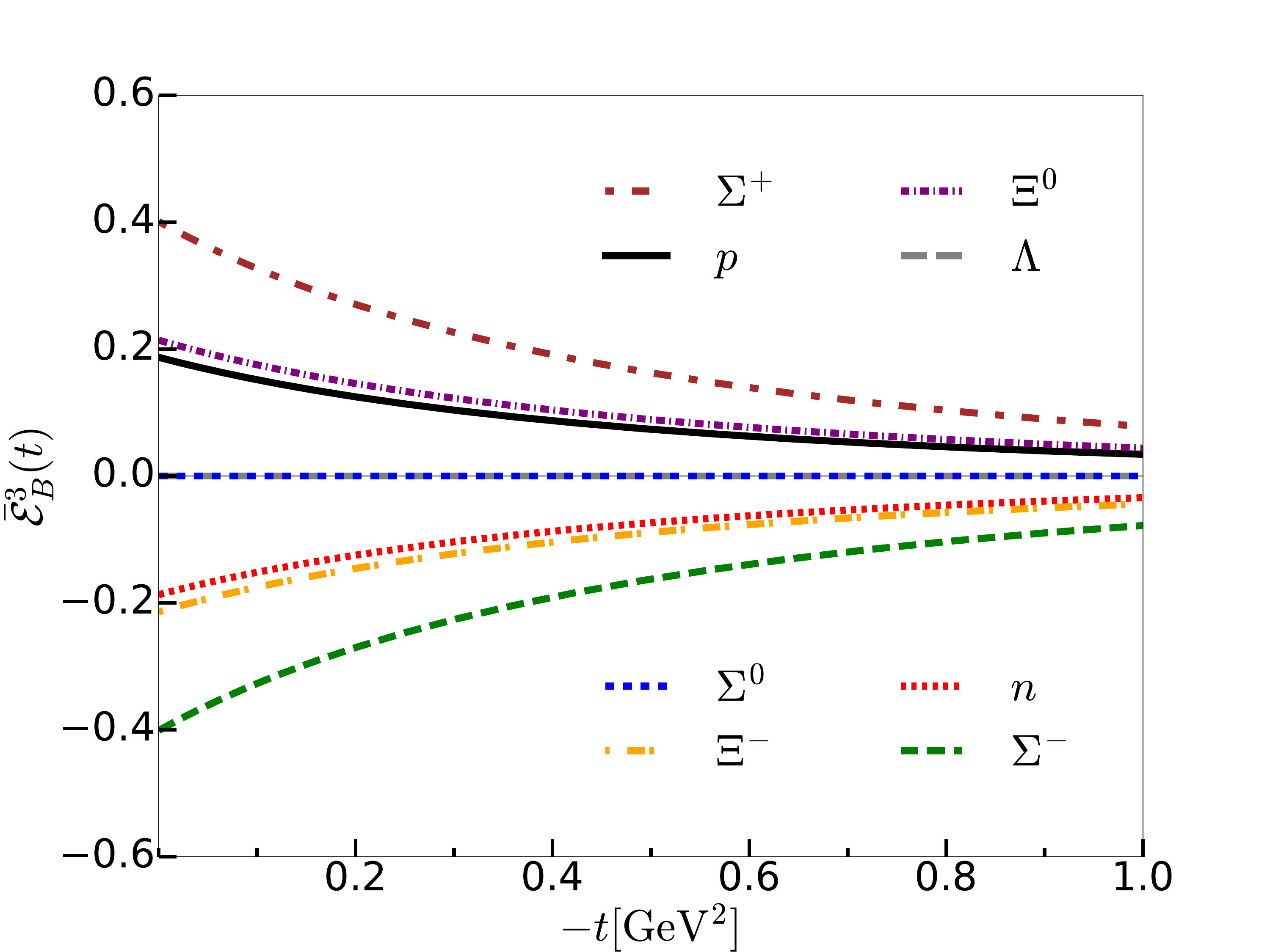}
\includegraphics[scale=0.147]{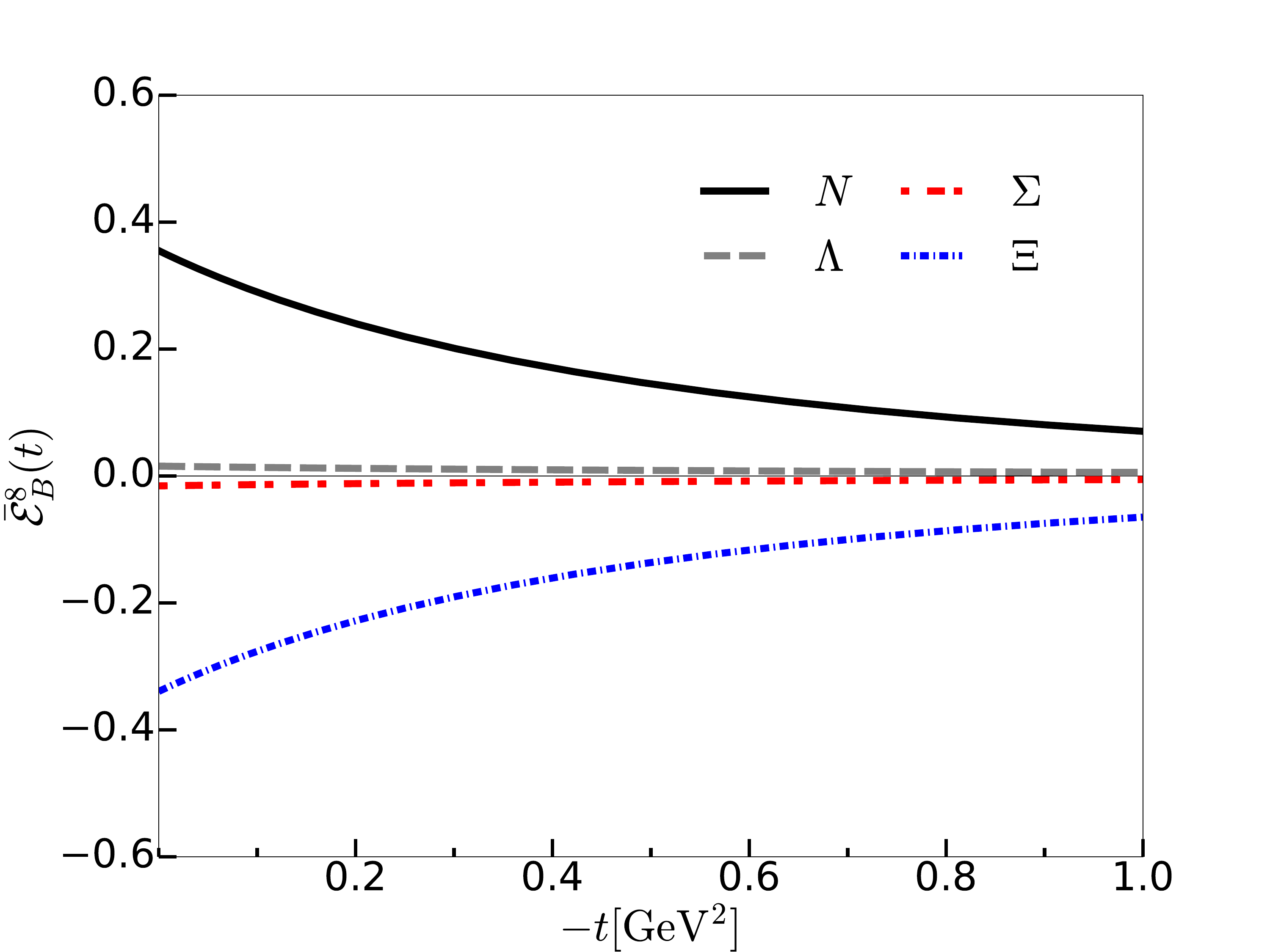}
  \caption{Flavor-singlet, -triplet, and -octet $\bar{\mathcal{E}}$ form factors for the octet baryons are drawn.}  
  \label{fig:7}
\end{figure}
Figure~\ref{fig:7} displays the $\bar{\mathcal{E}}^{0,3,8}$ form factors 
for the octet baryons. Firstly, we observe that the
$\bar{\mathcal{E}}^{0}$ form factors for the baryon octet are clearly
degenerate, which arises from the absence of $m_{s}$ corrections. 
Introducing these corrections would break the degeneracy 
among the mass form factors $\bar{\mathcal{E}}$ for the baryon octet, 
as discussed in Ref.~\cite{Won:2022cyy}. 
Secondly, the $\bar{\mathcal{E}}^{3}$ form factors for the baryon octet
are proportional to the third component of the isospin,
denoted as $\propto T^{3}$. This implies that
$\bar{\mathcal{E}}^{3}_{\Lambda^{0},\Sigma^{0}}(t)=0$. Additionally,
the sums of the $\bar{\mathcal{E}}^{3}$ form factors for the
iso-multiplets yield zero. Lastly, the $\bar{\mathcal{E}}^{8}$ form
factors for the iso-multiplet members are degenerate.   
Interestingly, we derive the following relations:
\begin{align}
\sum_{B=N,\Xi} \bar{\mathcal{E}}^{8}_{B} = 0, \quad \sum_{B=\Lambda,\Sigma}
  \bar{\mathcal{E}}^{8}_{B}=0. 
\end{align}
Consequently, Table~\ref{tab:1} provides the flavor-decomposed
GFFs. Notably, the $u$-, $d$-, and $s$-quarks carry almost equally the
momentum fraction of the $\Lambda^{0}$ and $\Sigma^{0}$ baryons. This
follows from the nearly vanishing values of the flavor-octet
components $A^{8}_{\Sigma^{0}, \Lambda^{0}}(0) \sim 0$. Comparing it
with the proton, the $\Sigma^{+}$ baryon contains one less $d$-quark
and one additional $s$-quark in the valence level, leading to an
exchange in the roles of $\bar{\mathcal{E}}^{d}$ and
$\bar{\mathcal{E}}^{s}$ between the proton and the $\Sigma^{+}$
baryon. Similar tendencies are observed for the $\Sigma^{-}$,
$\Xi^{0}$, and $\Xi^{-}$ baryons.  

\begin{figure}[htp]
\centering
\includegraphics[scale=0.147]{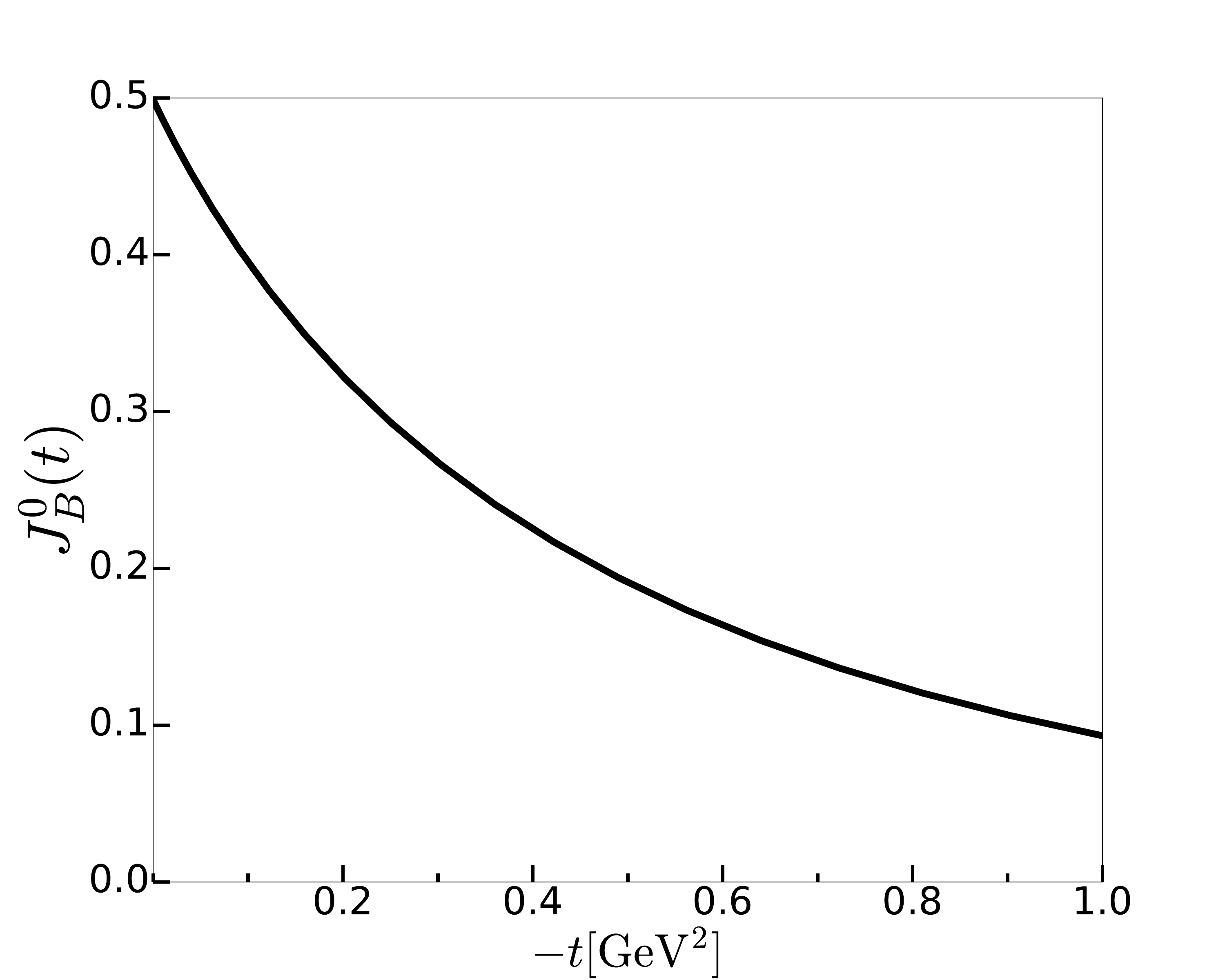}
\includegraphics[scale=0.147]{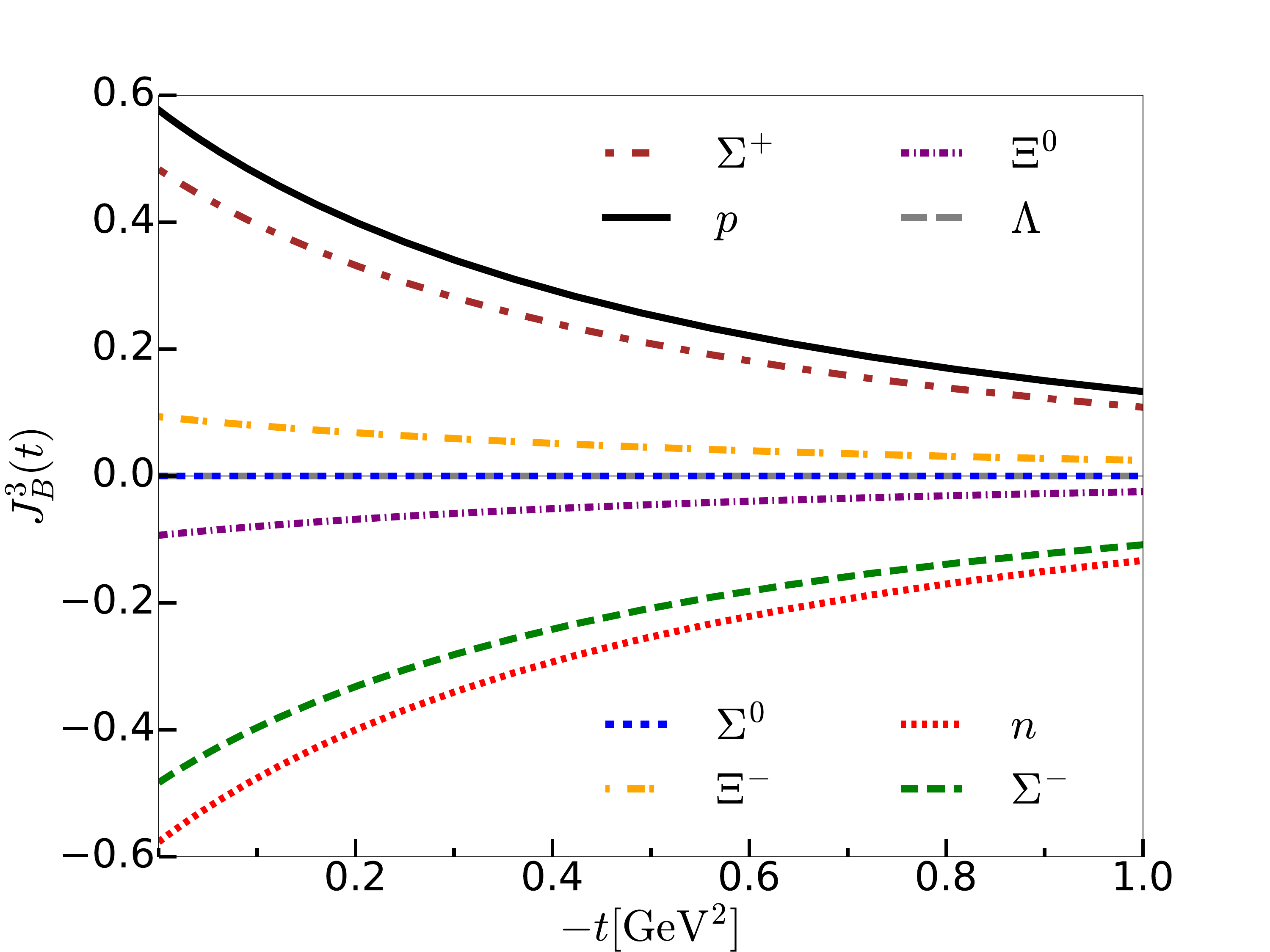}
\includegraphics[scale=0.147]{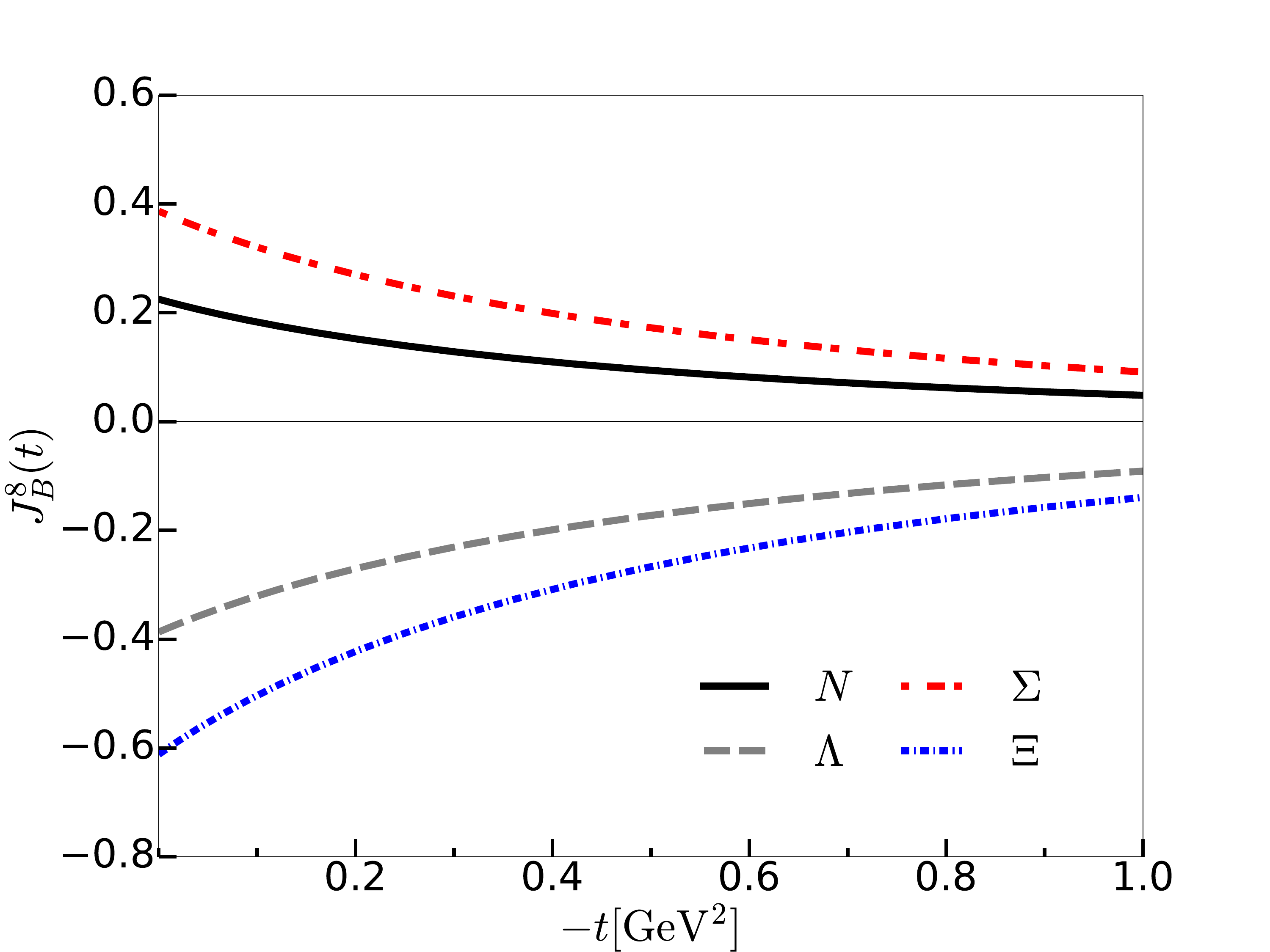}
  \caption{Flavor-singlet, -triplet, and -octet $J$ form factors for
    the baryon octet are drawn.}   
  \label{fig:8}
\end{figure}
Figure~\ref{fig:8} illustrates the $J^{0,3,8}$ form factors for the
octet baryons. Firstly, we observe that the $J^{0}$ form factors for
the octet baryons are also degenerate. However, this degeneracy can be
lifted by considering $m_{s}$ corrections, as discussed in
Ref.~\cite{Won:2022cyy}. Secondly, the $J^{3}$ form factor  
for the octet baryons is once again proportional to $T^{3}$, 
resulting in null results for the $\Sigma^{0}$ and $\Lambda^{0}$
baryons.  While the octet components $J^{8}$ for the iso-multiplet
members remain degenerate, we find the following relations:
\begin{align}
\sum_{B=N,\Xi} J^{8}_{B} \neq 0, \quad \sum_{B=\Lambda,\Sigma}
  J^{8}_{B}=0. 
\end{align}
Thus, in contrast to the $\bar{\mathcal{E}}^{8}$ form factors,  the
angular-momentum form factors have a different 
flavor structure. In Table~\ref{tab:1}, we provide the
flavor-decomposed $J$ form factors. It is interesting to note that 
the $\Lambda^{0}$ and $\Sigma^{0}$ baryons exhibit different quark
contributions despite having the same quark content.  This finding is
reminiscent of the flavor-decomposed axial charges presented  
in Ref.~\cite{Suh:2022atr}:
\begin{align}
    \Delta u_{\Lambda^{0}}
& = - 0.093, \quad 
    \Delta d_{\Lambda^{0}}
  = - 0.093, \quad 
    \Delta s_{\Lambda^{0}}
  = + 0.623, \cr
    \Delta u_{\Sigma^{0}}
& = + 0.384, \quad 
    \Delta d_{\Sigma^{0}}
  = + 0.384, \quad 
    \Delta s_{\Sigma^{0}}
  = - 0.332.
\end{align}
Thus, we conclude that the spin and OAM of the $s$-quark 
in the $\Lambda^{0}$ baryon are strongly polarized, 
while those in the $\Sigma^{0}$ baryon are relatively weakly polarized, 
despite both baryons having the same quark content. 
For the $p$, $n$, $\Sigma^{+}$, $\Sigma^{-}$, $\Xi^{0}$, and $\Xi^{-}$
baryons, we can easily obtain the flavor-decomposed $J^{q}(0)$ form
factors by considering the number of valence quarks. 
For instance, the flavor-decomposed $J^{q}(0)$ form factors 
for the $\Sigma^{+}$ baryon are found to be:
\begin{align}
    \mathrm{two \ quarks \ with \ the \ same \ flavor}~(u)& \to
                                                            J^{u}(0)=
                                                            +0.520,
                                                            \cr 
    \mathrm{one \ quark}~(s)& \to J^{s}(0)= -0.057, \cr
    \mathrm{non}\text{-}\mathrm{valence \ quark}~(d)& \to J^{d}(0)=
                                                      +0.036. 
\end{align}

\begin{figure}[htp]
\centering
\includegraphics[scale=0.147]{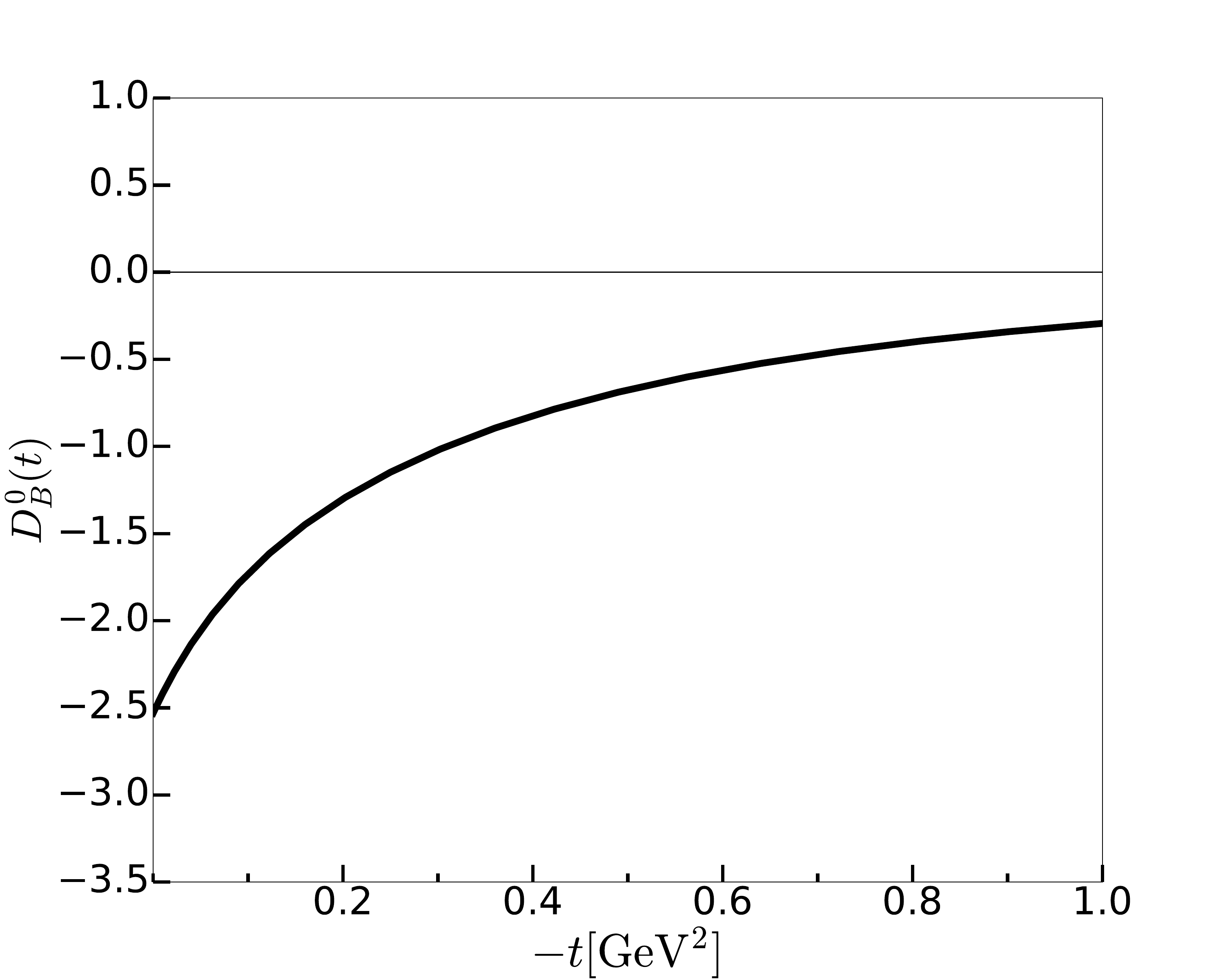}
\includegraphics[scale=0.147]{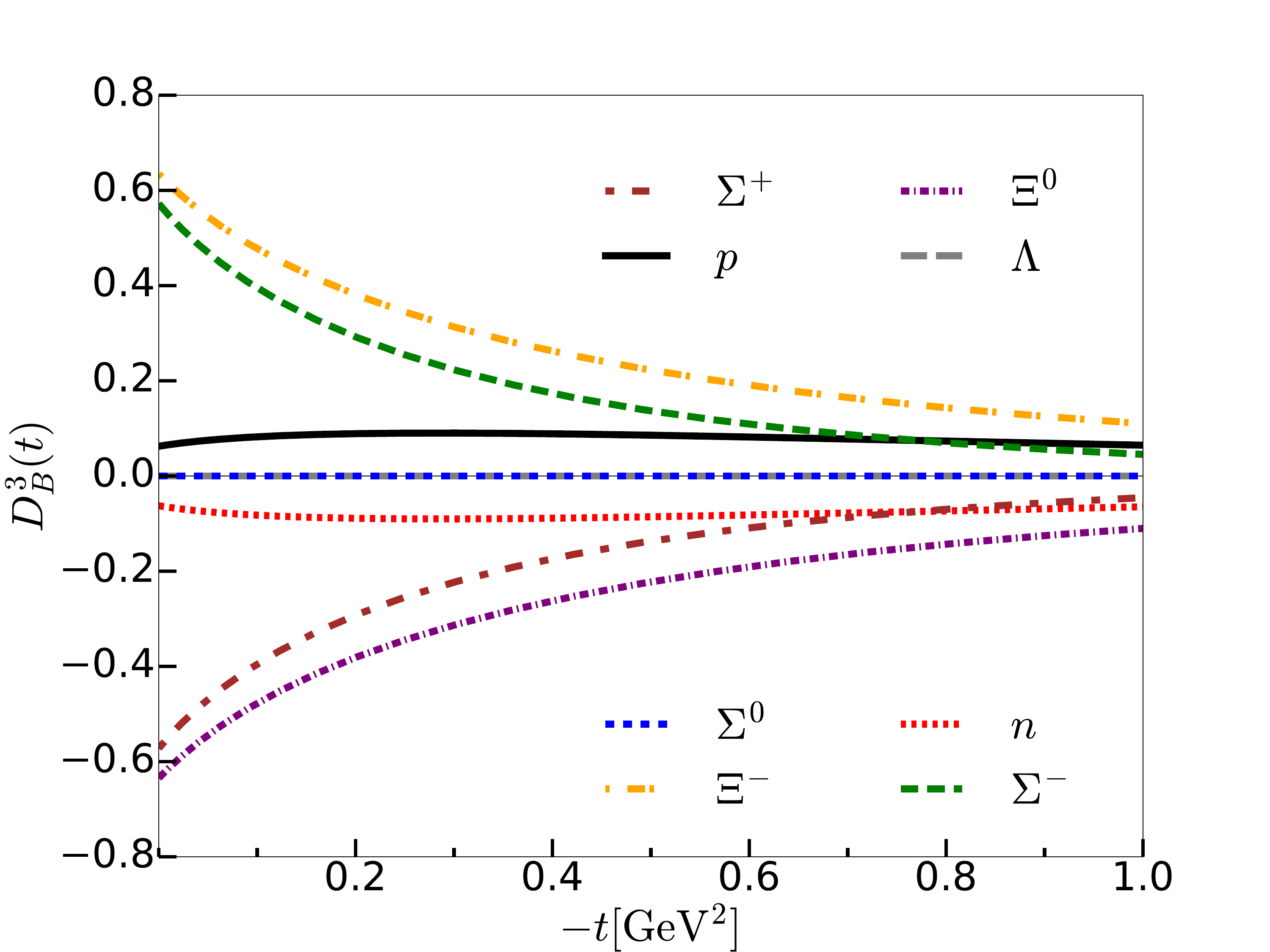}
\includegraphics[scale=0.147]{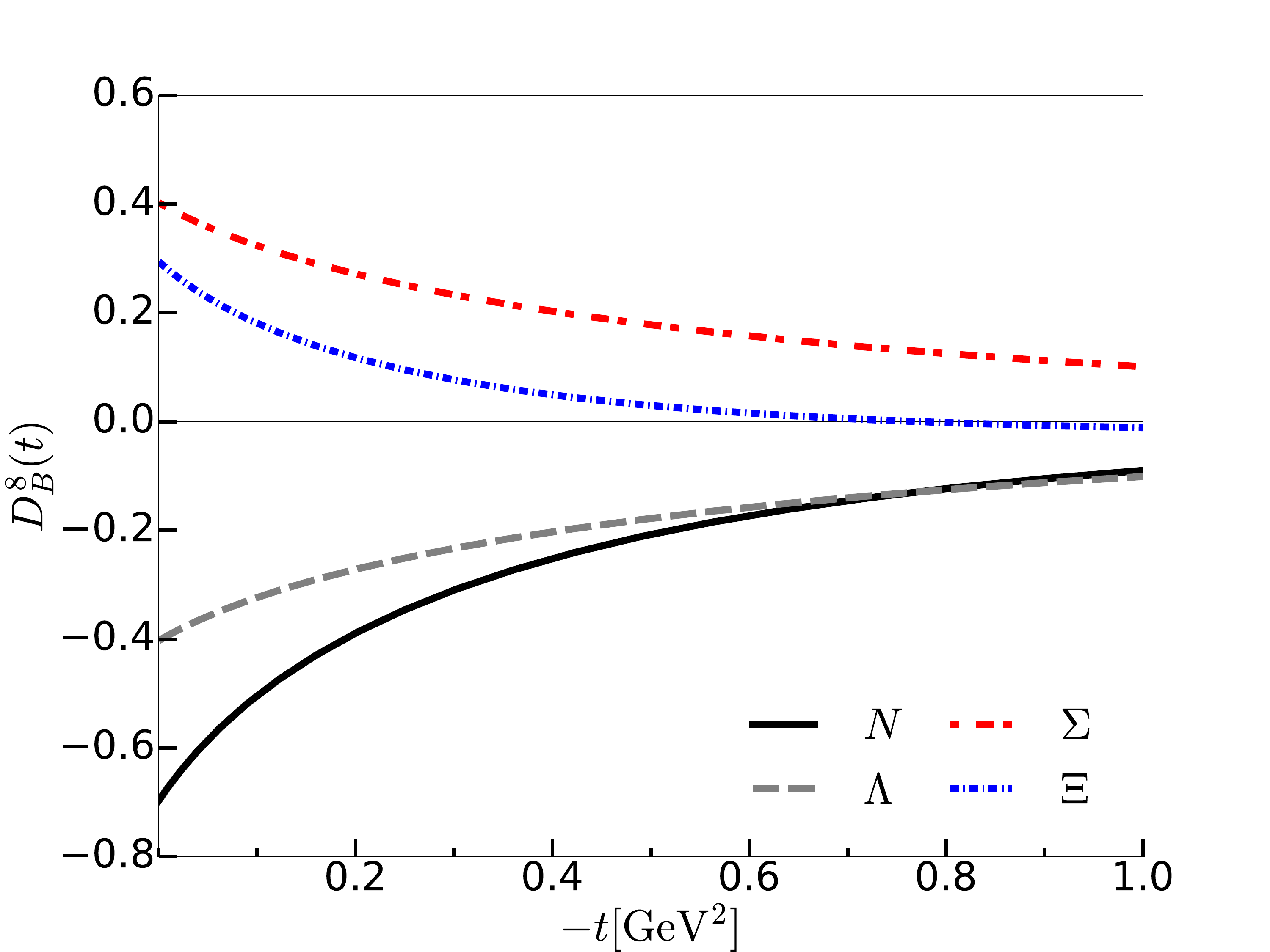}
  \caption{Flavor-singlet, -triplet, and -octet $D$-term form factors
    for the baryon octet are drawn.}   
  \label{fig:9}
\end{figure}
In Fig.~\ref{fig:9}, we present the $D^{0,3,8}$ form factors 
for the octet baryons. We observe that the $D^{0}(t)$ form factors 
for the octet baryons are degenerate, similar to the previous cases. 
Additionally, the $D^{3}$ form factors for the octet baryons 
are proportional to $T^{3}$, following the pattern we have seen before. 
Similarly to the $J^{8}$ form factor, the flavor-octet $D^{8}$ form factors 
for the iso-multiplet members are degenerate, leading to the following
relations: 
\begin{align}
\sum_{B=N,\Xi} D^{8}_{B} \neq 0, \quad \sum_{B=\Lambda,\Sigma}
  D^{8}_{B}=0. 
\end{align}
Next, we investigate the flavor-decomposed $D$-term form factors. 
Interestingly, unlike the $J(t)$ form factors for the $\Lambda^{0}$ and
$\Sigma^{0}$ baryons, we find that the $s$-quark contributions to the
$D$-term for the $\Sigma^{0}$ baryon are larger than those for the
$\Lambda^{0}$ baryon: 
\begin{align}
    D^{u}_{\Lambda^{0}}
  = - 0.960, \quad 
    D^{d}_{\Lambda^{0}}
  = - 0.960, \quad 
    D^{s}_{\Lambda^{0}}
  = - 0.611, \cr
    D^{u}_{\Sigma^{0}}
  = - 0.727, \quad 
    D^{d}_{\Sigma^{0}}
  = - 0.727, \quad 
    D^{s}_{\Sigma^{0}}
  = - 1.076.
\end{align}
Similarly, for the other octet baryons $p$, $n$, $\Sigma^{+}$,
$\Sigma^{-}$, $\Xi^{0}$, and $\Xi^{-}$, we find that the
flavor-decomposed $D^{q}$ form factors can be obtained  
by counting the number of valence quarks. For example, the
flavor-decomposed $D$-term form factors for the $\Sigma^{+}$ baryon
are given by: 
\begin{align}
    \mathrm{two \ quarks \ with \ the \ same \ flavor}~(u)& \to D^{u}(0)= -1.014, \cr
    \mathrm{one \ quark}~(s)& \to D^{s}(0)= -1.076, \cr
    \mathrm{non}\text{-}\mathrm{valence \ quark}~(d)& \to D^{d}(0)= -0.441.
\end{align}
These relations are exactly the same as the flavor-decomposed 
$J(t)$ form factors for the baryon octet. It is important 
to note that the contributions of non-valence quarks to the 
$D$-term form factors are rather significant. Therefore, 
these contributions should be considered in estimating 
the flavor-decomposed $D$-term form factors as they play 
an essential role alongside valence quarks.

\begin{figure}[htp]
\centering
\includegraphics[scale=0.147]{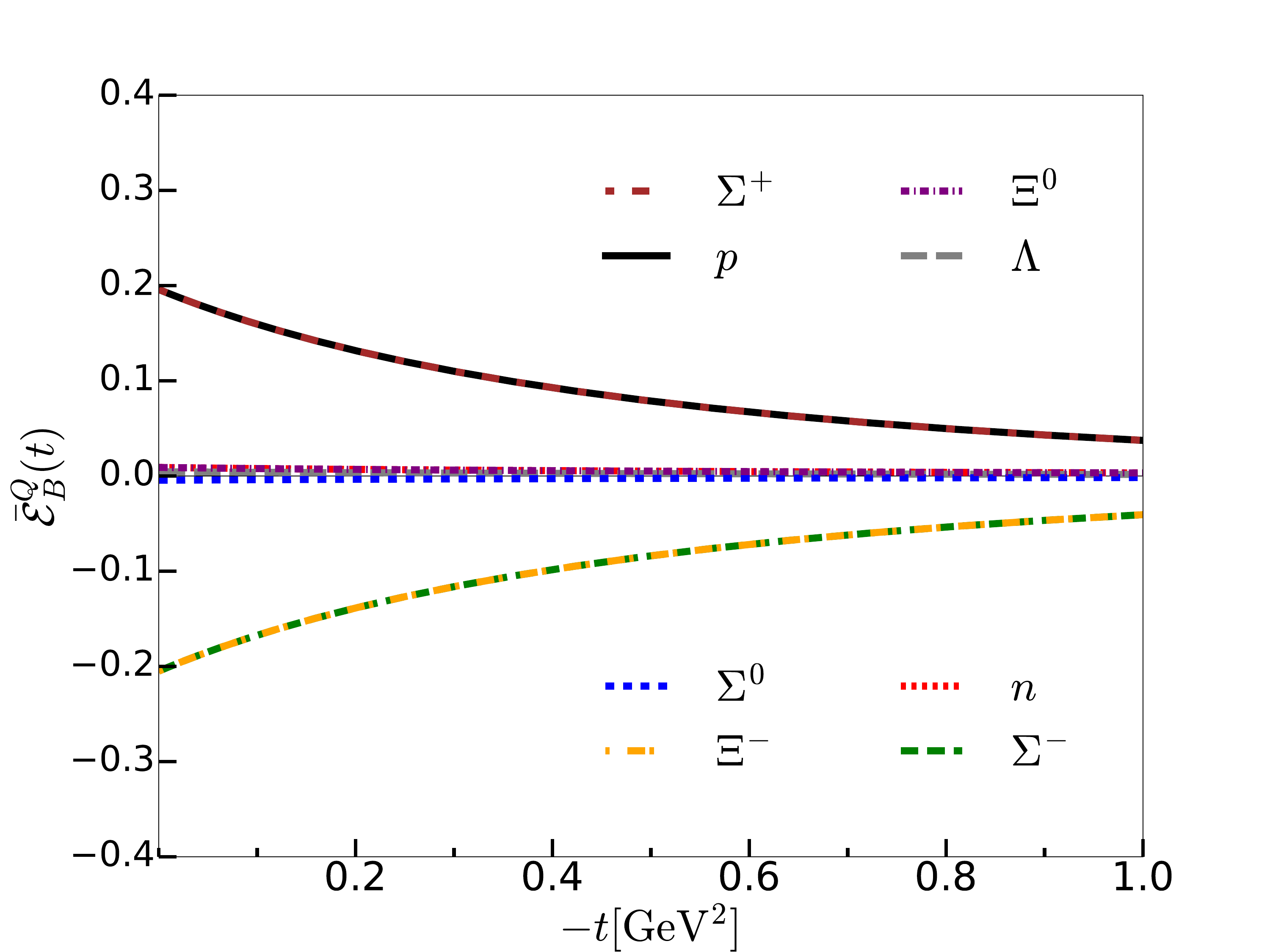}
\includegraphics[scale=0.147]{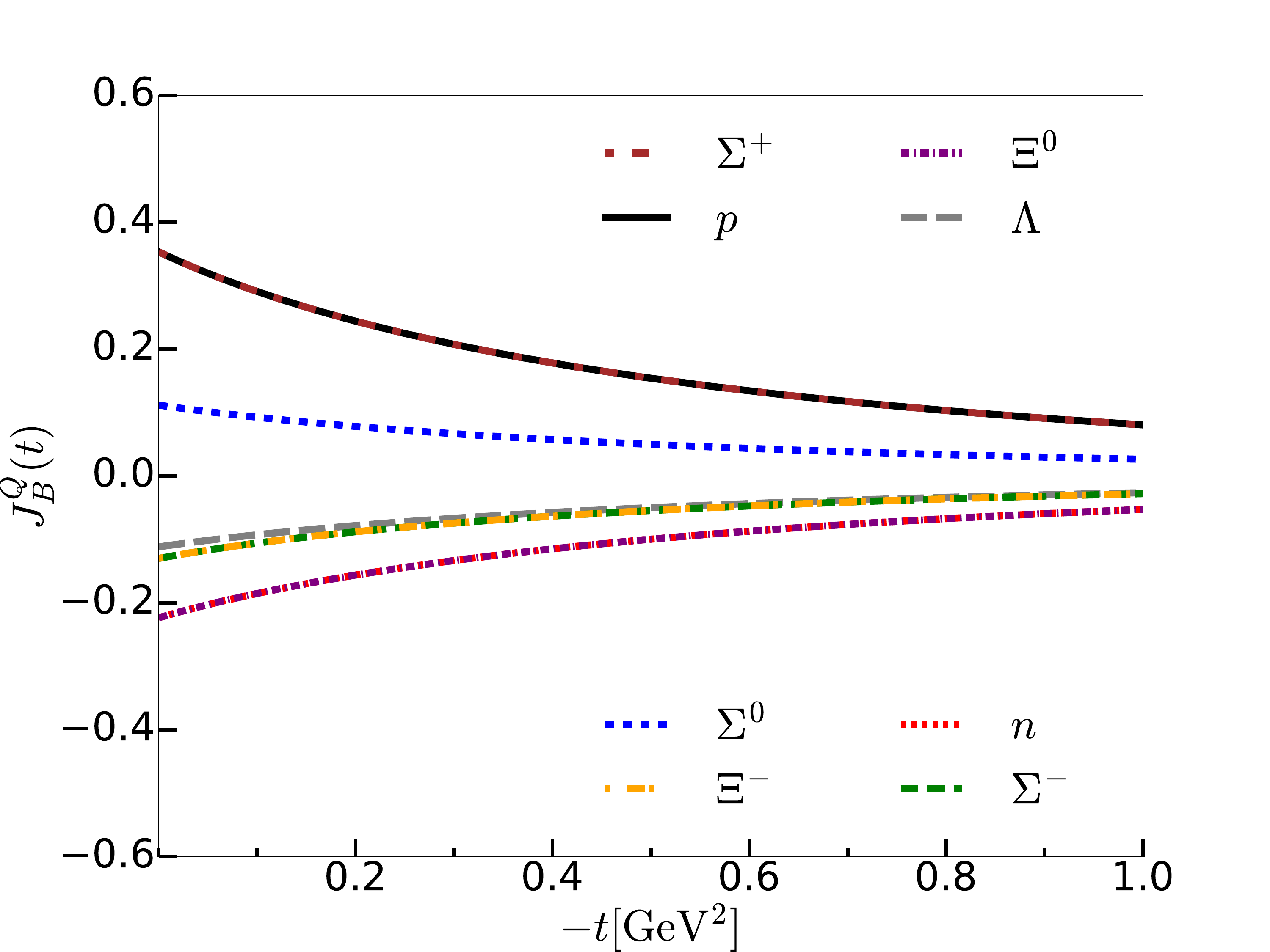}
\includegraphics[scale=0.147]{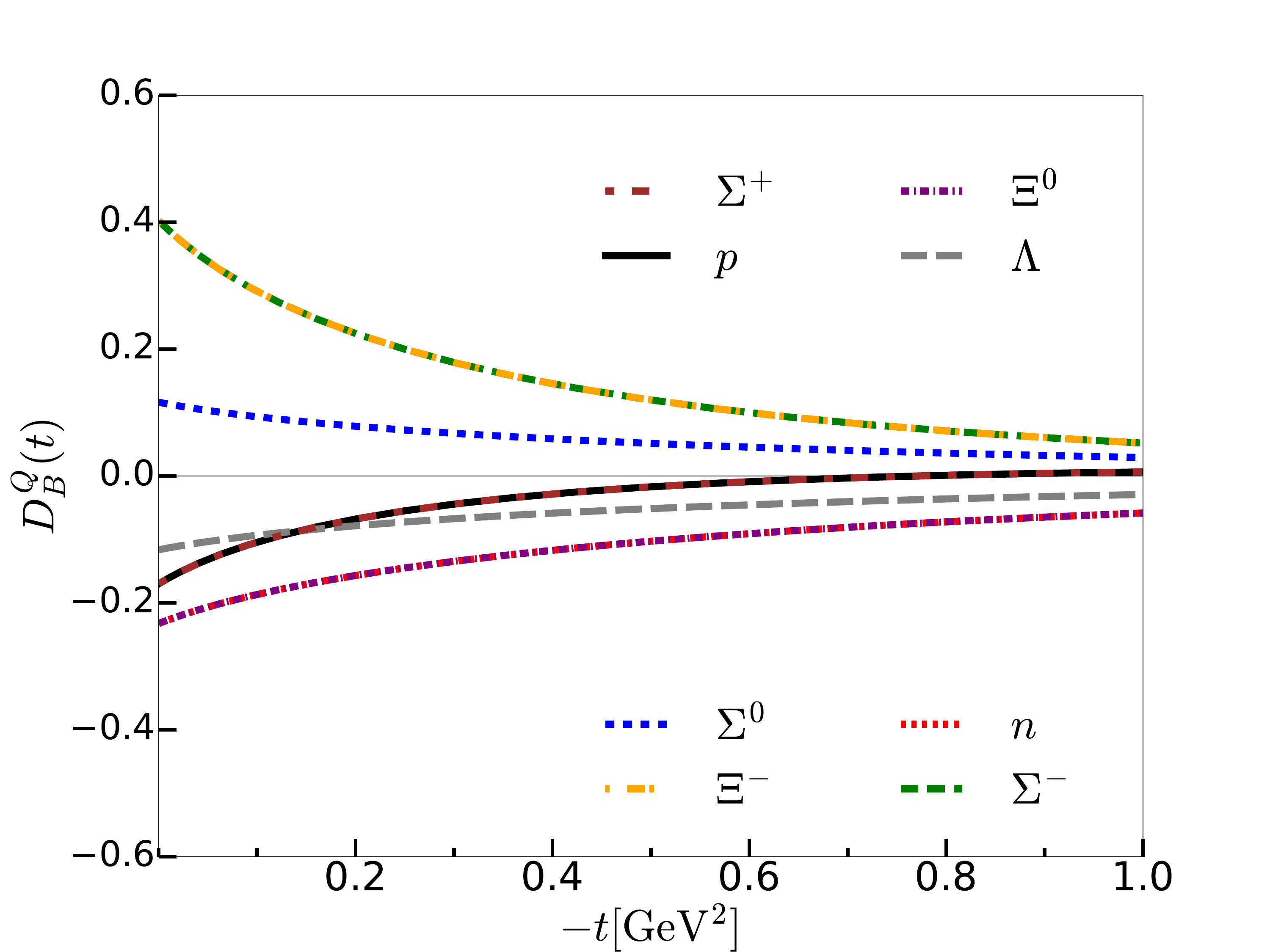}
  \caption{Generlized electromagnetic form factor for the octet
    baryons are drawn.}   
  \label{fig:10}
\end{figure}

Finally, we turn our attention to the generalized electromagnetic form
factors~(GEMFFs). The first Mellin moments of the GPDs are directly
related to the EMFFs. By retaining the flavor structure of the EMFFs,  
we can derive the GEMFFs through the second Mellin moments of the GPDs. 
The flavor structure of the electromagnetic current is given by the
matrix: 
\begin{align}
Q = \left( \begin{array}{c c c} 
\frac{2}{3} & 0 & 0 \\
0 & -\frac{1}{3} & 0 \\
1 & 0 & -\frac{1}{3} \\
\end{array} \right) = \frac{1}{2} \left(\lambda^{3} + \frac{1}{\sqrt{3}} \lambda^{8}\right),
\end{align}
where $\lambda^{3}$ and $\lambda^{8}$ are Gell-Mann matrices. 
By inserting this flavor operator into the equation 
governing the electromagnetic current, Eq.~\eqref{eq:EMT_current}, 
we obtain the GEMFFs. These GEMFFs can be expressed as 
a linear combination of the flavor-triplet and -octet GFFs, 
namely $F^{Q}_{B}= \frac{1}{2} \left(F^{3}_{B} +
  \frac{1}{\sqrt{3}}F^{8}_{B}\right)$. Similar to the EMFFs, the
GEMFFs satisfy the $U$-spin symmetry. This symmetry implies that the
GEMFFs for baryons with the same charge, except for the $\Lambda^{0}$
and $\Sigma^{0}$ baryons, are equivalent when the flavor SU(3)
symmetry is imposed. In other words, we have:
\begin{align}
    F^{Q}_{p}(t)
  = F^{Q}_{\Sigma^{+}}(t), \quad 
    F^{Q}_{n}(t)
  = F^{Q}_{\Xi^{0}}(t), \quad 
    F^{Q}_{\Sigma^{-}}(t) 
  = F^{Q}_{\Xi^{-}}(t), \quad 
    F^{Q}_{\Lambda^{0}}(t)
  = - F^{Q}_{\Sigma^{0}}(t).
\end{align}
This $U$-spin symmetry is observed numerically in Fig.~\ref{fig:10}.
The observed $U$-spin symmetry aligns with the spin-flavor relations  
found in the analysis of the separate flavor-singlet, -triplet, and
-octet GFFs. 

\section{Conclusions and summary \label{sec:5}}

In the current work, we focused on investigating the flavor-decomposed
gravitational form factors~(GFFs) for the nucleon and hyperons, and
their mechanical interpretations.

Once we perform the flavor decomposition of the GFFs and
distributions, the higher-twist form factor ($\bar{c}$) comes into
play in the mechanical interpretation and plays an essential
role. However, extracting the higher-twist form factor is challenging
in the various dynamical models and lattice QCD. Therefore, we isolate
this twist-4 part by performing the twist projection. We then obtain
the twist-projected EMT current and define the corresponding EMT
distributions.  

Using the chiral quark-soliton model derived from the QCD instanton
vacuum, we estimate the GFFs and the EMT distributions. Before
discussing the numerical results, we have highlighted two important
features:  
\begin{itemize}
  \item In the large $N_{c}$ limit of QCD, the collective motion of
      the chiral soliton is nonrelativistic, while the internal
      dynamics remains fully relativistic. Therefore, we naturally
      adopt three-dimensional (3D) mechanical interpretations of 
the gravitational form factors. 

\item In the effective chiral theory, the flavor-decomposed EMT
  current must be derived from the QCD operator which is consistent
  with the effective quark-gluon dynamics. In any models, such as NJL
  and bag models, once the effective action is given, the
  flavor-singlet EMT current can be derived from the global
  symmetry. However, there is no relevant global symmetry for the
  flavor triplet and octet currents. Thus, one must derive
  the corresponding effective operators from the QCD operator.
 This can be done by replacing the gluon fields by the effective quark
 fields through the QCD instanton vacuum.  
\end{itemize}
By adopting the 3D mechanical interpretation and the twist-2 effective
operator derived from the QCD instanton vacuum, we estimated the
twist-2 GFFs and understood the role of the strange quark in the
mechanics of the proton. 

We summarize the numerical results obtained in the current work:
Firstly, we obtained the flavor-decomposed twist-2 mass distribution
in the rest frame, which is influenced by the $A$. While the 
flavor-singlet component was properly normalized to unity, i.e.,
$\sum_{q} A^{q}_{p}(0) =1$, no constraints were imposed on the
flavor-triplet and -octet components. Our findings revealed that the
light-front momentum fraction carried by up, down, and strange quarks
in the proton was estimated to be 59\%, 35\%, and 6\%.  

Secondly, regarding the AM distribution, we observed that the $J$
form factor was appropriately normalized to the proton spin, i.e,
$\sum_{q} J^{q}(0)=1/2$. We determined the fraction of the proton spin
carried by up, down, and strange quarks, which were found to be
$J^{u}_{p}=0.52$, $J^{u}_{p}=-0.06$, and $J^{s}_{p}=0.04$,
respectively. Similar to the mass form factor, the strange quark
contributed minimally to the proton AM. For the flavor-singlet AM,   
it can be decomposed into OAM and intrinsic quark
spin. Each carries a half of the nucleon spin within the current
framework. However, in this work, we focused on the total AM
instead of the flavor-decomposed OAM due to ambiguities in matching
twist-3 QCD operators with the effective operators we employed.  

Thirdly, the mechanical properties of the proton were investigated. The
stress tensor was parameterized in terms of the twist-2 pressure and
shear-force distributions, which were obtained through
three-dimensional Fourier transforms of the pressure $\mathcal{P}$ and
$D$-term form factors. Notably, we observed that $D^{u-d}_{p} \sim 0$
in the flavor SU(3) sector: 
\begin{align}
  D^{u-d}_{p} \sim 0.3,  \quad [\mathrm{SU}(2)] \quad \mathrm{vs.} \quad
  D^{u-d}_{p} \sim 0, \quad [\mathrm{SU}(3)]. 
\end{align}
This suggests that the large $N_{c}$ assumption ($D^{u-d}\sim 0$) used
in extracting the $D$-term from the DVCS data is more
appropriate for the flavor SU(3) sector, rather than for the flavor
SU(2). Moreover, in the flavor SU(3), we have determined the
significant strange quark contributions to the $D$-term form factor,
i.e., $D^{s}_{p}=-0.44$. 

Lastly, we explored the GFFs for hyperons. We observed the interesting
spin-flavor symmetries and introduced the electromagnetic
flavor structure into the GFFs, resulting in the observation of
$U$-spin symmetries in the generalized electromagnetic form factors. 

\section{Acknowledgement}
Authors want to express gratitude to 
C\'edric Lorc\'e for the invaluable comments and discussions. JYK is
grateful to C. Weiss and J.L. Goity for discussions on the large
$N_{c}$ behavior of the GFFs and the effective operator formalism. The
work was supported by the Basic Science Research Program through the
National Research Foundation of Korea funded by the Korean government
(Ministry of Education, Science and Technology, MEST),
Grant-No. 2021R1A2C2093368 and 2018R1A5A1025563 (HYW and HChK).  
This work was also supported by the U.S.~Department of Energy, Office
of Science, Office of Nuclear Physics under contract
DE-AC05-06OR23177~(JYK) and by the France Excellence scholarship through Campus 
France funded by the French government 
(Minist\`ere de l’Europe et des Aﬀaires \'Etrang\`eres), 141295X (HYW).

\appendix

\section{EMT distributions and regularization functions \label{app:A}}
We provide explicit expressions for the EMT distributions. These
distributions are compiled below. We have mass distributions: 
\begin{align}
    \mathcal{M} ( \bm{r} )   
& =\frac{3}{4} N_{c}\Bigg{[} E_{v}
    \psi_{v}^{\dagger}  ( \bm{r}  ) \psi_{v}  ( \bm{r}  )
  + \sum_{n} \psi_{n}^{\dagger}  ( \bm{r}  ) \psi_{n}  ( \bm{r}  ) R_{0n} \Bigg{]}, \cr
    \mathcal{J}_{1} ( \bm{r}  )
& = \frac{3}{16}N_{c} \Bigg[  \sum_{n\neq v}  \frac{E_{n}+E_{v}}{E_{n}-E_{v}}
    \mel{n}{\tau_{3}}{v}  \psi_{v}^{\dagger}  ( \bm{r}  ) \tau_{3}  \psi_{n}  ( \bm{r}  ) \cr
&   \hspace{1cm}
  + \frac{1}{2} \sum_{n,m} ( E_{n} + E_{m} )
    \mel{n}{\tau_{3}}{m}  \psi_{m}^{\dagger}  ( \bm{r}  ) \tau_{3}  \psi_{n}  ( \bm{r}  ) R_{3nm}  \Bigg],  \cr
    \mathcal{J}_{2}  ( \bm{r}  )
& = \frac{3}{32}N_{c}  \Bigg{[} \sum_{n^{0}} \frac{E_{n^{0}}+E_{v}}{E_{n^{0}}-E_{v}}
    \braket{n^{0}}{v} \psi_{v}^{\dagger}  ( \bm{r}  ) \psi_{n^{0}}  ( \bm{r}  ) \cr
&   \hspace{1cm} 
  + \sum_{n^{0},m} ( E_{n^{0}}+E_{m} )
    \braket{n^{0}}{m} \psi_{m}^{\dagger}  ( \bm{r}  ) \psi_{n^{0}}  ( \bm{r}  ) R_{3 n^{0}m}  \Bigg{]}, 
\end{align}
and the AM distributions:
\begin{align}
&\hspace{-1cm}\mathcal{Q}_{0} ( \bm{r}  ) 
 = \frac{N_{c}}{4} \Bigg[ \psi_{v}^{\dagger}  ( \bm{r}  ) \Gamma^{J}_{vv3} \tau_{3}  \psi_{v}  ( \bm{r}  ) - \frac{1}{2} \sum_{n} \mathrm{sign} (E_{n}) \psi_{n}^{\dagger}  ( \bm{r}  ) \Gamma^{J}_{nn3} \tau_{3}  \psi_{n}  ( \bm{r}  )  \Bigg]  , \cr
&\hspace{-1cm}\mathcal{Q}_{1} ( \bm{r}  ) 
 = \frac{N_{c}}{4} i f_{ij3} \Bigg[  \sum_{n\neq v}  \frac{\mathrm{sign}(E_{n})}{E_{n}-E_{v}}  \mel{n}{\tau_{i}}{v}
  \psi_{v}^{\dagger}  ( \bm{r}  ) \tau_{j}  \Gamma^{J}_{vn3} \psi_{n}  ( \bm{r}  ) \cr
  &\hspace{1.3cm} + \frac{1}{2} \sum_{n,m} \mel{n}{\tau_{i}}{m}
  \psi_{m}^{\dagger}  ( \bm{r}  ) \tau_{j}  \Gamma^{J}_{mn3} \psi_{n}  ( \bm{r}  ) R_{6nm}  \Bigg], \cr
&\hspace{-1cm}\mathcal{I}_{1} ( \bm{r}  ) 
 = \frac{N_{c}}{4} \Bigg[  \sum_{n\neq v}  \frac{\mel{n}{\tau_{3}}{v}}{E_{n}-E_{v}} 
    \psi_{v}^{\dagger}  ( \bm{r}  ) \Gamma^{J}_{vn3} \psi_{n}  ( \bm{r}  )   + \frac{1}{2} \sum_{n,m}
  \mel{n}{\tau_{3}}{m}  \psi_{m}^{\dagger}  ( \bm{r}  ) \Gamma^{J}_{mn3} \psi_{n}  ( \bm{r}  ) R_{3nm}  \Bigg]  ,  \cr
&\hspace{-1cm}\mathcal{I}_{2}(\bm{r}) 
 = \frac{N_{c}}{4} \Bigg[  \sum_{n^{0}} \frac{\braket{n^{0}}{v}}{E_{n^{0}}-E_{v}}  \psi_{v}^{\dagger}  ( \bm{r}  )  \tau_{3} \Gamma^{J}_{vn^{0}3}  \psi_{n^{0}}  ( \bm{r}  ) \cr
 &\hspace{0.8cm}+ \sum_{n^{0},m} \braket{n^{0}}{m} \psi_{m}^{\dagger}  ( \bm{r}  )  \tau_{3} \Gamma^{J}_{mn^{0}3}  \psi_{n^{0}}  ( \bm{r}  ) R_{3n^{0}m}  \Bigg],
\end{align}
where $\Gamma^{J}_{3}(E_{n},E_{m})=\Gamma^{J}_{nm3} = \left(  2 \hat{L}_{3} + ( E_{n} + E_{m} ) \gamma_{5}  ( \bm{r}  \times  \bm{\sigma}  )_{3} \right)$ with $\hat{\bm{L}}=   \left[ \bm{r} \times \frac{i}{2} ( \overleftarrow{\bm{\nabla}}- \overrightarrow{\bm{\nabla}}) \right]$.
The quadrupole distributions $s(r)$ relevant for the $D$-term form
factors are given by 
\begin{align}
&\hspace{-1cm}\mathcal{N}_{1}(\bm{r})=
\frac{3}{2} N_{c}\left[\psi_{v}^{\dagger}(\bm{r})\Gamma^{s}\psi_{v}(\bm{r})+\sum_{n}\psi_{n}^{\dagger}(\bm{r})\Gamma^{s}\psi_{n}(\bm{r})R_{1n}\right]  , \cr
&\hspace{-1cm}\mathcal{J}_{3}(\bm{r})=
\frac{3}{4} N_{c}\left[\sum_{n\neq v}\frac{\mel{n}{\tau_{3}}{v}}{E_{n}-E_{v}}
\psi_{v}^{\dagger}(\bm{r})\tau_{3}\Gamma^{s}
\psi_{n}(\bm{r})+\frac{1}{2}\sum_{n,m}\mel{n}{\tau_{3}}{m}
\psi_{m}^{\dagger}(\bm{r})\tau_{3}\Gamma^{s}
\psi_{n}(\bm{r})R_{5nm}\right],  \cr
&\hspace{-1cm}\mathcal{J}_{4}(\bm{r})=
\frac{3}{8} N_{c}\left[\sum_{n^{0}}\frac{ \braket{n^{0}}{v} }{E_{n^{0}}-E_{v}}
\psi_{v}^{\dagger}(\bm{r})\Gamma^{s}
\psi_{n^{0}}(\bm{r}) + \sum_{n^{0},m}\braket{n^{0}}{m}
\psi_{m}^{\dagger}(\bm{r})\Gamma^{s}
\psi_{n^{0}}(\bm{r})R_{5n^{0}m}\right],
\end{align}
where $\Gamma^{s} =
\gamma^{0}\left(\bm{\hat{n}}\cdot\bm{p}\right) -
\frac{1}{3}\gamma^{0}\left(\bm{\gamma}\cdot\bm{p})\right) 
$. The moments of inertia $I_{1}$ and $I_{2}$ are written as follows:  
\begin{align}
     I_{1}
& = \frac{N_{c}}{2}
    \left[
    \sum_{n\neq v}
    \frac{1}{E_{n}-E_{v}}
    \mel{n}{\tau_{3}}{v}
    \mel{v}{\tau_{3}}{n} 
  + \frac{1}{2}\sum_{\substack{n,m}}
    \mel{n}{\tau_{3}}{m}
    \mel{m}{\tau_{3}}{n}
    R_{3nm}
    \right],\cr
    I_{2}
& = \frac{N_{c}}{4} 
    \left[  
    \sum_{n^{0}} 
    \frac{1}{E_{n^{0}}-E_{v}}  
    \braket{n^{0}}{v}
    \braket{v}{n^{0}} 
  + \sum_{n^{0},m} 
    \braket{n^{0}}{m} 
    \braket{m}{n^{0}}
    R_{3n^{0}m} 
    \right].
\label{eq:dynamical_para}
\end{align}
In addition, all distributions are regularized, and their
regularization functions are written as  
\begin{align}
  R_{0}(E_{n}) & := R_{0n} = \frac{1}{4\sqrt{\pi}} \int_{\Lambda^{-2}}  \frac{du}{u^{3/2}}  e^{-uE_{n}^{2}}, \cr
  R_{1}(E_{n})& := R_{1n}=-\frac{E_{n}}{2\sqrt{\pi}}\int_{\Lambda^{-2}}  \frac{du}{\sqrt{u}} e^{-uE_{n}^{2}},\cr
  R_{3}(E_{n},E_{m})& := R_{3nm} = \frac{1}{2\sqrt{\pi}}\int_{\Lambda^{-2}} \frac{du}{\sqrt{u}} 
  \bigg[\frac{1}{u}\frac{e^{-uE_{n}^{2}}-e^{-uE_{m}^{2}}}{E_{m}^{2}-E_{n}^{2}}
  -\frac{E_{n}e^{-uE_{n}^{2}}+E_{m}e^{-uE_{m}^{2}}}{E_{n}+E_{m}}\bigg],\cr
  R_{5}(E_{n},E_{m})& := R_{5nm}=\frac{1}{2}\frac{\mathrm{sign}(E_{n})-\mathrm{sign}(E_{m})}{E_{n}-E_{m}},\cr
  R_{6}(E_{n},E_{m})& := R_{6nm}=\frac{1-\mathrm{sign}(E_{n})\mathrm{sign}(E_{m})}{E_{n}-E_{m}},
\end{align}
with $\psi_{v}(\bm{r}):=\langle \bm{r}| v \rangle$ and $\psi_{n}(\bm{r}):=\langle \bm{r}| n \rangle$. 

\section{Matrix elements of the spin-flavor operators \label{app:B}}
In Appendix~\ref{app:B} we list the matrix elements of the
spin-flavor operators relevant to $T^{00}$ and $T^{ij}$ in
Table~\ref{tab:0}, and those relevant to $T^{0k}$ in Table~\ref{tab:01}. 
  \begin{table}[htb] 
    \centering
    \caption{The matrix elements of the spin-flavor operators relevant
      to $T^{00}$ and $T^{ij}$ are listed.}   
    \begin{center}
      \renewcommand{\arraystretch}{1.7}
    \scalebox{1}{%
    \begin{tabular}{rrrrrrrrr} 
      \hline
      \hline
      $B$        & $Y$   & $T$            & $D_{38}$                    & $D_{88}$        & $D_{3i}J_{i}$        & $D_{8i}J_{i}          $ & $D_{3a}J_{a}$        & $D_{8a}J_{a}$  \\
      \hline
      $N$        & $1$   & $\frac{1}{2}$  & $\frac{\sqrt{3}}{15}T_{3}$  & $\frac{3}{10}$  & $-\frac{7}{10}T_{3}$ & $-\frac{\sqrt{3}}{20} $ & $-\frac{1}{5}T_{3}$  & $-\frac{3\sqrt{3}}{10}$  \\
      $\Lambda$  & $0$   & $0$            & $0$                         & $\frac{1}{10}$  & $0$                  & $\frac{3\sqrt{3}}{20} $ & $0$                  & $-\frac{\sqrt{3}}{10}$  \\
      $\Sigma$   & $0$   & $1$            & $\frac{\sqrt{3}}{6}T_{3}$   & $-\frac{1}{10}$ & $-\frac{1}{4}T_{3}$  & $-\frac{3\sqrt{3}}{20}$ & $-\frac{1}{2}T_{3}$  & $\frac{\sqrt{3}}{10}$  \\
      $\Xi$      & $-1$  & $\frac{1}{2}$  & $\frac{4\sqrt{3}}{15}T_{3}$ & $-\frac{1}{5}$  & $\frac{1}{5}T_{3}$   & $\frac{\sqrt{3}}{5}   $ & $-\frac{4}{5}T_{3}$  & $\frac{\sqrt{3}}{5}$  \\
      \hline
      \hline
    \end{tabular}}
    \end{center}
    \label{tab:0}
    \end{table}
    \begin{table}[htb] 
\centering
\caption{The matrix elements of the spin-flavor operators relevant to
  $T^{0k}$ are listed.}  
\begin{center}
  \renewcommand{\arraystretch}{1.7}
\scalebox{1}{%
\begin{tabular}{rrrrrrrrrrrr} 
  \hline
  \hline
  $B$        & $Y$   & $T$            & $D_{33}$                    & $D_{83}$                    & $D_{38}J_{3}$ & $D_{88}J_{3}$     & $d_{ab3}D_{3a}J_{b}$  & $d_{ab3}D_{8a}J_{b}$          \\
  \hline
  $N$        & $1$   & $\frac{1}{2}$  & $-\frac{14}{15}T_{3}J_{3}$  & $-\frac{\sqrt{3}}{15}J_{3}$ & $\frac{\sqrt{3}}{15}T_{3}J_{3}$   & $\frac{3}{10}J_{3}$   & $\frac{7}{15}T_{3}J_{3}$  & $\frac{\sqrt{3}}{30}J_{3}$    \\
  $\Lambda$  & $0$   & $0$            & $0$                         & $\frac{\sqrt{3}}{5}J_{3}$   & $0$                               & $\frac{1}{10}J_{3}$   & $0$                       & $-\frac{\sqrt{3}}{10}J_{3}$   \\
  $\Sigma$   & $0$   & $1$            & $-\frac{1}{3}T_{3}J_{3}$    & $-\frac{\sqrt{3}}{5}J_{3}$  & $\frac{\sqrt{3}}{6}T_{3}J_{3}$    & $-\frac{1}{10}J_{3}$  & $\frac{1}{6}T_{3}J_{3}$   & $\frac{\sqrt{3}}{10}J_{3}$    \\
  $\Xi$      & $-1$  & $\frac{1}{2}$  & $\frac{4}{15}T_{3}J_{3}$    & $\frac{4\sqrt{3}}{15}J_{3}$ & $\frac{4\sqrt{3}}{15}T_{3}J_{3}$  & $-\frac{1}{5}J_{3}$   & $-\frac{2}{15}T_{3}J_{3}$ & $-\frac{2\sqrt{3}}{15}J_{3}$  \\
  \hline
  \hline
\end{tabular}}
\end{center}
\label{tab:01}
\end{table}

\newpage
\bibliographystyle{JHEP}
\bibliography{FlavorSU3EMT}

\end{document}